\documentstyle[epsf]{article}

\newcounter{subequation}[equation]

\makeatletter

\def\@aabuffer{}
\def\author #1{\expandafter\def\expandafter\@aabuffer\expandafter
{\@aabuffer \small\rm      #1\relax \par}}
\def\address#1{\expandafter\def\expandafter\@aabuffer\expandafter
{\@aabuffer \small\it #1\relax \par\vspace{1em}}}

\def\maketitle{
\begin{center}
   {\bf \@title \par}
   \vskip 2em                      
   \@aabuffer\relax
\end{center} \par
\gdef\@aabuffer{}
}

\def\abstracts#1{
\begin{center}
{\begin{minipage}{4.2truein}
                 \footnotesize
                 \parindent=0pt #1\par
                 \end{minipage}}\end{center}
                 \vskip 2em \par}

\fussy
\def\section{\@startsection {section}{1}{\z@}{-3.25ex plus -1ex minus
    -.2ex}{1.5ex plus .2ex}{\bf }}
\def\subsection{\@startsection{subsection}{2}{\z@}{-3.25ex plus -1ex minus
    -.2ex}{1.5ex plus .2ex}{\it }}

\def\thefootnote{\alph{footnote}}

\renewcommand{\@makefntext}[1]{\noindent{\@makefnmark}#1} 

\renewenvironment{thebibliography}[1]
	{\begin{list}{\arabic{enumi}.}
	{\usecounter{enumi}\setlength{\parsep}{0pt}
	 \setlength{\itemsep}{0pt}
         \settowidth
	{\labelwidth}{#1.}\sloppy}}{\end{list}}

\newcounter{arabiclistc}

\def\@citex[#1]#2{\if@filesw\immediate\write\@auxout
	{\string\citation{#2}}\fi
\def\@citea{}\@cite{\@for\@citeb:=#2\do
	{\@citea\def\@citea{,}\@ifundefined
	{b@\@citeb}{{\bf ?}\@warning
	{Citation `\@citeb' on page \thepage \space undefined}}
	{\csname b@\@citeb\endcsname}}}{#1}}

\newif\if@cghi
\def\cite{\@cghitrue\@ifnextchar [{\@tempswatrue
	\@citex}{\@tempswafalse\@citex[]}}
\def\citelow{\@cghifalse\@ifnextchar [{\@tempswatrue
	\@citex}{\@tempswafalse\@citex[]}}
\def\@cite#1#2{{$\!^{#1}$\if@tempswa\typeout
	{IJCGA warning: optional citation argument
	ignored: `#2'} \fi}}

\def\baselinestretch{1.0}
\ifx\selectfont\undefined
\@normalsize\else\let\glb@currsize=\relax\selectfont
\fi

\ifx\selectfont\undefined
\def\@singlespacing{%
\def\baselinestretch{1}\ifx\@currsize\normalsize\@normalsize\else\@currsize\fi%
}
\else
\def\@singlespacing{\def\baselinestretch{1}\let\glb@currsize=\relax\selectfont}
\fi

\long\def\@makecaption#1#2{
   \vskip 10pt
   \setbox\@tempboxa\hbox{\footnotesize #1: #2}
 \ifdim \wd\@tempboxa >\hsize \footnotesize #1: #2\par \else \hbox
 to\hsize{\hfil\box\@tempboxa\hfil}
   \fi}

\def\thesubequation{\theequation\@alph\c@subequation}
\def\@subeqnnum{{\rm (\thesubequation)}}
\def\slabel#1{\@bsphack\if@filesw {\let\thepage\relax
   \xdef\@gtempa{\write\@auxout{\string
      \newlabel{#1}{{\thesubequation}{\thepage}}}}}\@gtempa
   \if@nobreak \ifvmode\nobreak\fi\fi\fi\@esphack}
\def\subeqnarray{\stepcounter{equation}
\let\@currentlabel=\theequation\global\c@subequation\@ne
\global\@eqnswtrue
\global\@eqcnt\z@\tabskip\@centering\let\\=\@subeqncr
$$\halign to \displaywidth\bgroup\@eqnsel\hskip\@centering
  $\displaystyle\tabskip\z@{##}$&\global\@eqcnt\@ne
  \hskip 2\arraycolsep \hfil${##}$\hfil
  &\global\@eqcnt\tw@ \hskip 2\arraycolsep
  $\displaystyle\tabskip\z@{##}$\hfil
   \tabskip\@centering&\llap{##}\tabskip\z@\cr}
\def\endsubeqnarray{\@@subeqncr\egroup
                     $$\global\@ignoretrue}
\def\@subeqncr{{\ifnum0=`}\fi\@ifstar{\global\@eqpen\@M
    \@ysubeqncr}{\global\@eqpen\interdisplaylinepenalty \@ysubeqncr}}
\def\@ysubeqncr{\@ifnextchar [{\@xsubeqncr}{\@xsubeqncr[\z@]}}
\def\@xsubeqncr[#1]{\ifnum0=`{\fi}\@@subeqncr
   \noalign{\penalty\@eqpen\vskip\jot\vskip #1\relax}}
\def\@@subeqncr{\let\@tempa\relax
    \ifcase\@eqcnt \def\@tempa{& & &}\or \def\@tempa{& &}
      \else \def\@tempa{&}\fi
     \@tempa \if@eqnsw\@subeqnnum\refstepcounter{subequation}\fi
     \global\@eqnswtrue\global\@eqcnt\z@\cr}
\let\@ssubeqncr=\@subeqncr
\@namedef{subeqnarray*}{\def\@subeqncr{\nonumber\@ssubeqncr}\subeqnarray}
\@namedef{endsubeqnarray*}{\global\advance\c@equation\m@ne%
                           \nonumber\endsubeqnarray}

\@addtoreset{equation}{section}
\renewcommand{\theequation}{\thesection.\arabic{equation}}

\makeatother

\def\Journal#1#2#3#4{{#1} {\bf #2}, #3 (#4)}

\def\IJMPA{{\em Int. J. Mod. Phys.} A}
\def\NPB{{\em Nucl. Phys.} B}
\def\PLB{{\em Phys. Lett.}  B}
\def\PRL{\em Phys. Rev. Lett.}
\def\PRD{{\em Phys. Rev.} D}
\def\CQG{\em Class. Quantum Grav.}

\newcommand{\bm}{\bibitem}
\newcommand{\be}{\begin{equation}} 
\newcommand{\ee}{\end{equation}} 
\newcommand{\ba}{\begin{array}}
\newcommand{\ea}{\end{array}}
\newcommand{\bea}{\begin{eqnarray}} 
\newcommand{\eea}{\end{eqnarray}} 
\newcommand{\bsea}{\begin{subeqnarray}} 
\newcommand{\esea}{\end{subeqnarray}}
\newcommand{\nn}{\nonumber}

\def\ft#1#2{{\textstyle{{\scriptstyle #1}\over {\scriptstyle #2}}}}
\def\fft#1#2{{#1 \over #2}}
\def\bmath#1{\mbox{\boldmath{$#1$}}}
\def\st{\scriptstyle}
\def\sst{\scriptscriptstyle}
\def\gtlt{\mathrel{\raise4.5pt\hbox{\oalign{$\scriptstyle>$\crcr
$\scriptstyle<$}}}}
\def\ffrac#1#2{\leavevmode\kern.1em
\raise.3ex\hbox{\the\scriptfont0 #1}\kern-.1em/\kern-.15em
\lower.2ex\hbox{\the\scriptfont0 #2}}
\def\sffrac#1#2{\leavevmode\kern.1em
\raise.3ex\hbox{\the\scriptscriptfont0 #1}\kern-.1em/\kern-.15em
\lower.2ex\hbox{\the\scriptscriptfont0 #2}}
\def\gtlt{\mathrel{\raise4.5pt\hbox{\oalign{$\scriptstyle>$\crcr
$\scriptstyle<$}}}}
\def\cramp{\medmuskip = 2mu plus 1mu minus 2mu}

\def\crampest{\medmuskip = 1mu plus 1mu minus 1mu}
\def\uncramp{\medmuskip = 4mu plus 2mu minus 4mu}
\def\isomorphic{\mathrel{\raise4pt\hbox{\oalign{$\scriptstyle\sim$\crcr
$\scriptstyle=$}}}}
\def\R{\rlap{\rm I}\mkern3mu{\rm R}}
\def\C{\mkern1mu\raise2.2pt\hbox{$\scriptscriptstyle|$}\mkern-7mu{\rm C}}
\def\E{\rlap{\rm I}\mkern3mu{\rm E}}
\def\Z{\rlap{\sf Z}\mkern3mu{\sf Z}}
\def\for{\lower3pt\hbox{$\st|$}}
\def\oneone{\rlap 1\mkern4mu{\rm l}}

\def\ie{{\it i.e.\ }}
\def\eg{{\it e.g.\ }}

\def\im{{\rm i}}
\def\phs{\phantom{\Big]}}
\def\dalemb#1#2{{\vbox{\hrule height .#2pt
        \hbox{\vrule width.#2pt height#1pt \kern#1 pt
                \vrule width.#2 pt}
        \hrule height.#2 pt}}}
\def\square{\mathord{\dalemb{6.2}{6}\hbox{\hskip1pt}}}

\begin{document}
\renewcommand{\thefootnote}{\fnsymbol{footnote}}
\null
\vspace{-3.5cm}
\begin{flushright}
\hfill{CERN-TH/98-80}\\
\hfill{Imperial/TP/97-98/30}\\
\hfill{\tt hep-th/9803116}
\end{flushright}
\vspace{.5cm}

\title{BPS BRANES IN SUPERGRAVITY\,\footnote[1]{Based upon lectures given
at the 1996 and 1997 ICTP Summer Schools in High Energy Physics and Cosmology.}}

\author{K.S. STELLE}

\address{The Blackett Laboratory, Imperial College,\\
Prince Consort Road, London SW7 2BZ, UK\\
and\\
TH Division, CERN \\
CH-1211 Geneva 23, Switzerland}

\maketitle

\begin{center}
DEDICATED TO THE MEMORY OF ABDUS SALAM
\end{center}
\vspace{.35cm}

\abstracts{
This review considers the properties of classical solutions to
supergravity theories with partially unbroken supersymmetry. These solutions
saturate Bogomol'ny-Prasad-Sommerfield bounds on their energy densities and are
the carriers of the $p$-form charges that appear in the supersymmetry algebra.
The simplest such solutions have the character of $(p+1)$-dimensional
Poincar\'e-invariant hyperplanes in spacetime, \ie $p$-branes. Topics covered
include the relations between mass densities, charge densities and the
preservation of unbroken supersymmetry; interpolating-soliton structure; diagonal
and vertical Kaluza-Klein reduction families; multiple-charge solutions and the
four $D=11$ elements; duality-symmetry multiplets; charge quantisation;
low-velocity  scattering and the geometry of worldvolume supersymmetric
$\sigma$-models; and the target-space geometry of BPS instanton solutions
obtained by the  dimensional reduction of static $p$-branes.}
\renewcommand{\thefootnote}{\alph{footnote}}

\vspace{-.8cm}
\tableofcontents
\addtocontents{toc}{\protect\setcounter{tocdepth}{2}}
\vfill\eject

\section{Introduction}\label{sec:intro}

     Let us begin from the bosonic sector of $D=11$ supergravity,\cite{cjs}
\be
I_{11} = \int d^{11}x\left\{\sqrt{-g}(R-\ft1{48}F_{[4]}^2) + \ft16
F_{[4]}\wedge F_{[4]}\wedge A_{[3]}\right\}\ .\label{D11act}
\ee
In addition to the metric, one has a 3-form antisymmetric-tensor
gauge potential $A_{[3]}$ with a gauge transformation $\delta
A_{[3]}=d\Lambda_{[2]}$ and a field strength $F_{[4]}=dA_{[3]}$.
The third term in the Lagrangian is invariant under the
$A_{[3]}$ gauge transformation only up to a total derivative, so the action
(\ref{D11act}) is invariant under gauge transformations that
are continuously connected to the identity. This term is required, with the
coefficient given in (\ref{D11act}), by the $D=11$ local supersymmetry that is
required of the theory when the gravitino-dependent sector is included.

     The equation of motion for the $A_{[3]}$ gauge potential is
\be
d{\,}^\ast\!F_{[4]}+\ft12F_{[4]}\wedge F_{[4]}=0\ ;\label{Aeqmot}
\ee
this equation of motion gives rise to the conservation of an
``electric'' type charge~\cite{dp}
\be
U=\int_{\partial{\cal M}_8}({}^\ast\!F_{[4]} + \ft12 A_{[3]}\wedge
F_{[4]})\ ,\label{electric}
\ee
where the integral of the 7-form integrand is over the boundary at
infinity of an arbitrary infinite spacelike 8-dimensional 
subspace of $D=11$ spacetime. Another conserved charge relies on the Bianchi
identity $dF_{[4]}=0$ for its conservation,
\be
V=\int_{\partial\widetilde{\cal M}_5}F_{[4]}\ ,\label{magnetic}
\ee
where the surface integral is now taken over the boundary at infinity of a
spacelike 5-dimensional subspace.

     Charges such as (\ref{electric},\,\ref{magnetic}) can occur on the
right-hand side of the supersymmetry algebra,\footnote{Although formally reasonable, there is
admittedly something strange about this algebra. For objects such as black
holes, the total momentum terms on the right-hand side have a
well-defined meaning, but for extended objects such as
$p$-branes, the $U$ and $V$ terms on the right-hand side have
meaning only as intensive quantities taken per spatial unit worldvolume.
This forces a similar intensive interpretation also for the momentum, requiring
it to be considered as a momentum per spatial unit worldvolume. Clearly, a more
careful treatment of this subject would recognize a corresponding divergence in
the $[Q,Q]$ anticommutator on the left-hand side of (\ref{susyalg}) in such
cases. This would then require then an infinite normalization factor for the
algebra, whose removal requires the right-hand side to be reinterpreted in an
intensive (i.e.\ per spatial unit worldvolume) as opposed to an
extensive way.}\,\cite{vanhvanpr}
\be
\{Q,Q\} = C(\Gamma^{\sst A}P_{\sst A} + \Gamma^{\sst AB}U_{\sst AB} +
\Gamma^{\sst ABCDE}V_{\sst ABCDE})\ ,\label{susyalg}
\ee
where $C$ is the charge conjugation matrix, $P_{\sst A}$ is the
energy-momentum 11-vector and $U_{\sst AB}$ and $V_{\sst ABCDE}$ are
2-form and 5-form charges that we shall find to be related to the charges $U$
and $V$ (\ref{electric},\,\ref{magnetic}) above. Note that since the supercharge
$Q$ in $D=11$ supergravity is a 32-component Majorana spinor, the LHS of
(\ref{susyalg}) has 528 components. The symmetric spinor matrices
$C\Gamma^{\sst A}$, $C\Gamma^{\sst AB}$ and $C\Gamma^{\sst ABCDE}$ on the RHS of
(\ref{susyalg}) also have a total of 528 independent components: 11 for
the momentum $P_{\sst A}$, 55 for the ``electric'' charge $U_{\sst
AB}$ and 462 for the ``magnetic'' charge $V_{\sst ABCDE}$.

     Now the question arises as to the relation between the charges $U$
and $V$ in (\ref{electric},\,\ref{magnetic}) and the 2-form and 5-form
charges appearing in (\ref{susyalg}). One thing that immediately stands
out is that the Gauss' law integration surfaces in
(\ref{electric},\,\ref{magnetic}) are the boundaries of integration
volumes ${\cal M}_8$, $\widetilde{\cal M}_5$ that do not fill out a whole
10-dimensional spacelike hypersurface in spacetime, unlike the more familiar
situation for charges in ordinary electrodynamics. A rough idea about the
origin of the index structures on $U_{\sst AB}$ and $V_{\sst ABCDE}$ may be
guessed from the 2-fold and 5-fold ways that the corresponding 8 and 5
dimensional integration volumes may be embedded into a 10-dimensional spacelike
hypersurface. We shall see in Section \ref{sec:masschargesusy} that this is too
na\"\i ve, however: it masks an important topological aspect of both the
electric charge $U_{\sst AB}$ and the magnetic charge $V_{\sst ABCDE}$.
The fact that the integration volume does not fill out a full spacelike
hypersurface does not impede the conservation of the charges
(\ref{electric},\,\ref{magnetic}); this only requires that no electric or
magnetic currents are present at the boundaries
$\partial{\cal M}_8$, $\partial\widetilde{\cal M}_5$. Before we can discuss
such currents, we shall need to consider in some detail the supergravity
solutions that carry charges like (\ref{electric},\,\ref{magnetic}). The simplest
of these have the structure of $p+1$-dimensional Poincar\'e-invariant
hyperplanes in the supergravity spacetime, and hence have been termed
``$p$-branes'' (see, \eg Ref.\,\cite{2bror1}). In Sections \ref{sec:pbraneans}
and \ref{sec:examples}, we shall delve in some detail into the properties of
these solutions.

     Let us recall at this point some features of the relationship between 
supergravity theory and string theory. Supergravity theories originally arose
from the desire to include supersymmetry into the framework of gravitational
models, and this was in the hope that the resulting models might solve some of
the outstanding difficulties of quantum gravity. One of these difficulties was
the ultraviolet problem, on which early enthusiasm for supergravity's promise
gave way to disenchantment when it became clear that local supersymmetry is not
in fact sufficient to tame the notorious ultraviolet divergences
that arise in perturbation theory.\footnote{For a review of  ultraviolet
behavior in supergravity theories, see Ref.\,\cite{hs}} Nonetheless, supergravity
theories won much admiration for their beautiful mathematical structure, which
is due to the stringent constraints of their symmetries. These severely
restrict the possible terms that can occur in the Lagrangian. For the maximal
supergravity theories, such as those descended from the $D=11$ theory
(\ref{D11act}), there is simultaneously a great wealth of fields present and at
the same time an impossibility of coupling any independent external
field-theoretic ``matter.'' It was only occasionally noticed in this early
period that this impossibility of coupling to matter fields does not, however,
rule out coupling to ``relativistic objects'' such as black holes, strings and
membranes. 

     The realisation that supergravity theories do not by themselves constitute
acceptable starting points for a quantum theory of gravity came somewhat before
the realisation sunk in that string theory might instead be the sought-after
perturbative foundation for quantum gravity. But the approaches of supergravity
and of string theory are in fact strongly interrelated: supergravity
theories arise as long-wavelength effective-field-theory limits of string
theories. To see how this happens, consider the $\sigma$-model action~\cite{ft}
that describes a bosonic string moving in a background ``condensate'' of its own
massless modes ($g_{\sst MN}$, $A_{\sst MN}$, $\phi$):
\bea
\lefteqn{I={1\over 4\pi\alpha'}\int
d^2z\sqrt\gamma\,[\gamma^{ij}\partial_i x^{\sst M}\partial_j x^{\sst N}
g_{\sst MN}(x)}\hspace{2cm}\nonumber\\
&&+ \im\epsilon^{ij}\partial_ix^{\sst M}\partial_jx^{\sst
N}A_{\sst MN}(x) +
\alpha'R(\gamma)\phi(x)]\ .\label{isigma}
\eea
Every string theory contains a sector described by fields
($g_{\sst MN}$, $A_{\sst MN}$, $\phi$); these are the only fields that
couple directly to the string worldsheet. In superstring theories,
this sector is called the Neveu-Schwarz/Neveu-Schwarz (NS--NS) sector.

     The $\sigma$-model action (\ref{isigma}) is classically
invariant under the worldsheet Weyl symmetry
$\gamma_{ij}\rightarrow\Lambda^2(z)\gamma_{ij}$. Requiring cancellation
of the anomalies in this symmetry at the quantum level gives
differential-equation restrictions on the background fields
($g_{\sst MN}$, $A_{\sst MN}$, $\phi$) that may be viewed as effective
equations of motion for these massless modes.\cite{cfmp} This system of
effective equations may be summarized by the corresponding field-theory
effective action
\bea
\lefteqn{I_{\rm eff}=\int
d^Dx\sqrt{-g}e^{-2\phi}\Big[(D-26)-\ft32\alpha'(R+4\nabla^2\phi
-4(\nabla\phi)^2}\hspace{4cm}\nonumber\\
&&-\ft1{12}F_{\sst MNP}F^{\sst MNP}+{\cal O}(\alpha')^2\Big]\ ,\label{ieff}
\eea
where
$F_{\sst MNP} =\partial_{\sst M}A_{\sst NP} +
\partial_{\sst N}A_{\sst PM} + \partial_{\sst P}A_{\sst MN}$ is the
3-form field strength for the $A_{\sst MN}$ gauge potential. The $(D-26)$
term reflects the critical dimension for the bosonic string: flat space is
a solution of the above effective theory only for
$D=26$. The effective action for the superstring theories that we shall
consider in this review contains a similar (NS--NS) sector, but with the
substitution of $(D-26)$ by $(D-10)$, reflecting the different critical
dimension for superstrings.

     The effective action (\ref{ieff}) is written in the form
directly obtained from string $\sigma$-model calculations. It is not
written in the form generally preferred by relativists, which has a
clean Einstein-Hilbert term free from exponential prefactors like
$e^{-2\phi}$. One may rewrite the effective action in a different
{\em frame} by making a Weyl-rescaling field redefinition
$g_{\sst MN}\rightarrow e^{\lambda\phi}g_{\sst MN}$. $I_{\rm eff}$ as
written in (\ref{ieff}) is in the {\em string frame}; after an
integration by parts, it takes the form, specialising now to $D=10$,
\be
I^{\rm \,string}=\int d^{10}x\sqrt{-g^{{\rm (s)}}}e^{-2\phi}\Big[R(g^{{\rm
(s)}})+4\nabla_{\sst M}\phi\nabla^{\sst
M}\phi-\ft1{12}F_{\sst MNP}F^{\sst MNP}\Big]\ .\label{istring}
\ee
After making the transformation
\be
g^{\rm (e)}_{\sst MN}=e^{-\phi/2}g^{\rm (s)}_{\sst MN}\ ,\label{weyltransf}
\ee
one obtains the {\em Einstein frame} action,
\be
I^{\rm \,Einstein}=\int d^{10}x\sqrt{-g^{\rm (e)}}\Big[R(g^{\rm (e)}) -
\ft12\nabla_{\sst M}
\phi\nabla^{\sst M}\phi-\ft1{12}e^{-\phi}F_{\sst MNP}F^{\sst MNP}\Big]\ ,
\label{ieinstein}
\ee
where the indices are now raised and lowered with $g^{\rm (e)}_{\sst
MN}$. To understand how this Weyl rescaling works, note that under
$x$-independent rescalings, the connection $\Gamma_{\sst MN}{}^{\sst
P}$ is invariant. This carries over also to terms with $\phi$
undifferentiated, which emerge from the $e^{\lambda\phi}$ Weyl
transformation. One then chooses $\lambda$ so as to eliminate the
$e^{-2\phi}$ factor. Terms with $\phi$ undifferentiated do change,
however. As one can see in (\ref{ieinstein}), the Weyl transformation is
just what is needed to unmask the positive-energy sign of the kinetic term 
for the $\phi$ field, despite the apparently negative sign of its kinetic
term in $I^{{\rm \,string}}$.

     Now let us return to the maximal supergravities descended from
(\ref{D11act}). We shall discuss in Section \ref{sec:kkred}) the process of
Kaluza-Klein dimensional reduction that relates theories in different
dimensions of spacetime. For the present, we note that upon specifying the
Kaluza-Klein ansatz expressing $ds_{11}^2$ in terms of $ds_{10}^2$,
the Kaluza-Klein vector ${\cal A}_{\sst M}$ and the dilaton $\phi$,
\be
ds_{11}^2 = e^{-\phi/6}ds_{10}^2 + e^{4\phi/3}(dz+{\cal A}_{\sst M}dx^{\sst
M})^2\qquad {\st M}=0,1,\ldots,9,\label{11to10metric}
\ee
the bosonic $D=11$ action (\ref{D11act}) reduces to the Einstein-frame type IIA
bosonic action~\cite{IIA}
\bea
I_{\rm IIA}^{\rm Einstein} &=& \int d^{10}x\sqrt{-g^{\rm (e)}}
\Big\{\big[R(g^{\rm (e)}) -
\ft12\nabla_{\sst M}
\phi\nabla^{\sst M}\phi-\ft1{12}e^{-\phi}F_{\sst MNP}F^{\sst MNP}\big]\nn\\
&& - \fft1{48}e^{\phi/2}F_{\sst MNPQ}F^{MNPQ} - \fft14e^{3\phi/2}{\cal
F}_{\sst MN}{\cal F}^{\sst MN}\Big\} + {\cal L}_{FFA}\ ,\label{einsteinIIA}
\eea
where ${\cal F}_{\sst MN}$ is the field strength for the Kaluza-Klein vector
${\cal A}_{\sst M}$. 

     The top line in (\ref{einsteinIIA}) corresponds to the
NS--NS sector of the IIA theory; the bottom line corresponds the R--R sector
(plus the Chern-Simons terms, which we have not shown explicitly). In order to
understand better the distinction between these two sectors,  rewrite
(\ref{einsteinIIA}) in string frame using (\ref{weyltransf}). One finds
\bea
I_{\rm IIA}^{\rm string} &=& \int d^{10}x\sqrt{-g^{\rm (s)}}
\Big\{e^{-2\phi}\big[R(g^{\rm (e)}) + 4\nabla_{\sst M}
\phi\nabla^{\sst M}\phi-\ft1{12}F_{\sst MNP}F^{\sst MNP}\big]\nn\\
&& - \fft1{48}F_{\sst MNPQ}F^{\sst MNPQ} - \fft14{\cal F}_{\sst MN}{\cal F}^{\sst
MN}\Big\} + {\cal L}_{FFA}\ .\label{stringIIA}
\eea
Now one may see the distinguishing feature of the NS--NS sector as opposed to
the R--R sector: the dilaton coupling is a uniform $e^{-2\phi}$ in the NS--NS
sector, and it does not couple (in string frame) to the R--R sector field
strengths. Comparing with the familiar $g^{-2}$ coupling-constant factor for the
Yang-Mills action, one sees that the asymptotic value $e^{\phi_\infty}$ plays
the r\^ole of the string-theory coupling constant. Since in
classical supergravity theory, one will encounter transformations that have the
effect of flipping the sign of the dilaton, $\phi\rightarrow-\phi$, the study
of classical supergravity will contain decidedly non-perturbative information
about string theory. In particular, this will arise in the study of $p$-brane
solitons, to which we shall shortly turn.

     In this review, we shall mostly consider the descendants of the type IIA
action (\ref{einsteinIIA}). This leaves out one important case that we shall
have to consider separately: the chiral type IIB theory in $D=10$. In the type
IIB theory,\cite{IIB} one has $F_{[1]}=d\chi$, where $\chi$
is a R--R zero-form ({\it i.e.}\  a pseudoscalar field), $F^{\rm
R}_{[3]}=dA^{\rm R}_{[2]}$, a second 3-form field strength making a pair
together with $F^{\rm NS}_{[3]}$ from the NS--NS sector, and
$F_{[5]}=dA_{[4]}$, which is a {\em self-dual} 5-form in $D=10$,
$F_{[5]}={}^\ast\!F_{[5]}$.

     Thus one naturally encounters field strengths of ranks 1--5 in
the supergravity theories deriving from superstring theories. In addition, one
may use $\epsilon_{[10]}$ to dualize certain field strengths; \eg the
original $F_{[3]}$ may be dualized to the 7-form ${}^\ast\!F_{[7]}$. The
upshot is that antisymmetric-tensor gauge field strengths of diverse ranks need
to be taken into account when searching for solutions to string-theory
effective field equations. These field strengths will play an essential r\^ole
in supporting the $p$-brane solutions that we shall now describe.

\section{The $p$-brane ansatz}\label{sec:pbraneans}
\subsection{Single-charge action and field equations}\label{ssec:genact}

     We have seen that one needs to consider effective theories containing
gravity, various ranks of antisymmetric-tensor field strengths and various
scalars. To obtain a more tractable system to study, we shall make a {\em
consistent truncation} of the action down to a simple system in $D$ dimensions
comprising the metric $g_{\sst MN}$, a scalar field $\phi$ and a single
$(n-1)$-form gauge potential $A_{[n-1]}$ with corresponding field strength
$F_{[n]}$; the whole is described by the action
\be
I=\int D^Dx\sqrt{-g}\Big[R-\ft12\nabla_{\sst M}\phi\nabla^{\sst M}\phi
-\ft{1}{2n!}e^{a\phi}F^2_{[n]}\Big]\ .\label{igen}
\ee
We shall consider later in more detail how (\ref{igen}) may be obtained by
a consistent truncation from a full supergravity theory in $D$ dimensions.
The notion of a consistent truncation will play a central r\^ole in our
discussion of the BPS solutions of supergravity theories. A consistent
truncation is one for which solutions of the truncated theory are also
perfectly good, albeit specific, solutions of the original untruncated theory.
Truncation down to the system (\ref{igen}) with a single scalar $\phi$ and a
single field strength $F_{[n]}$ will be consistent except for certain special
cases when $n=D/2$ that we shall have to consider separately. In such cases,
one can have {\em dyonic} solutions, and in such cases it will generally be
necessary to retain an axionic scalar $\chi$ as well. Note that in (\ref{igen})
we have not included contributions coming from the $FFA$ Chern-Simons term in
the action. These are also consistently excluded in the truncation to the
single-charge action (\ref{igen}). The value of the important parameter $a$
controlling the interaction of the scalar field $\phi$ with the field strength
$F_{[n]}$ in (\ref{igen}) will vary according to the cases considered in the
following.

     Varying the action (\ref{igen}) produces the following
set of equations of motion:
\bsea
R_{MN} &=& \ft12\partial_{\sst M}\phi\partial_{\sst N}\phi +
S_{\sst MN}\\
S_{MN} &=& {1\over
2(n-1)!}e^{a\phi}(F_{{\sst M}\cdots}F_{\sst N}{}^{\cdots} -{n-1\over
n(D-2)}F^2 g_{\sst MN})\\
\makebox[0pt]{\hspace{2.5cm}$\nabla_{{\sst M}_1}(e^{a\phi}F^{{\sst
M}_1\cdots {\sst M}_n}) ~~ = ~~ 0$}\\
\square\phi &=& {a\over 2n!}e^{a\phi}F^2\ .\label{eqmots}
\esea

\subsection{Electric and magnetic ans\"atze}\label{ssec:ansatze}

     In order to solve the above equations, we shall make a simplifying
ansatz. We shall be looking for solutions preserving certain unbroken
supersymmetries, and these will in turn require unbroken
translational symmetries as well. For simplicity, we shall also require
isotropic symmetry in the directions ``transverse'' to the
translationally-symmetric ones. These restrictions can subsequently be
relaxed in generalizations of the basic class of $p$-brane solutions that
we shall discuss here. For this basic class of solutions, we make an
ansatz requiring $\hbox{ (Poincar\'e)}_d\times {\rm SO}(D-d)$ symmetry.
One may view the sought-for solutions as flat $d=p+1$ dimensional
hyperplanes embedded in the ambient $D$-dimensional spacetime; these
hyperplanes may in turn be viewed as the histories, or worldvolumes, of
$p$-dimensional spatial surfaces. Accordingly, let the spacetime
coordinates be split into two ranges: $x^{\sst M}=(x^\mu,y^m)$, where
$x^\mu$ ($\mu=0,1,\cdots,p=d-1$) are coordinates adapted to the
$\hbox{(Poincar\'e)}_d$ isometries on the worldvolume and where $y^m$
($m=d,\cdots,D-1$) are the coordinates ``transverse'' to the
worldvolume.

     An ansatz for the spacetime metric that respects the $\hbox{
(Poincar\'e)}_d\times {\rm SO}(D-d)$ symmetry is~\cite{dghrr}
\bea
&&ds^2 = e^{2A(r)}dx^\mu dx^\nu\eta_{\mu\nu} +
e^{2B(r)}dy^mdy^n\delta_{mn}\nn\\
&&\mu = 0,1,\ldots,p\qquad m=p+1,\ldots,D-1\ ,\label{ansatz}
\eea
where $r=\sqrt{y^my^m}$ is the isotropic radial coordinate in the
transverse space. Since the metric components depend only on $r$,
translational invariance in the worldvolume directions $x^\mu$ and ${\rm
SO}(D-d)$ symmetry in the transverse directions $y^m$ is guaranteed.

     The corresponding ansatz for the scalar field $\phi(x^M)$ is
simply $\phi=\phi(r)$.

     For the antisymmetric tensor gauge field, we face a
bifurcation of possibilities for the ansatz, the two possibilities
being related by duality. The first possibility is naturally expressed
directly in terms of the gauge potential $A_{[n-1]}$. Just as the
Maxwell 1-form naturally couples to the worldline of a charged
particle, so does $A_{[n-1]}$ naturally couple to the worldvolume of
a $p=d-1=(n-1)-1$ dimensional ``charged'' extended object. The
``charge'' here will be obtained from Gauss'-law surface
integrals involving $F_{[n]}$, as we shall see later. Thus, the first
possibility for $A_{[n-1]}$ is to support a $d_{\rm el}=n-1$
dimensional worldvolume. This is what we shall call the ``elementary,'' or
``electric'' ansatz:
\be
A_{\mu_1\cdots\mu_{n-1}} = \epsilon_{\mu_1\cdots\mu_{n-1}}e^{C(r)}
\ ,\hspace{.5cm}\mbox{others zero.}\label{elans}
\ee
${\rm SO}(D-d)$ isotropicity and $\hbox{(Poincar\'e)}_d$
symmetry are guaranteed here because the function $C(r)$ depends only on
the transverse radial coordinate $r$. Instead of the ansatz (\ref{elans}),
expressed in terms of
$A_{[n-1]}$, we could equivalently have given just the $F_{[n]}$ field
strength:
\be
F^{\rm(el)}_{m\mu_1\cdots\mu_{n-1}} =
\epsilon_{\mu_1\cdots\mu_{n-1}}\partial_me^{C(r)}\ ,
\hspace{.5cm}\mbox{others zero.}\label{felans}
\ee
The worldvolume dimension for the elementary ansatz
(\ref{elans},\,\ref{felans}) is clearly $d_{\rm el}=n-1$.

     The second possible way to relate the rank $n$ of $F_{[n]}$ to the
worldvolume dimension $d$ of an extended object is suggested by considering
the dualized field strength ${}^\ast\!F$, which is a $(D-n)$ form. If one
were to find an underlying gauge potential for ${}^\ast\!F$ (locally
possible by courtesy of a Bianchi identity), this would naturally couple
to a $d_{\rm so}=D-n-1$ dimensional worldvolume. Since such a dualized
potential would be nonlocally related to the fields appearing in the
action (\ref{igen}), we shall not explicitly follow this
construction, but shall instead take this reference to the
dualized theory as an easy way to identify the worldvolume dimension for
the second type of ansatz. This ``solitonic'' or ``magnetic'' ansatz for
the antisymmetric tensor field is most conveniently expressed in terms
of the field strength $F_{[n]}$, which now has nonvanishing values only
for indices corresponding to the transverse directions:
\be
F^{\rm (mag)}_{m_1\cdots m_n} = \lambda\epsilon_{m_1\cdots
m_np}{y^p\over r^{n+1}}\ ,\hspace{.5cm}
\mbox{others zero,}\label{magans}
\ee
where the magnetic-charge parameter $\lambda$ is a constant of
integration, the only thing left undetermined by this ansatz. The
power of $r$ in the solitonic/mag\-netic ansatz is determined by requiring
$F_{[n]}$ to satisfy the Bianchi identity.\footnote{Specifically, one
finds $\partial_qF_{m_1\cdots m_n} = r^{-(n+1)}\big(\epsilon_{m_1\cdots
m_nq}-(n+1)\epsilon_{m_1\cdots m_np}y^py_q/r^2\big)$; upon taking the
totally antisymmetrised combination $[qm_1\cdots m_n]$, the factor of
$(n+1)$ is evened out between the two terms and then one finds from
cycling a factor $\sum_my^my_m=r^2$, thus obtaining cancellation.} Note
that the worldvolume dimensions of the elementary and solitonic cases are
related by $d_{\rm so}= \tilde d_{\rm el}\equiv D-d_{\rm el}-2$; note
also that this relation is idempotent, {\it i.e.}\ $\widetilde{(\tilde
d)}=d$.

\subsection{Curvature components and $p$-brane
equations}\label{ssec:pbraneqs}

In order to write out the field equations after insertion of the
above ans\"atze, one needs to compute the Ricci tensor for the
metric.\cite{stainless} This is most easily done by introducing vielbeins,
{\it i.e.,} orthonormal frames,\cite{mtw} with tangent-space indices
denoted by underlined indices:
\be
g_{\sst MN} = e_{\sst M}{}^{\sst
\underline E}e_{\sst N}{}^{\sst\underline F}
\eta_{\sst\underline E\,\underline F}\ .\label{vielbs}
\ee
Next, one constructs the corresponding 1-forms:
$e^{\sst\underline E}=dx^{\sst M}e_{\sst M}{}^{\sst\underline E}$.
Splitting up the tangent-space indices
${\st\underline E}=(\underline\mu,\underline m)$ similarly to the
world indices ${\st M}=(\mu,m)$, we have for our ans\"atze the vielbein
1-forms
\be
e^{\underline\mu}=e^{A(r)}dx^\mu\ ,\hspace{2cm}
e^{\underline m}=e^{B(r)}dy^m\ .\label{1forms}
\ee

     The corresponding spin connection 1-forms are determined by the
condition that the torsion vanishes,
$de^{\sst\underline
E}+\omega^{\sst\underline E}{}_{\sst\underline F}\wedge
e^{\sst\underline F}=0$, which yields
\bea
\omega^{\underline\mu\,\underline\nu}&=&0\ ,\hspace{2cm}
\omega^{\underline\mu\,\underline n}
=e^{-B(r)}\partial_nA(r)e^{\underline\mu}\nonumber\\
\omega^{\underline m\,\underline n}
&=& e^{-B(r)}\partial_nB(r)e^{\underline m}\ 
-\ e^{-B(r)}\partial_mB(r)e^{\underline n}\ .\label{conn1forms}
\eea
The curvature 2-forms are then given by
\be
R_{[2]}^{\sst\underline E\,\underline F} =
d\omega^{\sst\underline E\,\underline F} +
\omega^{\sst\underline E\,\underline D}\wedge
\omega_{\sst\underline D}{}^{\sst\underline F}\ .\label{curv2f}
\ee
{}From the curvature components so obtained, one finds the Ricci tensor
components
\bea
R_{\mu\nu} &=& -\eta_{\mu\nu}e^{2(A-B)}(A''+d(A')^2+\tilde dA'B' +
{(\tilde d+1)\over r}A')\nonumber\\
R_{mn} &=& -\delta_{mn}(B''+dA'B'+\tilde d(B')^2+{(2\tilde d+1)\over
r}B' + {d\over r}A')\label{riccicomps}\\
&&-{y^my^n\over r^2}(\tilde d
B''+dA''-2dA'B'+d(A')^2-\tilde d(B')^2-{\tilde d\over r}B'-{d\over
r}A')\ ,\nonumber
\eea
where again, $\tilde d=D-d-2$, and the primes indicate
$\partial/\partial r$ derivatives.

     Substituting the above relations, one finds the set of equations
that we need to solve to obtain the metric and $\phi$:
\crampest
\be
\begin{array}{rclr}
\hspace{1cm}A''+d(A')^2+\tilde dA'B'+{(\tilde d+1)\over r}A'
&\makebox[0pt]{=}& {\tilde d\over 2(D-2)}S^2
&\phs\hspace{1.25cm}\makebox[0pt]{$\{\mu\nu\}$}\\
\hspace{1cm}B''+dA'B'+\tilde d(B')^2+{(2\tilde d+1)\over r}B'+{d\over r}A'
&\makebox[0pt]{=}& -{d\over
2(D-2)}S^2&\phs\hspace{1cm}\makebox[0pt]{$\{\delta_{mn}\}$}\\
\hspace{1.25cm}\tilde d B'' + dA''-2dA'B'+d(A')^2-\tilde d(B')^2\\
-{\tilde d\over r}B'-{d\over r}A'+\ft12(\phi')^2 &\makebox[0pt]{=}&
\ft12S^2&\phs\hspace{1.25cm}\makebox[0pt]{$\{y_my_n\}$}\\
\hspace{1cm}\phi''+dA'\phi'+\tilde dB'\phi'+{(\tilde d+1)\over r}\phi'
&\makebox[0pt]{=}& -\ft12\varsigma
aS^2&\phs\hspace{1.25cm}\makebox[0pt]{$\{\phi\}$}
\end{array}\label{pbraneqs}
\ee\uncramp
where $\varsigma=\pm 1$ for the elementary/solitonic cases and the source
appearing on the RHS of these equations is
\be
S = \left\{\begin{array}{ll}
(e^{\fft12 a\phi-dA+C})C'\hspace{1cm} &\mbox{electric: $d=n-1$,
$\varsigma=+1$}\\
\lambda(e^{\fft12 a\phi-\tilde dB})r^{-\tilde d-1}&\mbox{magnetic:
$d=D-n-1$, $\varsigma=-1$.}
\end{array}\right.\label{pbranesource}
\ee

\subsection{$p$-brane solutions}\label{ssec:pbranesols}

     The $p$-brane equations (\ref{pbraneqs},\,\ref{pbranesource}) are
still rather daunting. Before we embark on solving these equations, let us first
note a generalisation. Although Eqs (\ref{pbraneqs}) have been specifically
written for an isotropic $p$-brane ansatz, one may recognise more general
possibilities by noting the form of the Laplace operator, which for isotropic
scalar functions of $r$ is
\be
\nabla^2\phi = \phi'' + (\tilde d+1)r^{-1}\phi'\ .\label{laplacian}
\ee
We shall see later that more general solutions of the Laplace equation than the
simple isotropic ones considered here will also play important r\^oles in the
story.

     In order to reduce the complexity of Eqs (\ref{pbraneqs}), we shall refine
the $p$-brane ansatz (\ref{ansatz},\,\ref{felans},\,\ref{magans}) by looking ahead
a bit and taking a hint from the requirements for supersymmetry preservation,
which shall be justified in more detail later on in Section
\ref{sec:masschargesusy}. Accordingly, we shall look for solutions satisfying
the linearity condition
\be
dA'+\tilde d B' = 0\ .\label{dadtb}
\ee
After eliminating $B$ using (\ref{dadtb}), the independent equations
become~\cite{dilatonic}
\begin{subeqnarray}
\nabla^2\phi &=& -\ft12\varsigma aS^2\label{symeqsa}\\
\nabla^2A &=& {\tilde d\over 2(D-2)}S^2\label{symeqsb}\\
d(D-2)(A')^2 + \ft12\tilde d(\phi')^2 &=& \ft12\tilde dS^2
\ ,\label{symeqsc}
\end{subeqnarray}
where, for spherically-symmetric ({\it i.e.}\ isotropic) functions
in the transverse $(D-d)$ dimensions, the Laplacian is $\nabla^2\phi
= \phi'' + (\tilde d+1)r^{-1}\phi'$. 

     Equations (\ref{symeqsa}a,b) suggest that we now further refine
the ans\"atze by imposing another linearity condition:
\be
\phi' = {-\varsigma a(D-2)\over\tilde d}A'\ .\label{phirel}
\ee
At this stage, it is useful to introduce a new piece of notation,
letting
\be
a^2 = \Delta-{2d\tilde d\over(D-2)}\ .\label{Delta}
\ee
With this notation, equation (\ref{symeqsc}c) gives
\be
S^2 = {\Delta(\phi')^2\over a^2}\ ,\label{s2}
\ee
so that the remaining equation for $\phi$ becomes $\nabla^2\phi +
{\varsigma\Delta\over2a}(\phi')^2 = 0$, which can be re-expressed as a
Laplace equation,\footnote{Note that Eq.\ (\ref{laplace}) can also be
more generally derived; for example, it still holds if one relaxes the
assumption of isotropicity in the transverse space.}
\be
\nabla^2e^{{\varsigma\Delta\over2a}\phi} = 0\ .\label{laplace}
\ee
Solving this in the transverse $(D-d)$ dimensions with our assumption of
transverse isotropicity ({\it i.e.}\ spherical symmetry) yields
\be
e^{{\varsigma\Delta\over 2a}\phi} \equiv H(y) = 1 + {k\over r^{\tilde d}}
\hspace{1.5cm}k>0\ ,\label{phisol}
\ee
where the constant of integration $\phi\for_{r\rightarrow\infty}$ has
been set equal to zero here for simplicity: $\phi_\infty = 0$. The
integration constant $k$ in (\ref{phisol}) sets the mass scale of the
solution; it has been taken to be positive in order to ensure the absence
of naked singularities at finite $r$. This positivity restriction is
similar to the usual restriction to a positive mass parameter $M$ in the
standard Schwarzschild solution.

     In the case of the elementary/electric ansatz, with
$\varsigma=+1$, it still remains to find the function $C(r)$ that
determines the antisymmetric-tensor gauge field potential. In this
case, it follows from (\ref{pbranesource}) that
$S^2=e^{a\phi-2dA}(C'e^C)^2$. Combining this with (\ref{s2}), one
finds the relation
\be
{\partial\over\partial r}(e^C)={-\sqrt\Delta\over a}e^{-\fft12 a\phi
+ dA}\phi'\label{crel}
\ee
(where it should be remembered that $a<0$). Finally, it is
straightforward to verify that the relation (\ref{crel}) is consistent 
with the equation of motion for $F_{[n]}$:
\be
\nabla^2 C + C'(C' + \tilde dB' - dA' + a\phi') = 0\ .\label{ceqn}
\ee

     In order to simplify the explicit form of the solution, we now
pick values of the integration constants to make
$A_\infty=B_\infty=0$, so that the solution tends to flat empty space
at transverse infinity. Assembling the result, starting from the
Laplace-equation solution $H(y)$ (\ref{phisol}), one
finds~\cite{dkl,stainless}
\begin{subeqnarray}
ds^2 &=& H^{-4\tilde d\over\Delta(D-2)}dx^\mu dx^\nu\eta_{\mu\nu} +
H^{4d\over\Delta(D-2)}dy^mdy^m\\
e^\phi &=&
H^{2a\over\varsigma\Delta}\hspace{1cm}\varsigma
=\cases{+1,&elementary/electric\cr-1,&solitonic/magnetic\cr}\\
H(y) &=& 1+{k\over r^{\tilde d}}\ ,\label{pbranesol}
\end{subeqnarray}
and in the elementary/electric case, $C(r)$ is given by
\be
e^C = {2\over\sqrt\Delta}H^{-1}\ .\label{csol}
\ee
In the solitonic/magnetic case, the constant of integration is related
to the magnetic charge parameter $\lambda$ in the ansatz (\ref{magans})
by
\be
k = {\sqrt\Delta\over2\tilde d}\lambda\ .\label{klambdarel}
\ee
In the elementary/electric case, this relation may be taken to {\em
define} the parameter $\lambda$.

     The harmonic function $H(y)$ (\ref{phisol}) determines all of the features
of a $p$-brane solution (except for the choice of gauge for the $A_{[n-1]}$
gauge potential). It is useful to express the electric and magnetic field
strengths directly in terms of $H$:
\bsea
F_{m\mu_1\ldots\mu_{n-1}} &=&
{2\over\sqrt\Delta}\epsilon_{\mu_1\ldots\mu_{n-1}}\partial_m(H^{-1})\quad
m=d,\ldots,D-1\quad \hbox{electric}\quad\\
F_{m_1\ldots m_n} &=& -{2\over\sqrt\Delta}\epsilon_{m_1\ldots
m_nr}\partial_rH\quad
m=d,\ldots,D-1\quad \hbox{magnetic,}\quad\label{fieldstrengths}
\esea
with all other independent components vanishing in either case.

\section{$D=11$ examples}\label{sec:examples}

     Let us now return to the bosonic sector of $D=11$ supergravity, which
has the action (\ref{D11act}). In searching for $p$-brane solutions to this
action, there are two particular points to note. The first is
that no scalar field is present in (\ref{D11act}). This follows from the
supermultiplet structure of the $D=11$ theory, in which all fields are gauge
fields. In lower dimensions, of course, scalars do appear; {\it e.g.}\ the
dilaton in $D=10$ type IIA supergravity emerges out of the $D=11$ metric upon
dimensional reduction from $D=11$ to $D=10$. The absence of the scalar that we
had in our general discussion may be handled here simply by identifying the
scalar coupling parameter $a$ with zero, so that the scalar may be consistently
truncated from our general action (\ref{igen}). Since $a^2=\Delta-2d\tilde
d/(D-2)$, we identify $\Delta=2\cdot3\cdot6/9=4$ for the $D=11$ cases.

     Now let us consider the consistency of dropping contributions arising
from the $FFA$ Chern-Simons term in (\ref{D11act}). Note that for $n=4$, the
$F_{[4]}$ antisymmetric tensor field strength supports either an
elementary/electric solution with $d=n-1=3$ (\ie a $p=2$
membrane) or a solitonic/magnetic solution with $\tilde d=11-3-2=6$
(\ie a $p=5$ brane). In both these elementary and solitonic
cases, the $FFA$ term in the action (\ref{D11act}) vanishes and hence
this term does not make any non-vanishing contribution to the metric
field equations for our ans\"atze. For the antisymmetric tensor field
equation, a further check is necessary, since there one requires
the {\it variation} of the $FFA$ term to vanish in order to consistently
ignore it. The field equation for $A_{[3]}$ is (\ref{Aeqmot}), which when
written out explicitly becomes
\be
\partial_{\sst M}\left(\sqrt{-g}F^{\sst MUVW}\right) +
{1\over2(4!)^2}\epsilon^{{\sst
UVW}x_1x_2x_3x_4y_1y_2y_3y_4}F_{x_1x_2x_3x_4}F_{y_1y_2y_3y_4}=0\
.\label{Feq}
\ee 
By direct inspection, one sees that the second term in this equation
vanishes for both ans\"atze.

Next, we shall consider the elementary/electric and the solitonic/magnetic
$D=11$ cases in detail. Subsequently, we shall explore how these
particular solutions fit into wider, ``black,'' families of $p$-branes.

\subsection{$D=11$ Elementary/electric 2-brane}\label{ssec:electric}

     From our general discussion in Sec.\ \ref{sec:pbraneans}, we have
the elementary-ansatz solution~\cite{ds}
\be\begin{array}{rcl}
ds^2 &\makebox[0pt]{=}& (1+{k\over r^6})^{-\sffrac23}dx^\mu
dx^\nu\eta_{\mu\nu} + (1+{k\over
r^6})^{\sffrac13}dy^mdy^m\\
A_{\mu\nu\lambda} &\makebox[0pt]{=}& \epsilon_{\mu\nu\lambda}
(1+{k\over r^6})^{-1}\ ,\hspace{.75cm}
\mbox{other components zero.}\phs\\
&&\multicolumn{1}{r}{\mbox{\underline{electric 2-brane: isotropic
coordinates}\phs}}\end{array}\label{isoel2br}
\ee
At first glance, this solution looks like it might be singular at
$r=0$. However, if one calculates the invariant components of the
curvature tensor $R_{\sst MNPQ}$ and of the field strength
$F_{m\mu_1\mu_2\mu_3}$, subsequently referred to an orthonormal frame by
introducing vielbeins as in (\ref{1forms}), one finds these
invariants to be nonsingular. Moreover, although the proper distance
to the surface $r=0$ along a $t=x^0=\mbox{const.}$ geodesic diverges,
the surface $r=0$ can be reached along null geodesics in finite
affine parameter.\cite{dgt}

     Thus, one may suspect that the metric as given in
(\ref{isoel2br}) does not in fact cover the entire spacetime, and so one
should look for an analytic extension of it. Accordingly, one may consider
a change to ``Schwarzschild-type'' coordinates by setting $r=(\tilde
r^6-k)^{\sffrac16}$. The solution then becomes:\,\cite{dgt}
\be\begin{array}{rclr}
ds^2 &\makebox[0pt]{=}& (1-{k\over\tilde r^6})^{\sffrac23}(-dt^2 + d\sigma^2
+d\rho^2) + (1-{k\over\tilde r^6})^{-2}d\tilde r^2 + \tilde
r^2d\Omega_7^2\\ 
A_{\mu\nu\lambda} &\makebox[0pt]{=}&
\epsilon_{\mu\nu\lambda} (1-{k\over\tilde r^6})\ ,
&\llap{other components zero,}\phs\\
&&\multicolumn{2}{r}{\mbox{\underline{electric 2-brane:
Schwarzschild-type coordinates}\phs}}
\end{array}\label{shwel2br}
\ee
where we have supplied explicit worldvolume coordinates
$x^\mu=(t,\sigma,\rho)$ and where $d\Omega_7^2$ is the line element on
the unit 7-sphere, corresponding to the boundary $\partial{\cal
M}_{8\rm T}$ of the $11-3=8$ dimensional transverse space.

     The Schwarzschild-like coordinates make the surface $\tilde
r=k^{\sffrac16}$ (corresponding to $r=0$) look like a horizon. One may
indeed verify that the normal to this surface is a {\em null} vector,
confirming that $\tilde r=k^{\sffrac16}$ is in fact a horizon. This
horizon is {\em degenerate}, however. Owing to the \ffrac23 exponent
in the $g_{00}$ component, curves along the $t$ axis for $\tilde
r<k^{\sffrac16}$ remain timelike, so that light cones do not ``flip
over'' inside the horizon, unlike the situation for the classic
Schwarzschild solution.

     In order to see the structure of the membrane spacetime more
clearly, let us change coordinates once again, setting $\tilde
r=k^{\sffrac16}(1-R^3)^{-\sffrac16}$. Overall, the transformation from
the original isotropic coordinates to these new ones is effected by
setting $r=k^{\sffrac16}R^{\sffrac12}/(1-R^3)^{\sffrac16}$. In these
new coordinates, the solution becomes~\cite{dgt}
\be\begin{array}{rclr}
ds^2 &\makebox[0pt]{=}& \left\{R^2(-dt^2 + d\sigma^2 + d\rho^2) +
\ft14 k^{\sffrac13}R^{-2}dR^2\right\} +
k^{\sffrac13}d\Omega_7^2&\hspace{1cm}\hfill\mbox{(a)}\\
&&+ \ft14 k^{\sffrac13}[(1-R^3)^{-\sffrac73}-1]R^{-2}dR^2 + k^{\sffrac13}[(1-R^3)^{-\sffrac13}-1]
d\Omega_7^2 &\hfill\mbox{(b)}\\
A_{\mu\nu\lambda} &\makebox[0pt]{=}&
R^3\epsilon_{\mu\nu\lambda}\ ,
\hspace{.5cm}\mbox{other components zero.}\phs\\
&&\multicolumn{1}{r}{\mbox{\underline{electric 2-brane: interpolating
coordinates}\phs}}
\end{array}\label{intel2br}
\ee

     This form of the solution makes it clearer that the light-cones
do not ``flip over'' in the region inside the horizon (which is now at
$R=0$, with $R<0$ being the interior). The main usefulness of the third
form (\ref{intel2br}) of the membrane solution, however, is that it
reveals how the solution {\em interpolates} between other ``vacuum''
solutions of $D=11$ supergravity.\cite{dgt} As $R\rightarrow1$, the
solution becomes flat, in the asymptotic exterior transverse region.
As one approaches the horizon at $R=0$, line (b) of the metric in
(\ref{intel2br}) vanishes at least linearly in $R$. The residual
metric, given in line (a), may then be recognized as a standard form of
the metric on $(\mbox{AdS})_4\times{\cal S}^7$, generalizing the
Robinson-Bertotti solution on  $(\mbox{AdS})_2\times{\cal S}^2$ in
$D=4$. Thus, the membrane solution interpolates between flat space as
$R\rightarrow1$ and  $(\mbox{AdS})_4\times{\cal S}^7$ as
$R\rightarrow0$ at the horizon.

     Continuing on inside the horizon, one eventually encounters a
true singularity at $\tilde r=0$ ($R\rightarrow-\infty$). Unlike the
singularity in the classic Schwarzschild solution, which is spacelike
and hence unavoidable, the singularity in the membrane spacetime is
{\em timelike.} Generically, geodesics do not intersect the
singularity at a finite value of an affine parameter value. Radial null
geodesics do intersect the singularity at finite affine parameter,
however, so the spacetime is in fact genuinely singular. The timelike
nature of this singularity, however, invites one to consider coupling a
$\delta$-function {\em source} to the solution at $\tilde r=0$. Indeed,
the $D=11$ supermembrane action,\cite{bst} which generalizes the
Nambu-Goto action for the string, is the unique  ``matter'' system that
can consistently couple to $D=11$ supergravity.\cite{bst,dhis} Analysis
of this coupling yields a relation between the parameter $k$ in the
solution (\ref{isoel2br}) and the {\em tension} $T$ of the supermembrane
action:\,\cite{ds}
\be
k={\kappa^2T\over3\Omega_7}\ ,\label{ktens}
\ee
where $1/(2\kappa^2)$ is the coefficient of $\sqrt{-g}R$ in the
Einstein-Hilbert Lagrangian and $\Omega_7$ is the volume of the unit
7-sphere ${\cal S}^7$, {\it i.e.}\ the solid angle subtended by the
boundary at transverse infinity.

\begin{figure}[ht]
\leavevmode\centering
\epsfbox{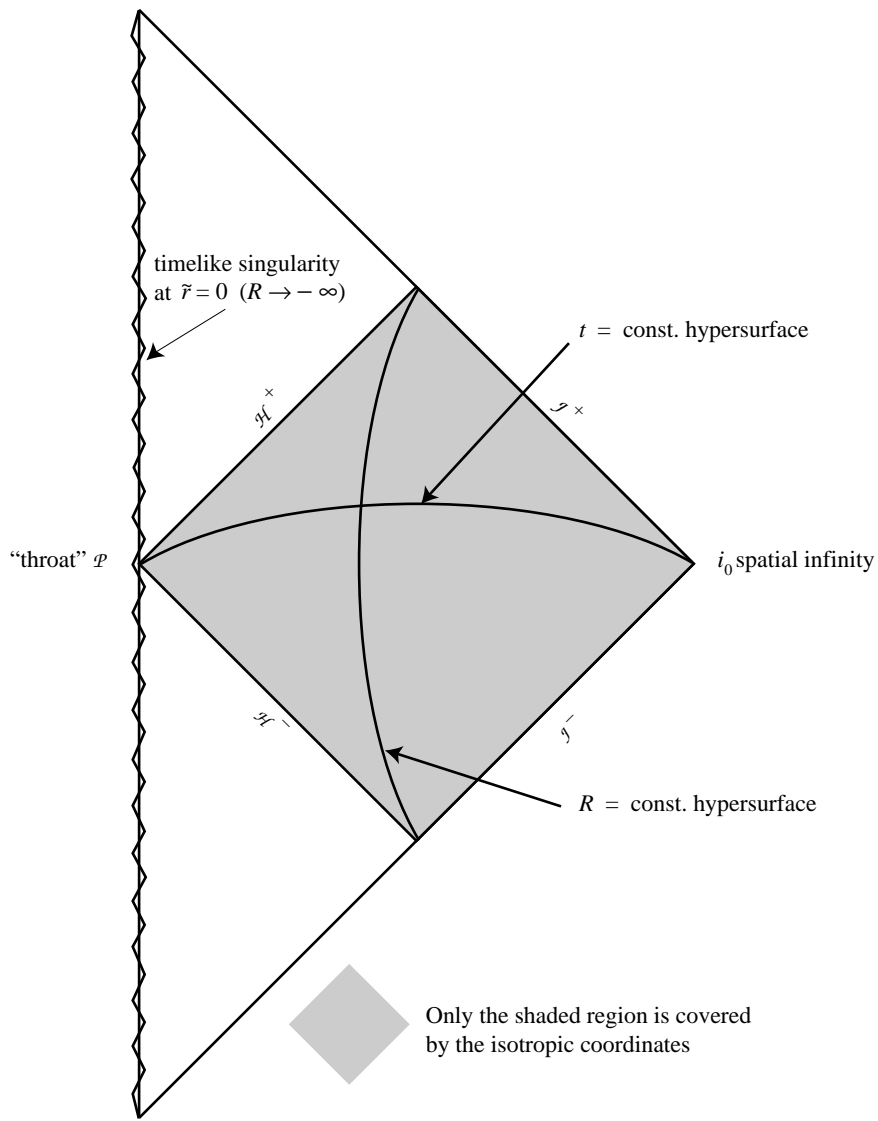}
\caption{Carter-Penrose diagram for the $D=11$ elementary/electric 2-brane
solution.\label{fig:D11elcp}}
\end{figure}

     The global structure of the membrane spacetime~\cite{dgt} is
similar to the extreme Reissner-Nordstrom solution of General
Relativity.\cite{hawkel} This global structure is summarized by a
Carter-Penrose diagram as shown in Figure \ref{fig:D11elcp}, in which
the angular coordinates on ${\cal S}^7$ and also two ignorable
worldsheet coordinates have been suppressed. As one can see, the region
mapped by the isotropic coordinates does not cover the whole spacetime.
This region, shaded in the diagram, is geodesically incomplete, since
one may reach its boundaries ${\cal H}^+$, ${\cal H}^-$ along radial
null geodesics at a finite affine-parameter value. These boundary
surfaces are not singular, but, instead, constitute future and past
horizons (one can see from the form (\ref{shwel2br}) of the solution
that the normals to these surfaces are null). The ``throat'' $\cal P$ in
the diagram should be thought of as an exceptional point at infinity,
and not as a part of the central singularity.

\begin{figure}[ht]
\leavevmode\centering
\epsfbox{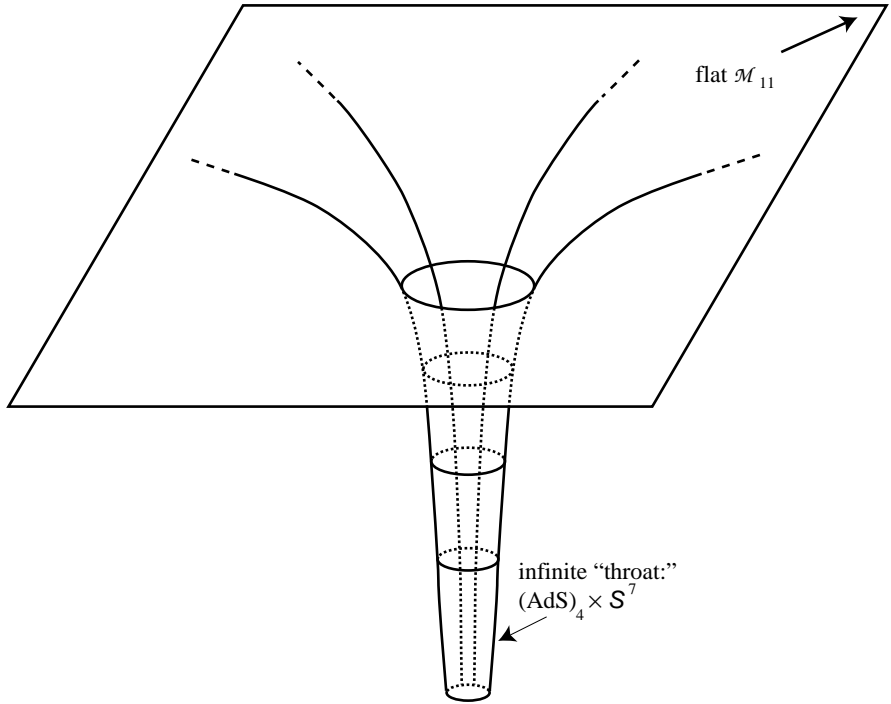}
\caption{The $D=11$ elementary/electric 2-brane solution interpolates
between flat space at ${\cal J}^\pm$ and $(\mbox{AdS})_4\times{\cal
S}^7$ at the horizon.\label{fig:funnel}}
\end{figure}

     The region exterior to the horizon interpolates between flat
regions ${\cal J}^\pm$ at future and past null infinities and a
geometry that asymptotically tends to $(\mbox{AdS})_4\times{\cal
S}^7$ on the horizon. This interpolating portion of the spacetime,
corresponding to the shaded region of Figure \ref{fig:D11elcp}
which is covered by the isotropic coordinates, may be sketched as shown
in Figure \ref{fig:funnel}.

\clearpage

\subsection{$D=11$ Solitonic/magnetic 5-brane}\label{ssec:magnetic}

     Now consider the 5-brane solution to the $D=11$ theory given by
the solitonic ansatz for $F_{[4]}$. In isotropic
coordinates, this solution is a magnetic 5-brane:\,\cite{guv}
\crampest\be\begin{array}{rclr}
ds^2 &\makebox[0pt]{=}& (1+{k\over r^3})^{-\sffrac13}dx^\mu
dx^\nu\eta_{\mu\nu} + (1+{k\over
r^3})^{\sffrac23}dy^mdy^m\hspace{.6cm}&\mu,\nu=0,\cdots,5\\ 
F_{m_1\cdots m_4} &\makebox[0pt]{=}& 3k\epsilon_{m_1\cdots m_4 p}{y^p\over r^5}
&\llap{other components zero.}\\
&&&\llap{\underline{magnetic 5-brane: isotropic coordinates}}
\end{array}\label{mag5br}
\ee\uncramp

     As in the case of the elementary/electric membrane, this solution
interpolates between two ``vacua'' of $D=11$ supergravity. Now,
however, these asymptotic geometries consist of the flat region
encountered as $r\rightarrow\infty$ and of $(\mbox{AdS})_7\times{\cal
S}^4$ as one approaches $r=0$, which once again is a degenerate
horizon. Combining two coordinate changes analogous to those of
the elementary case, $r=(\tilde r^3-k)^{\sffrac13}$ and $\tilde r =
k^{\sffrac13}(1-R^6)^{-\sffrac13}$, one has an overall transformation
\be
r = {k^{\sffrac13}R^2\over(1-R^6)^{\sffrac13}}\ .\label{coordtransf}
\ee
After these coordinate changes, the metric becomes
\be\begin{array}{rcl}
ds^2 &\makebox[0pt]{=}& R^2dx^\mu dx^\nu\eta_{\mu\nu} +
k^{\sffrac23}\left[{4R^{-2}\over(1-R^6)^{\sffrac83}}dR^2 +
{d\Omega_4^2\over(1-R^6)^{\sffrac23}}\right]\ .\\
&&\multicolumn{1}{r}{\mbox{\underline{magnetic 5-brane: interpolating
coordinates}\hspace{.3cm}}}
\end{array}\label{symmag5br}
\ee

     Once again, the surface $r=0 \leftrightarrow R=0$ may be 
seen from (\ref{symmag5br}) to be a nonsingular degenerate horizon. In this
case, however, not only do the light cones maintain their timelike
orientation when crossing the horizon, as already happened in the
electric case (\ref{intel2br}), but now the magnetic solution
(\ref{symmag5br}) is in fact fully {\em symmetric}~\cite{ght} under a
discrete isometry $R\rightarrow -R$.\cite{ght}

\begin{figure}[ht]
\leavevmode\centering
\epsfbox{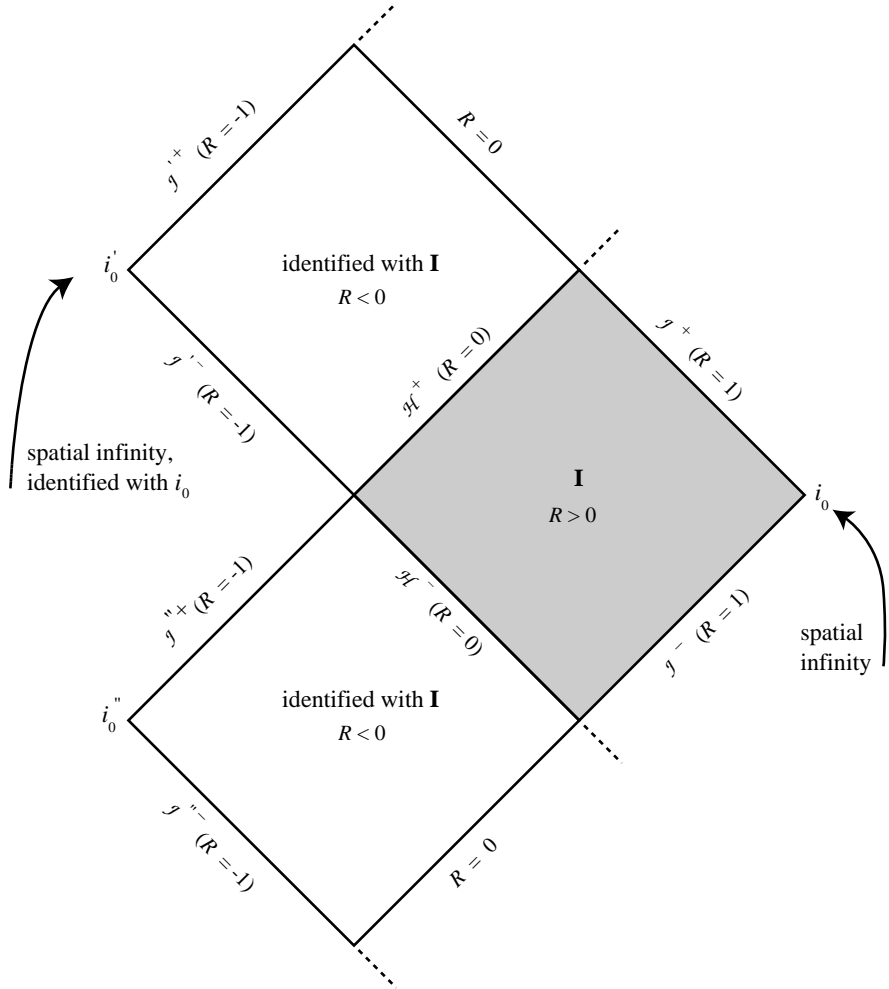}
\caption{Carter-Penrose diagram for the solitonic/magnetic
5-brane solution.\label{fig:D11cpmag}}
\end{figure}

     Given this isometry $R\rightarrow -R$, one can {\em identify}
the spacetime region $R\le 0$ with the region $R\ge 0$. This
identification is analogous\,\footnote{In considering this analogy, one should
also take into account the possibility of conical singularities. In the case of
flat space, a conical singularity with deficit angle $\theta=2\pi$ arises at
the origin $R=0$ if one chooses not to make a discrete identification of the
two regions $R\gtlt0$. This is most easily seen by considering a 
combined pair of forward and backward cones with deficit angle $\theta<2\pi$,
then taking the limit $\theta\rightarrow2\pi$. In this case, as in the case of
the magnetic 5-brane geometry, the elimination of conical singularities
actually {\em requires} making the discrete identification.} to the
identification one naturally makes for flat space when written in polar
coordinates, with the metric $ds^2_{\hbox{flat}}=-dt^2+dr^2+r^2d^2$. Ostensibly,
in these coordinates there appear to be separate regions of flat space with
$r\gtlt 0$, but, owing to the existence of the isometry $r\rightarrow
-r$, these regions may be identified. Accordingly, in the
solitonic/magnetic 5-brane spacetime, we identify the region $-1<R\le
0$ with the region $0\le R<1$. In the asymptotic limit where
$R\rightarrow -1$, one finds                           an asymptotically
flat geometry that is indistinguishable from the region where
$R\rightarrow +1$, {\it i.e.}\ where $r\rightarrow\infty$. Thus, there
is no singularity at all in the solitonic/magnetic 5-brane geometry.
There is still an infinite ``throat,'' however, at the horizon, and the
region covered by the isotropic coordinates might again be sketched as
in Figure \ref{fig:funnel}, except now with the asymptotic geometry down
the ``throat'' being $(\mbox{AdS})_7\times{\cal S}^4$ instead of
$(\mbox{AdS})_4\times{\cal S}^7$ as for the elementary/electric
solution. The Carter-Penrose diagram for the solitonic/magnetic
5-brane solution is given in Figure \ref{fig:D11cpmag}, where the full
diagram extends indefinitely by ``tiling'' the section shown. Upon
using the $R\rightarrow -R$ isometry to make discrete
identifications, however, the whole of the spacetime may be
considered to consist of just region {\bf I}, which is the region
covered by the isotropic coordinates (\ref{mag5br}).

\clearpage

     After identification of the $R\gtlt 0$ regions, the
5-brane spacetime (\ref{mag5br}) is {\em geodesically complete.}
Unlike the case of the elementary membrane solution
(\ref{isoel2br},\,\ref{intel2br}), one finds in the solitonic/magnetic
case that the null geodesics passing through the horizon at $R = 0$
continue to evolve in their affine parameters without bound as
$R\rightarrow -1$. Thus, the solitonic 5-brane solution is {\em
completely non-singular.}

     The electric and magnetic $D=11$ solutions discussed here and in
the previous subsection are special in that they do not involve a
scalar field, since the bosonic sector of $D=11$ supergravity
(\ref{D11act}) does not even contain a scalar field. Similar solutions
occur in other situations where the parameter $a$ (\ref{Delta}) for a
field strength supporting a $p$-brane solution vanishes, in which cases
the scalar fields may consistently be set to zero; this happens for
$(D,d)=(11,3)$, (11,5), (10,4), (6,2), (5,1), (5,2) and (4,1). In these
special cases, the solutions are nonsingular at the horizon and so one
may analytically continue through to the other side of the horizon. When
$d$ is even for ``scalarless'' solutions of this type, there exists a
discrete isometry analogous to the $R\rightarrow -R$ isometry of the
$D=11$ 5-brane solution (\ref{symmag5br}), allowing the outer and inner
regions to be identified.\cite{ght} When $d$ is odd in such cases, the
analytically-extended metric eventually reaches a timelike curvature
singularity at $\tilde r=0$.

     When $a\ne0$ and the scalar field associated to the field strength
supporting a solution cannot be consistently set to zero, then the
solution is singular at the horizon, as can be seen directly in the
scalar solution (\ref{phisol}) itself (where we recall that in
isotropic coordinates, the horizon occurs at $r=0$)

\subsection{Black branes}\label{ssec:blackbranes}

     In order to understand better the family of
supergravity solutions that we have been discussing, let us now
consider a generalization that lifts the degenerate nature of the
horizon. Written in Schwarzschild-type coordinates, one finds the
generalized ``black brane'' solution~\cite{hsdl,dlp}
\be\begin{array}{rcl}
ds^2 &\makebox[0pt]{=}& -{\Sigma_+\over\Sigma_-^{\left\{1-
{4\tilde d\over\Delta(D-2)}\right\}}}dt^2 + \Sigma_-^{{4\tilde
d\over\Delta(D-2)}}dx^idx^i\\
&&\hspace{.5cm} + {\Sigma_-^{\left\{{2a^2\over\Delta\tilde
d}-1\right\}}\over\Sigma_+}d\tilde r^2 + \tilde
r^2\Sigma_-^{2a^2\over\Delta\tilde d}d\Omega^2_{D-d-1}\\
e^{{\varsigma\Delta\over2a}\phi} &\makebox[0pt]{=}& \Sigma_-^{-1}
\hspace{2cm}\Sigma_\pm = 1-\left(r_\pm\over\tilde r\right)^{\tilde
d}\ .\phs\\
&&\multicolumn{1}{r}{\mbox{\underline{black brane:
Schwarzschild-type coordinates}\phs}}
\end{array}\label{blackbrane}
\ee
The antisymmetric tensor field strength for this solution
corresponds to a charge parameter $\lambda = 2\tilde
d/\sqrt\Delta(r_+r_-)^{\tilde d/2}$, either electric or magnetic.

     The characteristic feature of the above ``blackened'' $p$-branes is
that they have a nondegenerate, nonsingular outer horizon at $\tilde
r=r_+$, at which the light cones ``flip over.'' At $\tilde r =r_-$, one
encounters an inner horizon, which, however, coincides in general with a
curvature singularity. The singular nature of the solution at $\tilde r
= r_-$ is apparent in the scalar $\phi$ in (\ref{blackbrane}). For
solutions with $p\ge1$, the singularity at the inner horizon persists
even in cases where the scalar $\phi$ is absent. 

     The {\em extremal} limit of the black brane solution occurs for
$r_+=r_-$. When $a=0$ and scalars may consistently be set to zero, the
singularity at the horizon $r_+=r_-$ disappears and then one may
analytically continue through the horizon. In this case, the light cones
do not ``flip over'' at the horizon because one is really crossing two
coalesced horizons, and the coincident ``flips'' of the light cones cancel
out.

     The generally singular nature of the inner horizon of the non-extreme
solution (\ref{blackbrane}) shows that the ``location'' of the $p$-brane
in spacetime should normally be thought to coincide with the inner
horizon, or with the degenerate horizon in the extremal case.

\section{Charges, Masses and Supersymmetry}\label{sec:masschargesusy}
\addtocontents{toc}{\protect\nopagebreak}

     The $p$-brane solutions that we have been studying are supported by
antisymmetric tensor gauge field strengths that fall off at transverse
infinity like $r^{-(\tilde d+1)}$, as one can see from
(\ref{felans},\,\ref{csol},\,\ref{magans}). This asymptotic falloff is slow
enough to give a nonvanishing total charge density from a Gauss' law flux
integral at transverse infinity, and we shall see that, for the
``extremal'' class of solutions that is our main focus, the mass density
of the solution saturates a ``Bogomol'ny bound'' with respect to the
charge density. In this Section, we shall first make more precise the relation
between the geometry of the $p$-brane solutions, the $p$-form charges $U_{\sst
AB}$ and $V_{\sst ABCDE}$ and the scalar charge magnitudes $U$ and $V$
(\ref{electric},\,\ref{magnetic}); we shall then discuss the relations
between these charges, the energy density and the preservation of unbroken
supersymmetry.

\subsection{$p$-form charges}\label{ssec:charges}
\addtocontents{toc}{\protect\nopagebreak}

     Now let us consider the inclusion of sources into the supergravity
equations. The harmonic function (\ref{phisol}) has a singularity which
has for simplicity been placed at the origin of the transverse coordinates
$y^m$. As we have seen in Sections \ref{ssec:electric} and \ref{ssec:magnetic},
whether or not this gives rise to a physical singularity in a solution depends
on the global structure of that solution. In the electric 2-brane case, the
solution does in the end have a singularity.\cite{ght} This singularity is
unlike the Schwarzschild singularity, however, in that it is a {\em timelike}
curve, and thus it may be considered to be the wordvolume of a
$\delta$-function source. The electric source that couples to $D=11$
supergravity is the fundamental supermembrane action,\cite{bst} whose bosonic
part is
\bea
&&I_{\rm source} = Q_{\rm e}\int_{{\cal
W}_3}d^3\xi\Big[\sqrt{-\det(\partial_\mu x^M\partial_\nu
x^Ng_{MN}(x))}\nn\\
&& \phantom{I_{\rm source} = Q_{\rm e}\int_{{\cal
W}_3}d^3\xi}\qquad+
\fft1{3!}\epsilon^{\mu\nu\rho}\partial_\mu x^M\partial_\nu
x^N\partial_\rho x^R A_{MNR}(x)\Big]\ .\label{smemb}
\eea
The source strength $Q_{\rm e}$ will shortly be found to be equal to the
electric charge $U$ upon solving the coupled equations of motion for the
supergravity fields and a single source of this type. Varying  the source action
(\ref{smemb}) with \ffrac{$\delta$}{$\delta A_{[3]}$}, one obtains the
$\delta$-function current
\be
J^{MNR}(z) = Q_{\rm e}\int_{{\cal W}_3}\delta^3(z-x(\xi))dx^M\wedge dx^N\wedge
dx^R\ .\label{current}
\ee

     This current now stands on the RHS of the $A_{[3]}$ equation of motion:
\be
d({\,}^\ast\!F_{[4]} + \ft12 A_{[3]}\wedge F_{[4]}) = {}^\ast\!J_{[3]}\ .
\label{sourcedeqn}
\ee
Thus, instead of the Gauss' law expression for the charge, one may
instead rewrite the charge as a volume integral of the source,
\be
U = \int_{{\cal M}_8}{}^\ast\!J_{[3]} = \fft1{3!}\int_{{\cal
M}_8}J^{0MN}d^8S_{MN}\ ,\label{volumecharge}
\ee
where $d^8S_{MN}$ is the 8-volume element on ${\cal M}_8$, 
specified within a $D=10$ spatial section of the supergravity
spacetime by a 2-form. The charge derived in this way from a single
2-brane source is thus $U=Q_{\rm e}$ as expected.

\begin{figure}[ht]
\leavevmode\centering
\epsfbox{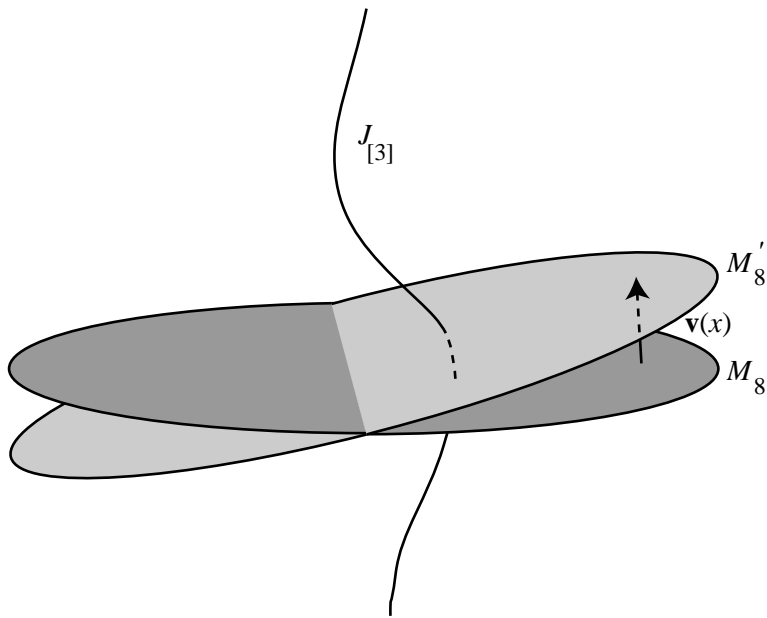}
\caption{Different choices of charge integration
volume ``capturing'' the current $J_{[3]}$.\label{fig:intvols}}
\end{figure}

     Now consider the effect of making different choices of the ${\cal
M}_8$ integration volume within the $D=10$ spatial spacetime section, as
shown in Figure \ref{fig:intvols}. Let the difference between the surfaces
${\cal M}_8$ and ${\cal M}'_8$ be infinitesimal and be given by a vector field
$v^N(x)$. The difference in the electric charges obtained is then given by
\be
\delta U = \int_{{\cal M}_8}{\cal L}_v{}^\ast\!J_{[3]}
 = \fft1{3!}\int_{\partial{\cal M}_8}J^{0MN}v^Rd^7S_{MNR}\ ,\label{deltaU}
\ee
where ${\cal L}_v$ is the Lie derivative along the vector field $v$. The
second equality in (\ref{deltaU}) follows using Stokes' theorem and the
conservation of the current $J_{[3]}$.

     Now a topological nature of the charge integral
(\ref{electric}) becomes apparent; similar considerations apply to the
magnetic charge (\ref{magnetic}). As long as the current
$J_{[3]}$ vanishes on the boundary $\partial{\cal M}_8$, the difference
(\ref{deltaU}) between the charges calculated using the integration
volumes ${\cal M}_8$ and ${\cal M}'_8$ will vanish. This divides the
electric-charge integration volumes into two topological classes
distinguishing those for which $\partial{\cal M}_8$ ``captures'' the
$p$-brane current, as shown in Figure \ref{fig:intvols} and giving
$U=Q_{\rm e}$, from those that do not capture the current,
giving $U$=0.

     The above discussion shows that the orientation-dependence of the $U$
charges (\ref{electric}) is essentially topological. The topological
classes for the charge integrals are naturally labeled by the asymptotic
orientations of the $p$-brane spatial surfaces; an integration volume
${\cal M}_8$ extending out to infinity flips from the ``capturing'' class
into the ``non-capturing'' class when $\partial{\cal M}_8$ crosses the
$\delta$-function surface defined by the current $J_{[3]}$. The charge
thus naturally has a magnitude $|Q_{[p]}|=Q_{\rm e}$ and a unit $p$-form
orientation $Q_{[p]}/|Q_{[p]}|$ that is proportional to the {\em
asymptotic spatial volume form} of the $p$-brane. Both the magnitude and the
orientation of this $p$-form charge are conserved using the
supergravity equations of motion.

     The necessity of considering asymptotic $p$-brane volume forms
arises because the notion of a $p$-form charge is not limited to 
static, flat $p$-brane solutions such as
(\ref{ansatz},\,\ref{felans},\,\ref{magans}). Such charges can also be
defined for any solution whose energy differs from that of a flat, static
one by a finite amount. The charges for such solutions will also
appear in the supersymmetry algebra (\ref{susyalg}) for such backgrounds, but
the corresponding energy densities will not in general saturate the BPS
bounds. For a finite energy difference with respect to a flat, static
$p$-brane, the asymptotic orientation of the $p$-brane volume form must
tend to that of a static flat solution, which plays the r\^ole of a 
``BPS vacuum'' in a given $p$-form charge sector of the theory.

\begin{figure}[ht]
\leavevmode\centering
\epsfbox{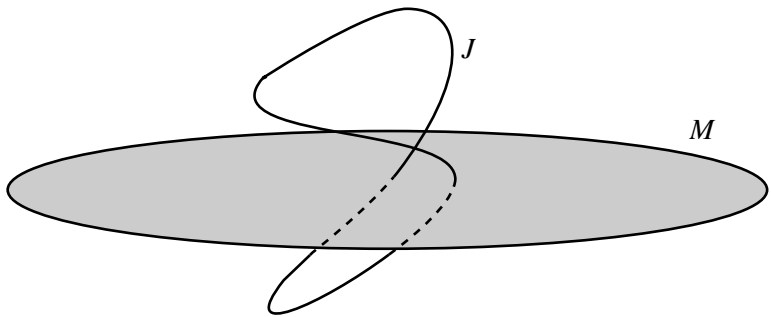}
\caption{Finite $p$-brane not captured by  $\partial{\cal
M}$, giving zero charge.\label{fig:finitebrane}}
\end{figure}

     In order to have a non-vanishing value for a charge
(\ref{electric}) or (\ref{magnetic}) occurring in the supersymmetry
algebra (\ref{susyalg}), the $p$-brane must be either {\em infinite} or
{\em wrapped} around a compact spacetime dimension. The case of a finite
$p$-brane is sketched in Figure \ref{fig:finitebrane}. Since the
boundary $\partial{\cal M}$ of the infinite integration volume
$\cal M$ does not capture the locus where the $p$-brane current is
non-vanishing, the current calculated using $\cal M$ will vanish as a result.
Instead of an infinite $p$-brane, one may alternately have a $p$-brane wrapped
around a compact dimension of spacetime, so that an integration-volume
boundary $\partial{\cal M}_8$ is still capable of capturing the $p$-brane
locus (if one considers this case as an infinite, but periodic, solution,
this case may be considered simultaneously with that of the infinite
$p$-branes). Only in such cases do the $p$-form charges occurring in the
supersymmetry algebra (\ref{susyalg}) take non-vanishing values.\footnote{If
one considers integration volumes that do not extend out to infinity, then
one can construct integration surfaces that capture finite
$p$-branes. Such charges do not occur in the supersymmetry algebra
(\ref{susyalg}), but they are still of importance in determining the possible
intersections of $p$-branes.\cite{branesurgery}}

\clearpage

\subsection{$p$-brane mass densities}\label{ssect:masses}
\addtocontents{toc}{\protect\nopagebreak}

     Now let us consider the mass density of a $p$-brane solution. Since the
$p$-brane solutions have translational symmetry in their $p$ spatial worldvolume
directions, the total energy as measured by a surface integral at spatial
infinity diverges, owing to the infinite extent. What is thus more appropriate
to consider instead is the value of the density, energy/(unit $p$-volume).
Since we are considering solutions in their rest frames, this will also
give the value of mass/(unit $p$-volume), or {\em tension} of the
solution. Instead of the standard spatial $d^{D-2}\Sigma^a$ surface
integral, this will be a $d^{(D-d-1)}\Sigma^m$ surface integral over the
boundary $\partial{\cal M}_{\rm\sst T}$ of the transverse space.

     The ADM formula for the energy density written as a Gauss'-law
integral (see, {\it e.g.,} Ref.\,\cite{mtw}) is, dropping the
divergent spatial $d\Sigma^{\mu=i}$ integral,
\be
{\cal E} = \int_{\partial{\cal M}_{\rm
T}}d^{D-d-1}\Sigma^m(\partial^nh_{mn}-\partial_mh^b_b)\ ,\label{admenerg}
\ee
written for $g_{\sst MN} = \eta_{\sst MN} + h_{\sst MN}$ tending
asymptotically to flat space in Cartesian coordinates, and with $a,b$
{\em spatial} indices running over the values $\mu=i=1,\ldots,d-1$;
$m=d,\ldots,D-1$. For the general $p$-brane solution (\ref{pbranesol}), one
finds
\be
h_{mn}={4kd\over\Delta(D-2)r^{\tilde d}}\delta_{mn}\ ,\hspace{2cm} h^b_b
= {8k(d+\ft12 \tilde d)\over\Delta(D-2)r^{\tilde d}}\ ,\label{hasymp}
\ee
and, since $d^{(D-d-1)}\Sigma^m = r^{\tilde d}y^md\Omega^{(D-d-1)}$, one
finds
\be
{\cal E} = {4k\tilde d\Omega_{D-d-1}\over\Delta}\ ,\label{energy}
\ee
where $\Omega_{D-d-1}$ is the volume of the ${\cal S}^{D-d-1}$ unit sphere.
Recalling that $k=\sqrt\Delta\lambda/(2\tilde d)$, we consequently have a
relation between the mass per unit $p$ volume and the charge parameter of the
solution
\be
{\cal E} = {2\lambda\Omega_{D-d-1}\over\sqrt\Delta}\ .\label{masschargerel}
\ee

     By contrast, the black brane solution (\ref{blackbrane}) has ${\cal
E}>2\lambda\Omega_{D-d-1}/\sqrt\Delta$, so the extremal $p$-brane solution
(\ref{pbranesol}) is seen to {\em saturate} the inequality ${\cal E}\ge
2\lambda\Omega_{D-d-1}/\sqrt\Delta$.

\subsection{$p$-brane charges}\label{ssect:charges}
\addtocontents{toc}{\protect\nopagebreak}

     As one can see from (\ref{energy},\,\ref{masschargerel}), the relation
(\ref{klambdarel}) between the integration constant $k$ in the solution
(\ref{pbranesol}) and the charge parameter $\lambda$ implies a deep link
between the energy density and certain electric or magnetic
charges. In the electric case, this charge is a quantity conserved by
virtue of the equations of motion for the antisymmetric tensor gauge
field $A_{[n-1]}$, and has generally become known as a ``Page charge,''
after its first discussion in Ref.\,\cite{dp} To be specific, if we once
again consider the bosonic sector of $D=11$ supergravity theory
(\ref{D11act}), for which the antisymmetric tensor field equation was
given in (\ref{Feq}), one finds the Gauss'-law form conserved
quantity~\cite{dp} $U$ (\ref{electric}).

     For the $p$-brane solutions (\ref{pbranesol}), the $\int A\wedge F$
term in (\ref{electric}) vanishes. The $\int{}^\ast\!F$ term does,
however, give a contribution in the elementary/electric case,
provided one picks ${\cal M}_8$ to coincide with the transverse space to the
$d=3$ membrane worldvolume, ${\cal M}_{8{\rm T}}$. The surface element for
this transverse space is $d\Sigma^m_{(7)}$, so for the $p=2$ elementary
membrane solution (\ref{isoel2br}), one finds
\be
U = \int_{\partial M_{8{\rm T}}}d\Sigma^m_{(7)}F_{m012} =
\lambda\Omega_7\ .\label{membcharge}
\ee
Since the $D=11$ $F_{[4]}$ field strength supporting this solution has
$\Delta=4$, the mass/charge relation is
\be
{\cal E} = U = \lambda\Omega_7\ .\label{elmasschargerel}
\ee
Thus, like the classic extreme Reissner-Nordstrom black-hole solution
to which it is strongly related (as can be seen from the Carter-Penrose
diagram given in Figure \ref{fig:D11elcp}), the $D=11$ membrane solution
has equal mass and charge densities, saturating the inequality ${\cal
E}\ge U$.

     Now let us consider the charge carried by the solitonic/magnetic 5-brane
solution (\ref{mag5br}). The field strength in (\ref{mag5br}) is purely
transverse, so no electric charge (\ref{electric}) is present. The magnetic
charge (\ref{magnetic}) is carried by this solution, however. Once again,
let us choose the integration subsurface so as to coincide with the transverse
space to the $d=6$ worldvolume, \ie $\widetilde{\cal M}_5={\cal M}_{5{\rm T}}$.
Then, we have
\be
V = \int_{\partial{\cal M}_{5{\rm T}}}d\Sigma^m_{(4)}\epsilon_{mnpqr}F^{npqr} =
\lambda\Omega_4\ .\label{5brcharge}
\ee
Thus, in the solitonic/magnetic 5-brane case as well, we have a
saturation of the mass-charge inequality:
\be
{\cal E} = V = \lambda\Omega_4\ .\label{magmasschargerel}
\ee

\subsection{Preserved supersymmetry}\label{ssect:supersymmetry}

     Since the bosonic solutions that we have been considering are {\em
consistent truncations} of $D=11$ supergravity, they must also possess
another conserved quantity, the {\em supercharge.} Admittedly, since the
supercharge is a Grassmanian (anticommuting) quantity, its value will
clearly be zero for the class of purely bosonic solutions that we have
been discussing. However, the functional form of the supercharge is still
important, as it determines the form of the asymptotic supersymmetry
algebra. The Gauss'-law form of the supercharge is given as
an integral over the boundary of the spatial hypersurface. For the $D=11$
solutions, this surface of integration is the boundary at
infinity $\partial{\cal M}_{10}$ of the $D=10$ spatial hypersurface; the
supercharge is then~\cite{cjs}
\be
Q = \int_{\partial{\cal M}_{10}}\Gamma^{0bc}\psi_cd\Sigma_{(9)b}\ .
\label{supercharge}
\ee
One can also rewrite this in fully Lorentz-covariant form, where
$d\Sigma_{(9)b} = d\Sigma_{(9)0b}\rightarrow d\Sigma_{(9){\sst AB}}$:
\be
Q = \int_{\partial{\cal M}_{10}}\Gamma^{\sst ABC}\psi_{\sst
C}d\Sigma_{(9){\sst AB}}\ .\label{lorentzsupercharge}
\ee

     After appropriate definitions of Poisson brackets, the
$D=11$ supersymmetry algebra for the supercharge
(\ref{supercharge},\,\ref{lorentzsupercharge}) is found to be given~\cite{agit}
by (\ref{susyalg}) Thus, the supersymmetry algebra wraps together
all of the conserved Gauss'-law type quantities that we have discussed.

     The positivity of the $Q^2$ operator on the LHS of the algebra
(\ref{susyalg}) is at the root of the {\em Bogomol'ny
bounds}~\cite{gh,ght,lp}
\begin{subeqnarray}
{\cal E} &\ge& (2/\sqrt\Delta)U\hspace{2cm}
\mbox{\underline{electric bound}}\\
{\cal E} &\ge& (2/\sqrt\Delta)V\hspace{2cm}
\mbox{\underline{magnetic bound}}\label{bogbounds}
\end{subeqnarray}
that are saturated by the $p$-brane solutions.

     The saturation of the Bogomol'ny inequalities by the $p$-brane
solutions is an indication that they fit into special types of
supermultiplets. All of these bound-saturating solutions share the
important property that they {\em leave some portion of the
supersymmetry unbroken}. Within the family of $p$-brane solutions that
we have been discussing, it turns out~\cite{lp} that the $\Delta$ values
of such ``supersymmetric'' $p$-branes are of the form $\Delta=4/N$, where
$N$ is the number of antisymmetric tensor field strengths participating
in the solution (distinct, but of the same rank). The different charge
contributions to the supersymmetry algebra occurring for different
values of $N$ (hence different $\Delta$) affect the Bogomol'ny bounds as
shown in (\ref{bogbounds}).

     In order to see how a purely bosonic solution may leave some
portion of the supersymmetry unbroken, consider specifically once again the
membrane solution of $D=11$ supergravity.\cite{ds} This theory~\cite{cjs}
has just one spinor field, the gravitino $\psi_{\sst M}$. Checking for
the consistency of setting $\psi_{\sst M}=0$ with the supposition of some
residual supersymmetry with parameter
$\epsilon(x)$ requires solving the equation
\be
\delta\psi_{\sst A}\for_{\psi=0} = \tilde D_{\sst A}\epsilon = 0\
,\label{deltapsi}
\ee
where $\psi_{\sst A}=e_{\sst A}{}^{\sst M}\psi_{\sst M}$ and
\bea
\tilde D_{\sst A}\epsilon &=& D_{\sst A}\epsilon -
{1\over288}\left(\Gamma_{\sst A}{}^{\sst BCDE} - 8\delta_{\sst
A}{}^{\sst B}\Gamma^{\sst CDE}\right)F_{\sst BCDE}\epsilon\nonumber\\
D_{\sst A}\epsilon &=& (\partial_{\sst A} + \ft14 \omega_{\sst
A}{}^{\sst BC}\Gamma_{\sst BC})\epsilon\ .\label{dtildeps}
\eea
Solving the equation $\tilde D_{\sst A}\epsilon = 0$ amounts to finding
a {\em Killing spinor} field in the presence of the bosonic background.
Since the Killing spinor equation (\ref{deltapsi}) is linear in
$\epsilon(x)$, the Grassmanian (anticommuting) character of this
parameter is irrelevant to the problem at hand, which thus reduces
effectively to solving (\ref{deltapsi}) for a commuting quantity.

     In order to solve the Killing spinor equation (\ref{deltapsi}) in a
$p$-brane background, it is convenient to adopt an appropriate basis for
the $D=11$ $\Gamma$ matrices. For the $d=3$ membrane background, one
would like to preserve ${\rm SO}(2,1)\times {\rm SO}(8)$ covariance. An
appropriate basis that does this is
\be
\Gamma_{\sst A} =
(\gamma_\mu\otimes\Sigma_9,\oneone_{(2)}\otimes\Sigma_m)\
,\label{gammasplit}
\ee
where $\gamma_\mu$ and $\oneone_{(2)}$ are $2\times2$ ${\rm SO}(2,1)$
matrices; $\Sigma_9$ and $\Sigma_m$ are $16\times16$ ${\rm SO}(8)$
matrices, with $\Sigma_9 = \Sigma_3\Sigma_4\ldots\Sigma_{10}$, so
$\Sigma_9^2 = \oneone_{(16)}$. The most general spinor field consistent
with $\hbox{(Poincar\'e)}_3\times {\rm SO}(8)$ invariance in this spinor
basis is of the form
\be
\epsilon(x,y) =\epsilon_2\otimes\eta(r)\ ,\label{epsilonform}
\ee
where $\epsilon_2$ is a {\em constant} ${\rm SO}(2,1)$ spinor and
$\eta(r)$ is an ${\rm SO}(8)$ spinor depending only on the isotropic
radial coordinate $r$; $\eta$ may be further decomposed into $\Sigma_9$
eigenstates by the use of $\ft12(\oneone\pm\Sigma_9)$ projectors.

     Analysis of the the Killing spinor condition (\ref{deltapsi}) in
the above spinor basis leads to the following requirements~\cite{dghrr,ds} on
the background and on the spinor field $\eta(r)$:
\begin{enumerate}
\item[1)] The background must satisfy the conditions $3A' + 6B' = 0$ and
$C'e^C = 3A'e^{3A}$. The first of these conditions is, however, precisely
the linearity-condition refinement (\ref{dadtb}) that we made in the
$p$-brane ansatz; the second condition follows from the ansatz refinement
(\ref{phirel}) (considered as a condition on $\phi'/a$) and from
(\ref{crel}). Thus, what appeared previously to be simplifying
specializations in the derivation given in Section \ref{sec:pbraneans}
turn out in fact to be conditions {\em required} for supersymmetric
solutions.
\item[2)] $\eta(r) = H^{-1/6}(y)\eta_0=e^{C(r)/6}\eta_0$, where $\eta_0$ is a
constant ${\rm SO}(8)$ spinor. Thus, the
surviving local supersymmetry parameter $\epsilon(x,y)$ must take the form
$\epsilon(x,y)=H^{-1/6}\epsilon_\infty$, where
$\epsilon_\infty=\epsilon_2\otimes\eta_0$. Note that, after imposing this
requirement, at most a finite number of parameters can remain unfixed in the
product spinor
$\epsilon_2\otimes\eta_0$; \ie the {\em local} supersymmetry of the
$D=11$ theory is almost entirely broken by any particular solution. So far, the
requirement (\ref{deltapsi}) has cut down the amount of surviving supersymmetry
from $D=11$ local supersymmetry (\ie effectively an infinite number of
components) to the finite number of independent components present in
$\epsilon_2\otimes\eta_0$. The maximum number of such {\em rigid} unbroken
supersymmetry components is achieved for
$D=11$ flat space, which has a full set of 32 constant components.
\item[3)] $(\oneone+\Sigma_9)\eta_0 = 0$, so the constant ${\rm SO}(8)$
spinor $\eta_0$ is also required to be {\em chiral}.\footnote{The
specific chirality indicated here is correlated with the sign choice
made in the elementary/electric form ansatz (\ref{elans}); one may
accordingly observe from (\ref{D11act}) that a $D=11$  parity
transformation requires a sign flip of $A_{[3]}$.} This cuts the number
of surviving parameters in the product $\epsilon_\infty=\epsilon_2\otimes\eta_0$
by {\em half:} the total number of surviving rigid supersymmetries in
$\epsilon(x,y)$ is thus $2\cdot 8 =16$ (counting real spinor components). Since
this is half of the maximum rigid number (\ie half of the 32 for flat
space), one says that the membrane solution preserves ``half'' of the
supersymmetry.
\end{enumerate}

     In general, the procedure for checking how much supersymmetry is preserved
by a given BPS solution follows steps analogous to points 1) -- 3) above: first
a check that the conditions required on the background fields are satisfied,
then a determination of the functional form of the supersymmetry parameter in
terms of some finite set of spinor components, and finally the imposition of
projection conditions on that finite set. In a more telegraphic partial
discussion, one may jump straight to the projection conditions 3). These must,
of course, also emerge from a full analysis of equations like (\ref{deltapsi}).
But one can also see more directly what they will be simply by considering the
supersymmetry algebra (\ref{susyalg}), specialised to the BPS background.
Thus, for example, in the case of a $D=11$ membrane solution oriented in the
$\{012\}$ directions, one has, after normalising to a unit 2-volume,
\be
{1\over\hbox{2-vol}}\{Q_\alpha,Q_\beta\} = -(C\Gamma^0)_{\alpha\beta}{\cal
E} + (C\Gamma^{12})_{\alpha\beta}U_{12}\ .\label{2bralg}
\ee
Since, as we have seen in (\ref{elmasschargerel}), the membrane solution
saturating the Bogomol'ny bound (\ref{bogbounds}a) with ${\cal
E}=U=U_{12}$, one may rewrite (\ref{2bralg}) as
\be
{1\over\hbox{2-vol}}\{Q_\alpha,Q_\beta\} = 2{\cal E}P_{012}\qquad\qquad 
P_{012} = \ft12(\oneone + \Gamma^{012})\ ,\label{2brproj}
\ee
where $P_{012}$ is a projection operator (\ie $P_{012}^2=P_{012}$) whose trace
is ${\rm tr}P_{012}=\ft12\cdot32$; thus, half of its eigenvalues are zero, and
half are unity. Any surviving supersymmetry transformation must give zero when
acting on the BPS background fields, and so the anticommutator
$\{Q_\alpha,Q_\beta\}$ of the generators must give zero when contracted with a
surviving supersymmetry parameter $\epsilon^\alpha$. From (\ref{2brproj}), this
translates to
\be
P_{012}\,\epsilon_\infty=0\ ,\label{projepsilon}
\ee
which is equivalent to condition 3) above, $(\oneone+\Sigma_9)\eta_0 = 0$.
Thus, we once again see that the $D=11$ supermembrane solution
(\ref{isoel2br}) preserves half of the maximal rigid $D=11$ supersymmetry. When
we come to discuss the cases of ``intersecting'' $p$-branes in Section
\ref{sec:intersectingbranes}, it will be useful to have quick
derivations like this for the projection conditions that must be satisfied
by surviving supersymmetry parameters.

     More generally, the positive semi-definiteness of the operator
$\{Q_\alpha,Q_\beta\}$ is the underlying principle in the
derivation~\cite{gh,ght,lp} of the Bogomol'ny bounds (\ref{bogbounds}). A
consequence of this positive semi-definiteness is that zero eigenvalues
correspond to solutions that saturate the Bogomol'ny inequalities
(\ref{bogbounds}), and these solutions preserve one component of unbroken
supersymmetry for each such zero eigenvalue.

     Similar consideration of the solitonic/magnetic 5-brane
solution~\cite{guv} (\ref{mag5br}) shows that it also preserves half the rigid
$D=11$ supersymmetry. In the 5-brane case, the analogue of condition 2) above
is $\epsilon(x,y)=H^{-1/12}(y)\epsilon_\infty$, and the
projection condition following from the algebra of preserved supersymmetry
generators for a 5-brane oriented in the \{012345\} directions is
$P_{012345}\,\epsilon_\infty=0$, where
$P_{012345}=\ft12(\oneone+\Gamma^{012345})$.

\section{Kaluza-Klein dimensional reduction}\label{sec:kkred}

     Let us return now to the arena of purely bosonic field theories,
and consider the relations between various bosonic-sector theories and the
corresponding relations between $p$-brane solutions. It is
well-known that supergravity theories are related by dimensional
reduction from a small set of basic theories, the largest of which being
$D=11$ supergravity. The spinor sectors of the theories are equally well
related by dimensional reduction, but in the following, we shall
restrict our attention to the purely bosonic sector.

     In order to set up the procedure, let us consider a theory in
$(D+1)$ dimensions, but break up the metric in $D$-dimensionally
covariant pieces:
\be
d\hat s^2 = e^{2\alpha\varphi}ds^2 + e^{2\beta\phi}(dz + {\cal A}_{\sst
M}dx^{\sst M})^2\label{kkmetricans}
\ee
where carets denote $(D+1)$-dimensional quantities corresponding to the
$(D+1)$-dimensional coordinates $x^{\hat{\sst M}}=(x^{\sst M},z)$;
$ds^2$ is the line element in $D$ dimensions and $\alpha$ and $\beta$
are constants. The scalar $\varphi$ in $D$ dimensions emerges from the
metric in $(D+1)$ dimensions as $(2\beta)^{-1}\ln g_{zz}$. Adjustment of the
constants $\alpha$ and $\beta$ is necessary to obtain desired structures in $D$
dimensions. In particular, one should pick $\beta=-(D-2)\alpha$ in order to
arrange for the Einstein-frame form of the gravitational action in $(D+1)$
dimensions to go over to the Einstein-frame form of the action in $D$
dimensions.

     The essential step in a Kaluza-Klein dimensional reduction is a
{\em consistent truncation} of the field variables, generally made
by choosing them to be independent of the reduction coordinate $z$. By a
consistent truncation, we always understand a restriction on the variables that
commutes with variation of the action to produce the field equations, {\it
i.e.}\ a restriction such that solutions to the equations for the
restricted variables are also solutions to the equations for the
unrestricted variables. This ensures that the lower-dimensional solutions
which we shall obtain are also particular solutions to higher-dimensional
supergravity equations as well. Making the parameter choice
$\beta=-(D-2)\alpha$ to preserve the Einstein-frame form of the action,
one obtains
\be
\sqrt{-\hat g}R(\hat g) = \sqrt{-g}\Big(R(g) -
(D-1)(D-2)\alpha^2\nabla_{\sst M}\varphi\nabla^{\sst M}\varphi -
\ft14e^{-2(D-1)\alpha\varphi}{\cal F}_{\sst MN}{\cal F}^{\sst MN}\Big)
\label{actionred}
\ee
where ${\cal F}=d{\cal A}$. If one now chooses $\alpha^2 =
[2(D-1)(D-2)]^{-1}$, the $\varphi$ kinetic term becomes
conventionally normalized.

     Next, one needs to establish the reduction ansatz for the
$(D+1)$-dimen\-sional antisymmetric tensor gauge field $\hat F_{[n]} =
d\hat A_{[n-1]}$. Clearly, among the $n-1$ antisymmetrised indices of
$\hat A_{[n-1]}$ at most one can take the value $z$, so we have the
decomposition
\be
\hat A_{[n-1]} = B_{[n-1]} + B_{[n-2]}\wedge dz\ .\label{kkantisymtens}
\ee
All of these reduced fields are to be taken to be functionally
independent of $z$. For the corresponding field strengths, first define
\begin{subeqnarray}
G_{[n]} &=& dB_{[n-1]}\\
G_{[n-1]} &=& dB_{[n-2]}\ .\label{redfstrengths}
\end{subeqnarray}
However, these are not exactly the most convenient quantities to
work with, since a certain ``Chern-Simons'' structure appears upon
dimensional reduction. The metric in $(D+1)$ dimensions couples to all
fields, and, consequently, dimensional reduction will produce some terms
with undifferentiated Kaluza-Klein vector fields ${\cal A}_{\sst M}$
coupling to $D$-dimensional antisymmetric tensors. Accordingly, it
is useful to introduce
\be
G'_{[n]} = G_{[n]}-G_{[n-1]}\wedge{\cal A}\ ,\label{kkchsimG}
\ee
where the second term in (\ref{kkchsimG}) may be viewed as a
Chern-Simons correction from the reduced $D$-dimensional point of view.

     At this stage, we are ready to perform the dimensional reduction of
our general action (\ref{igen}). We find that
\be
\hat I =: \int d^{D+1}x\sqrt{-\hat g}\Big[R(\hat g) - \ft12
\nabla_{\hat{\sst M}}\phi\nabla^{\hat{\sst M}}\phi - {1\over2n!}e^{\hat
a\phi}\hat F^2_{n]}\Big]\label{Ihat}
\ee
reduces to
\bea
I &=& \int d^Dx\sqrt{-g}\Big[R - \ft12\nabla_{\sst M}\phi\nabla^{\sst
M}\phi - \ft12\nabla_{\sst M}\varphi\nabla^{\sst M}\varphi -
\ft14e^{-2(D-1)\alpha\varphi}{\cal F}^2_{[2]}\nonumber\\
&&~~-{1\over2n!}e^{-2(n-1)\alpha\varphi + \hat\alpha\phi}G'^2_{[n]} -
{1\over2(n-1)!}e^{2(D-n)\alpha\varphi+\hat a\phi}G^2_{[n-1]}\Big]\
.\label{Ired}
\eea
Although the dimensional reduction (\ref{Ired}) has
produced a somewhat complicated result, the important point to note is
that each of the $D$-dimensional antisymmetric-tensor field strength
terms $G'^2_{[n]}$ and $G^2_{[n-1]}$ has an exponential prefactor of the
form $e^{a_r\tilde\phi_r}$, where the $\tilde\phi_r$, $r=(n,n-1)$ are
$SO(2)$-rotated combinations of $\varphi$ and $\phi$. Now, keeping just one
while setting to zero the other two of the three gauge fields $({\cal
A}_{[1]}, B_{[n-2]}, B_{n-1]})$, but retaining at the same time the
scalar-field combination appearing in the corresponding exponential
prefactor, is a {\em consistent truncation.} Thus, any one of the three
field strengths $({\cal F}_{[2]}, G_{[n-1]}, G'_{[n]})$, retained alone
together with its corresponding scalar-field combination, can support
$p$-brane solutions in $D$ dimensions of the form that we have been
discussing.

     An important point to note here is that, in each of the
$e^{a_r\tilde\phi}$ prefactors, the coefficient $a_r$ satisfies
\be
a_r^2 = \Delta - {2d_r\tilde d_r\over(D-2)} = \Delta -
{2(r-1)(D-r-1)\over(D-2)}\label{deltared}
\ee
with the {\em same} value of $\Delta$ as for the ``parent'' coupling
parameter $\hat a$, satisfying
\be
\hat a_r^2 = \Delta - {2d_{(n)}\tilde d_{(n)}\over((D+1)-2)} = \Delta -
{2(n-1)(D-n)\over(D-1}\label{deltahigh}
\ee
in $D+1$ dimensions. Thus, although the individual parameters $a_r$ are
both $D$- and $r$-dependent, the quantity $\Delta$ is {\em preserved} under
Kaluza-Klein reduction for both of the ``descendant'' field-strength
couplings (to $G'^2_{[n]}$ or to $G^2_{[n-1]}$) coming from the
original term $e^{\hat a\phi}\hat F_{[n]}^2$. The 2-form field strength
${\cal F}_{[2]} = d{\cal A}$, on the other hand, emerges out of the
gravitational action in $D+1$ dimensions; its coupling parameter
corresponds to $\Delta=4$.

     If one retains in the reduced theory only one of the field
strengths (${\cal F}_{[2]}$, $G_{[n-1]}$, $G'_{[n]}$), together with its
corresponding scalar-field combination, then one finds oneself back in the
situation described by our general action (\ref{igen}), and then the $p$
brane solutions obtained for the general case in Sec.\ \ref{sec:pbraneans}
immediately become applicable. Moreover, since retaining only one field
strength \& scalar combination in this way effects a consistent truncation
of the theory, solutions to this simple truncated system are {\em also}
solutions to the untruncated theory, and indeed are also solutions to
the original $(D+1)$-dimensional theory, since the Kaluza-Klein
dimensional reduction is also a consistent truncation.

\subsection{Multiple field-strength solutions and
the single-charge truncation}\label{ssec:multifsols}

     After repeated single steps of Kaluza-Klein dimensional reduction from
$D=11$ down to $D$ dimensions, the metric takes the form~\cite{lpsol,lp}
\bsea
ds_{11}^2 &=& e^{\ft23 \vec a\cdot\vec\phi} \, ds_{\sst D}^2 +
\sum_i e^{(\ft23\vec a-\vec a_i)\cdot\vec\phi}\, (h^i)^2\\
h^i &=& dz^i + {\cal A}_{[1]}^i + {\cal A}_{[0]}^{ij}\, dz^j\ ,\label{met}
\esea
where the ${\cal A}^i_{\sst M}$ are a set of $(11-D)$ Kaluza-Klein vectors
generalising the vector ${\cal A}_{\sst M}$ in (\ref{kkmetricans}),
emerging from the higher-dimensional metric upon dimensional reduction. Once
such Kaluza-Klein vectors have appeared, subsequent dimensional reduction also
gives rise to the zero-form gauge potentials ${\cal A}_{[0]}^{ij}$
appearing in (\ref{met}b) as a consequence of the usual one-step reduction
(\ref{kkantisymtens}) of a 1-form gauge potential.

     We shall also need the corresponding reduction of the $\hat F_{[4]}$ field
strength\,\footnote{Note that the lower-dimensional field strengths $F_{[n]}$
include ``Chern-Simons'' corrections similar to those in (\ref{kkchsimG}).}
(where hatted quantities refer to the original, higher, dimension) and, for
later reference, we shall also give the reduction of its Hodge
dual ${}^{\hat \ast}\!\hat F_{[4]}$:
\crampest
\bsea
\hat F_{[4]} &=& F_{[4]} + F_{[3]}^i\wedge h^i + \ft12
F_{[2]}^{ij}\wedge h^i\wedge h^j +
\phs\ft16 F_{[1]}^{ijk}\wedge h^i\wedge h^j\wedge h^k\\
{}^{\hat \ast}\!\hat F_{[4]} &=& e^{\vec
a\cdot\vec\phi}\, {}^\ast\!F_{[4]}
\wedge v + e^{\vec a_i\cdot\vec\phi}\, {}^\ast\!F_{[3]}^i \wedge v^i +
\ft12 e^{\vec a_{ij}\cdot\vec\phi}\, {}^\ast\!F_{[2]}^{ij} \wedge v^{ij}
+\ft16 e^{\vec a_{ijk}\cdot\vec\phi}\, {}^\ast\!F_{[1]}^{ijk} \wedge
v^{ijk}\ ,\nonumber\\
&&\label{Fred}
\esea\uncramp
(noting that, since the Hodge dual is a metric-dependent construction,
exponentials of the dilatonic vectors $\vec\phi$ appear in the reduction of
${}^{\hat \ast}\!\hat F_{[4]}$) where the forms $v$, $v_i$, $v_{ij}$ and
$v_{ijk}$ appearing in (\ref{Fred}b) are given by
\bea
v &=& \ft1{(11-D)!} \epsilon_{i_1\cdots i_{\sst{11-D}}}\, h^{i_1}\wedge
\cdots
\wedge h^{i_{\sst{11-D}}}\nn\\
v_i &=& \ft1{(10-D)!} \epsilon_{ii_2\cdots i_{\sst{11-D}}}\,
h^{i_2}\wedge\cdots
\wedge h^{i_{\sst{11-D}}}\nn\\
v_{ij} &=& \ft1{(9-D)!} \epsilon_{iji_3\cdots i_{\sst{11-D}}}\, 
h^{i_3}\wedge \cdots
\wedge h^{i_{\sst{11-D}}}\nn\\
v_{ijk} &=& \ft1{(8-D)!} \epsilon_{ijki_4\cdots i_{\sst{11-D}}}\, 
h^{i_4} \wedge\cdots
\wedge h^{i_{\sst{11-D}}}\ .\label{vdef}
\eea

     Using (\ref{met},\,\ref{Fred}a), the bosonic sector of maximal
supergravity (\ref{D11act}) now reduces to~\cite{lpsol,lp}
\bea
I_D &=& \int d^Dx\sqrt{-g}\Big[R-\ft12(\partial\vec \phi)^2 -
\ft1{48}e^{\vec a\cdot\vec\phi}F_{[4]}^2 - \ft1{12}\sum_ie^{\vec
a_i\cdot\vec\phi}(F^i_{[3]})^2\nonumber\\
&&\hspace{1.5cm}-\ft14\sum_{i<j}e^{\vec
a_{ij}\cdot\vec\phi}(F^{ij}_{[2]})^2 -
\ft14\sum_ie^{\vec b_i\cdot\phi}({\cal F}^i_{[2]})^2\label{I_D}\\
&&\hspace{1.5cm}-\ft12\sum_{i<j<k}e^{\vec
a_{ijk}\cdot\vec\phi}(F^{ijk}_{[1]})^2 -
\ft12\sum_{ij}e^{\vec b_{ij}\cdot\phi}({\cal F}^{ij}_{[1]})^2\Big] + {\cal
L}_{FFA}\ ,\nonumber
\eea
where $i,j = 1,\ldots,11-D$, and field strengths with multiple $i,j$
indices may be taken to be antisymmetric in those indices since these
``internal'' indices arise in the stepwise reduction procedure, and two
equal index values never occur in a multi-index sum. From (\ref{Fred}), one
sees that the ``straight-backed'' field strengths $F_{[4]}$, $F^i_{[3]}$,
$F^{ij}_{[2]}$ and $F^{ijk}_{[1]}$ are descendants from $F_{[4]}$ in $D=11$. The
``calligraphic'' field strengths ${\cal F}^i_{[2]}$, on the other hand, are the
field strengths for the Kaluza-Klein vectors ${\cal A}^i_{\sst M}$ appearing in
(\ref{met}b). Similarly, one also has a set of 1-form field strengths ${\cal
F}^{ij}_{[1]}$ for the Kaluza-Klein zero-form gauge potentials ${\cal
A}_{[0]}^{ij}$ appearing in (\ref{met}b).

     The nonlinearity of the original $D=11$ action (\ref{D11act}) in the
metric tensor produces a consequent nonlinearity in the $(11-D)$ dilatonic
scalar fields $\vec\phi$ appearing in the exponential prefactors of the
antisymmetric-tensor kinetic terms in (\ref{I_D}). For each field-strength
kinetic term in (\ref{I_D}), there is a corresponding ``dilaton vector'' of
coefficients determining the linear combination of the dilatonic scalars
appearing in its exponential prefactor. For the 4-, 3-, 2- and 1-form
``straight-backed'' field strengths emerging from $F_{[4]}$ in $D=11$, these
coefficients are denoted correspondingly $\vec a$, $\vec a_i$, $\vec a_{ij}$
and $\vec a_{ijk}$; for the ``calligraphic'' field strengths corresponding to
Kaluza-Klein vectors and zero-form gauge potentials emerging out of the metric,
these are denoted $\vec b_i$ and $\vec b_{ij}$. However, not
all of these dilaton vectors are independent; in fact, they may all be
expressed in terms of the 4-form and 3-form dilaton vectors $\vec a$ and
$\vec a_{ij}$:\,\cite{lpsol,lp}
\be\begin{array}{rclrcl}
\vec a_{ij} &\makebox[0pt]{=}& \vec a_i+\vec a_j-\vec a&\hspace{1cm} b_i
&\makebox[0pt]{=}& -\vec a_i + \vec a\\
\vec a_{ijk}&\makebox[0pt]{=}&\vec a_i + \vec a_j +\vec a_k -
2\vec a&\hspace{1cm}\vec b_{ij}&\makebox[0pt]{=}&-\vec a_i + \vec a_j
\ .
\end{array}\label{dilvecrels}
\ee
Another important feature of the dilaton vectors is that they satisfy the
following dot-product relations:
\bea
\vec a\cdot\vec a &=& {2(11-D)\over D-2}\nonumber\\
\vec a\cdot\vec a_i &=& {2(8-D)\over D-2}\label{dilvecprods}\\
\vec a_i\cdot\vec a_j &=& 2\delta_{ij} + {2(6-D)\over D-2}\ .\nonumber
\eea

     Throughout this discussion, we have emphasized {\em consistent
truncations} in making simplifying restrictions of complicated systems of
equations, so that the solutions of a simplified system are nonetheless
perfectly valid solutions of the more complicated untruncated system. With the
equations of motion following from (\ref{I_D}) we face a complicated system
that calls for analysis in simplified subsectors. Accordingly, we now seek a
consistent truncation down to a simplified system of the form  (\ref{igen}),
retaining just one dilatonic scalar combination $\phi$ and one rank-$n$ field
strength combination $F_{[n]}$ constructed out of a certain number $N$ of
``retained'' field strengths $F_{\alpha\,[n]}$, $\alpha=1,\ldots,N$, (this
could possibly be a straight-backed/calligraphic mixture) selected from
those appearing in (\ref{I_D}), with all the rest being set to zero.\cite{lp}
Thus, we let
\be
\vec\phi=\vec n\phi + \vec\phi_\perp\ ,\label{phidecomp}
\ee
where $\vec n\cdot\vec\phi_\perp = 0$; in the truncation we then seek to set
consistently $\vec\phi_\perp = 0$.

     We shall see that consistency for the retained field strengths
$F_{\alpha\,[n]}$ requires them all to be proportional.\cite{lp} We shall let
the dot product matrix for the dilaton vectors of the retained field strengths
be denoted $M_{\alpha\beta}=:\vec a_\alpha\cdot \vec a_\beta$. Consistency of
the truncation requires that the
$\phi_\perp$ field equation be satisfied:
\be
\square\vec\phi_\perp - \sum_\alpha \Pi_\perp\cdot\vec
a_\alpha(F_{\alpha\,[n]})^2 = 0\ ,\label{phiperpeq}
\ee
where $\Pi_\perp$ is the projector into the dilaton-vector subspace
orthogonal to the retained dilaton direction $\vec n$. Setting
$\vec\phi_\perp = 0$ in (\ref{phiperpeq}) and letting the retained
$F_{\alpha\,[n]}$ be proportional, one sees that achieving consistency is
hopeless unless all the $e^{\vec a_\alpha\cdot\vec\phi}$ prefactors are
the same, thus requiring
\be
\vec a_\alpha\cdot\vec n = a \hspace{.5cm}\forall\alpha=1,\ldots,N\ ,
\label{anprod}
\ee
where the constant $a$ will play the role of the dilatonic scalar
coefficient in the reduced system (\ref{igen}). Given a set of dilaton
vectors for retained field strengths satisfying (\ref{anprod}),
consistency of (\ref{phiperpeq}) with the imposition of $\vec\phi_\perp=0$
requires
\be
\Pi_\perp\cdot\sum_\alpha\vec a_\alpha(F_{\alpha\,[n]})^2 = 0\ .
\ee
This equation requires, for every point $x^{\sst M}$ in spacetime, that
the combination $\sum_\alpha\vec a_\alpha(F_{\alpha\,[n]})^2$ be parallel
to $\vec n$ in the dilaton-vector space. Combining this with the
requirement (\ref{anprod}), one has
\be
\sum_\alpha\vec a_\alpha(F_{\alpha\,[n]})^2 = a\vec
n\sum_\alpha(F_{\alpha\,[n]})^2\ .
\ee
Taking then a dot product of this with $\vec a_\beta$, one has
\be
\sum_\alpha
M_{\beta\alpha}(F_{\alpha\,[n]})^2=a^2\sum_\alpha(F_{\alpha\,[n]})^2\ .
\label{MF2}
\ee
Detailed analysis~\cite{lp} shows it to be sufficient to consider the
cases where $M_{\alpha\beta}$ is invertible, so by applying
$M^{-1}_{\alpha\beta}$ to (\ref{MF2}), one finds
\be
(F_{\alpha\,[n]})^2 = a^2\sum_\beta
M^{-1}_{\alpha\beta}\sum_\gamma(F_{\gamma\,[n]})^2\ ,
\ee
and, indeed, we find that the $F_{\alpha\,[n]}$ must all be proportional.
Summing on $\alpha$, one has
\be
a^2=(\sum_{\alpha,\beta}M^{-1}_{\alpha\beta})^{-1}\ ;\label{arel}
\ee
one then defines the retained field-strength combination $F_{[n]}$ so that
\be
(F_{\alpha\,[n]})^2=a^2\sum_\beta M^{-1}_{\alpha\beta}(F_{[n]})^2\
.\label{Fprop}
\ee

     The only remaining requirement for consistency of the truncation down to
the simplified ($g_{\sst MN}$, $\phi$, $F_{[n]}$) system (\ref{igen})
arises from the necessity to ensure that the variation of the ${\cal
L}_{FFA}$ term in (\ref{I_D}) is not inconsistent with setting to zero the
discarded dilatonic scalars and gauge potentials. In general, this imposes
a somewhat complicated requirement. In the present review,
however, we shall concentrate mainly on either purely-electric cases
satisfying the elementary ansatz (\ref{elans}) or purely-magnetic
cases satisfying the solitonic ansatz (\ref{magans}). As one can see
by inspection, for pure electric or magnetic solutions of these sorts, the
terms that are dangerous for consistency arising from the variation of
${\cal L}_{FFA}$ all vanish. Thus, for such solutions one may safely
ignore the complications of the ${\cal L}_{FFA}$ term. This restriction to
pure electric or magnetic solutions does, however, leave out the very
interesting cases of dyonic solutions that exist in $D=8$ and $D=4$, upon
which we shall comment later on in Section \ref{sec:dualities}.

     After truncating down to the system (\ref{igen}), the analysis proceeds as
in Section~\ref{sec:pbraneans}. It turns out~\cite{lp} that {\em
supersymmetric} $p$-brane solutions arise when the matrix
$M_{\alpha\beta}$ for the retained $F_{\alpha\,[n]}$ satisfies
\be
M_{\alpha\beta}=4\delta_{\alpha\beta} - {2d\tilde d\over D-2}\ ,
\ee
and the corresponding $\Delta$ value for $F_{[n]}$ is
\be
\Delta={4\over N}\ ,
\ee
where we recall that $N$ is the number of retained field strengths. A
generalization of this analysis leads to a classification of solutions
with more than one independent retained scalar-field combination.\cite{lp} We
shall see in Section \ref{sec:intersectingbranes} that the $N>1$ solutions to
single-charge truncated systems (\ref{igen}) may also be interpreted as special
solutions of the full reduced action (\ref{I_D}) containing $N$ constituent
$\Delta=4$ brane components that just happen to have coincident charge centers.
Consequently, one may consider only the $N=1$, $\Delta=4$ solutions to be
fundamental.

\subsection{Diagonal dimensional reduction of $p$-branes}\label{ssec:diagred}

     The family of $p$-brane solutions is ideally suited to interpretation as
solutions of Kaluza-Klein reduced theories, because they are naturally
independent of the ``worldvolume'' $x^\mu$ coordinates. Accordingly, one may
let the reduction coordinate $z$ be one of the $x^\mu$. Consequently, the only
thing that needs to be done to such a solution in order to {\em reinterpret} it
as a solution of a reduced system (\ref{Ired}) is to perform a Weyl rescaling
on it in order to be in accordance with the form of the metric chosen in the
Kaluza-Klein ansatz (\ref{met}), which was adjusted so as to maintain the
Einstein-frame form of the gravitational term in the dimensionally
reduced action.

     Upon making such a reinterpretation, elementary/solitonic $p$-branes
in $(D+1)$ dimensions give rise to elementary/solitonic $(p-1)$-branes in
$D$ dimensions, corresponding to the {\em same} value of $\Delta$, as one
can see from (\ref{deltared},\,\ref{deltahigh}). Note that in this process,
the quantity $\tilde d$ is conserved, since both $D$ and $d$ reduce by one.
Reinterpretation of $p$ brane solutions in this way, corresponding to
standard Kaluza-Klein reduction on a worldvolume coordinate, proceeds 
diagonally on a $D$ versus $d$ plot, and hence is referred to as {\em
diagonal} dimensional reduction. This procedure is the analogue, for
supergravity field-theory solutions, of the procedure of {\em double
dimensional reduction}~\cite{dhis} for $p$-brane worldvolume actions,
which can be taken to constitute the $\delta$-function {\em sources} for
singular $p$-brane solutions, coupled in to resolve the singularities, as we
discussed in subsection \ref{ssec:charges}.

\subsection{Multi-center solutions and vertical dimensional
reduction}\label{ssec:vertical}

     As we have seen, translational Killing symmetries of $p$-brane solutions
allow a simultaneous interpretation of these field configurations as solutions
belonging to several different supergravity theories, related one to another by
Kaluza-Klein dimensional reduction. For the original single $p$-brane solutions
(\ref{pbranesol}), the only available translational Killing symmetries are
those in the worldvolume directions, which we have exploited in describing
diagonal dimensional reduction above. One may, however, generalize the basic
solutions (\ref{pbranesol}) by replacing the harmonic function $H(y)$ in
(\ref{phisol}) by a different solution of the Laplace equation (\ref{laplace}).
Thus, one can easily extend the family of $p$-brane solutions to {\em
multi-center} $p$-brane solutions by taking the harmonic function to be
\be
H(y) = 1 + \sum_\alpha {k_\alpha\over|\vec y-\vec y_\alpha|^{\tilde
d}}\hspace{1cm}k_\alpha>0\ .
\label{multiharmonic}
\ee
Once again, the integration constant has been adjusted to make
$H\for_\infty = 1 \leftrightarrow \phi\for_\infty = 0$. The generalized
solution (\ref{multiharmonic}) corresponds to {\em parallel} and {\em
similarly-oriented} $p$-branes, with all charge parameters $\lambda_\alpha =
2\tilde d k_\alpha/\sqrt\Delta$ required to be positive in order to avoid naked
singularities. The ``centers'' of the individual ``leaves'' of this
solution are at the points $y=y_\alpha$, where $\alpha$ ranges over any
number of centers. The metric and the electric-case antisymmetric tensor
gauge potential corresponding to (\ref{multiharmonic}) are given again in
terms of $H(y)$ by (\ref{pbranesol}a,\ref{csol}). In the solitonic case,
the ansatz  (\ref{magans}) needs to be modified so as to accommodate the
multi-center form of the solution:
\be
F_{m_1\ldots m_n} = -\tilde d^{-1}\epsilon_{m_1\ldots
m_np}\partial_p\sum_\alpha{\lambda_\alpha\over|\vec y-\vec
y_\alpha|^{\tilde d}}\ ,\label{multimagsol}
\ee
which ensures the validity of the Bianchi identity just as well
as (\ref{magans}) does. The mass/(unit $p$-volume) density is now
\be
{\cal E} = {2\Omega_{D-d-1}\over\sqrt\Delta}\sum_\alpha\lambda_\alpha\ ,
\ee
while the total electric or magnetic charge is given by
$\Omega_{D-d-1}\sum\lambda_\alpha$, so the Bogomol'ny bounds (\ref{bogbounds})
are saturated just as they are for the single-center solutions
(\ref{pbranesol}). Since the multi-center solutions given by
(\ref{multiharmonic}) satisfy the same supersymmetry-preservation
conditions on the metric and antisymmetric tensor as (\ref{pbranesol}), the
multi-center solutions leave the same amount of supersymmetry unbroken as
the single-center solution.

     From a mathematical point of view, the multi-center solutions
(\ref{multiharmonic}) exist owing to the properties of the Laplace
equation (\ref{laplace}). From a physical point of view, however, these
static solutions exist as a result of {\em cancellation} between {\em
attractive} gravitational and scalar-field forces against {\em repulsive}
antisymmetric-tensor forces for the similarly-oriented $p$-brane
``leaves.''

     The multi-center solutions given by (\ref{multiharmonic}) can now be
used to prepare solutions adapted to dimensional reduction in the {\em
transverse} directions. This combination of a modification of the solution
followed by dimensional reduction on a transverse coordinate is called
{\em vertical} dimensional reduction~\cite{vertical} because it relates
solutions vertically on a $D$ versus $d$ plot.\footnote{Similar procedures
have been considered in a number of articles in the literature; see,
\eg Refs.\cite{kghl}} In order to do this, we need first to
develop translation invariance in the transverse reduction coordinate.
This can be done by ``stacking'' up identical
$p$ branes using (\ref{multiharmonic}) in a periodic array, \ie by
letting the integration constants $k_\alpha$ all be equal, and aligning
the ``centers'' $y_\alpha$ along some axis, \eg the $z$ axis.
Singling out one ``stacking axis'' in this way clearly destroys the overall
isotropic symmetry of the solution, but, provided the centers are all in
a line, the solution will nonetheless remain isotropic in the $D-d-1$
dimensions orthogonal to the stacking axis.  Taking the limit of a
densely-packed infinite stack of this sort, one has
\begin{subeqnarray}
\sum_\alpha{k_\alpha\over|\vec y - \vec y_\alpha|^{\tilde
d}} &\longrightarrow& \int_{-\infty}^{+\infty}{kdz\over(\hat r^2 +
z^2)^{\tilde d/2}} = {\tilde k\over\tilde r^{\tilde d-1}}\\
\hat r^2 &=& \sum_{m=d}^{D-2}y^my^m\\
\hat k &=& {\sqrt\pi k\Gamma(\tilde d - \ft12)\over2\Gamma(\tilde d)}\ ,
\label{stackint}
\end{subeqnarray}
where $\hat r$ in (\ref{stackint}b) is the radial coordinate for the
$D-d-1$ residual isotropic transverse coordinates. After a conformal
rescaling in order to maintain the Einstein frame for the solution, one can
finally reduce on the coordinate $z$ along the stacking axis.

     After stacking and reduction in this way, one obtains a $p$-brane solution
with the {\em same} worldvolume dimension as the original higher-dimensional
solution that was stacked up. Since the same antisymmetric tensors are used here
to support both the stacked and the unstacked solutions, and since $\Delta$ is
preserved under dimensional reduction, it follows that vertical dimensional
reduction from $D$ to $D-1$ spacetime dimensions preserves the value of $\Delta$
just like the diagonal reduction discussed in the previous subsection. Note that
under vertical reduction, the worldvolume dimension
$d$ is preserved, but $\tilde d=D-d-2$ is reduced by one with each reduction step.

\begin{figure}[ht]
\leavevmode\centering
\epsfbox{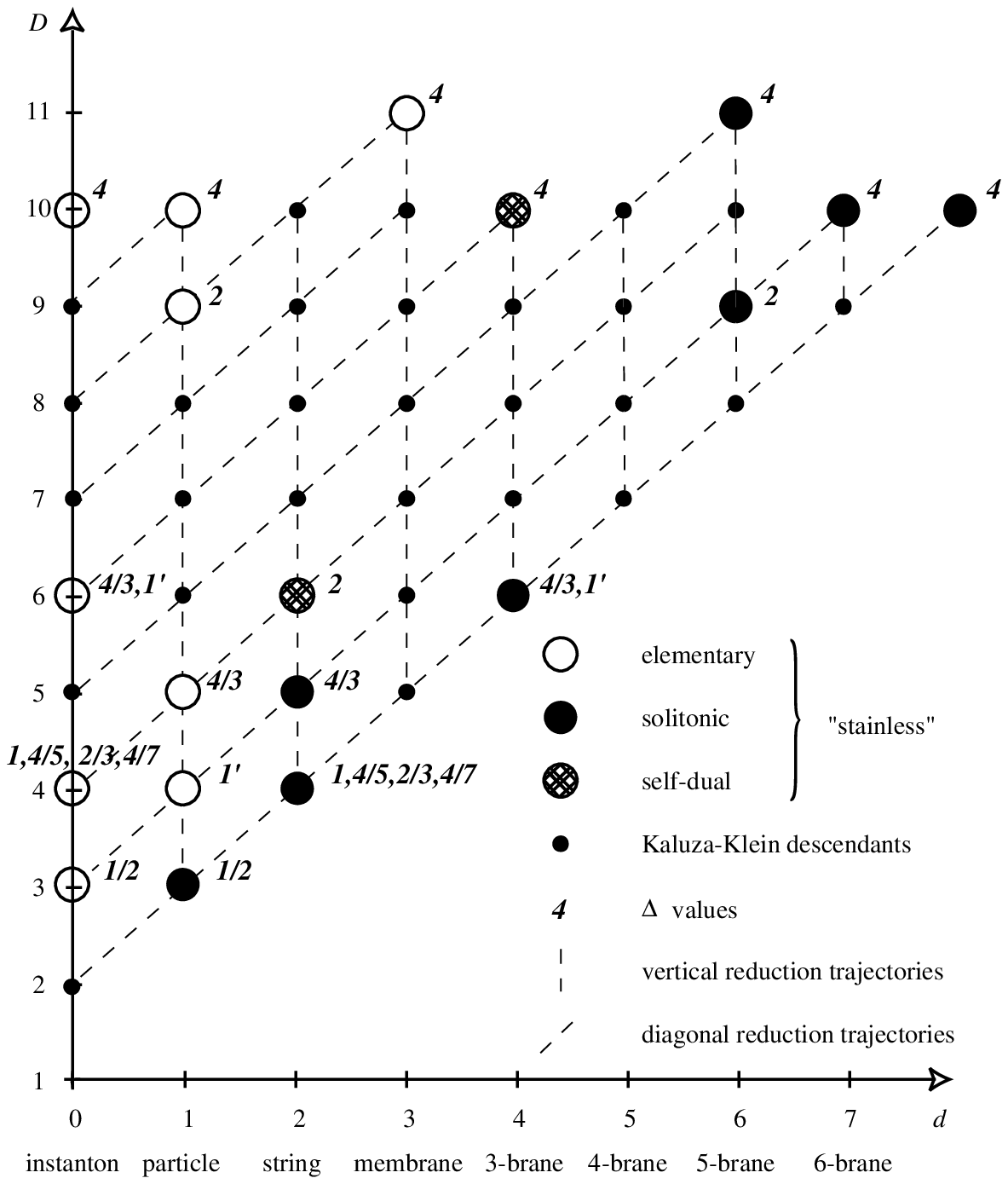}
\caption{Brane-scan of supergravity $p$-brane
solutions ($p\le (D-3)$)\label{fig:nbscan}}
\end{figure}

     Combining the diagonal and vertical dimensional reduction trajectories of
``descendant'' solutions, one finds the general picture given in the plot of
Figure \ref{fig:nbscan}. In this plot of spacetime dimension $D$ versus
worldvolume dimension $d$, reduction families  emerge from certain basic solutions
that cannot be ``oxidized'' back up to higher-dimensional isotropic $p$-brane
solutions, and hence can be called ``stainless'' $p$-branes.\cite{stainless} In
Figure \ref{fig:nbscan}, these solutions are indicated by the large circles, with
the corresponding $\Delta$ values shown adjacently. The indication of elementary
or solitonic type relates to solutions of supergravity theories in versions with
the lowest possible choice of rank ($n\le D/2$) for the supporting field strength,
obtainable by appropriate dualization. Of course, every solution to a theory
obtained by dimensional reduction from $D=11$ supergravity (\ref{D11act}) may be
oxidised back up to some solution in $D=11$. We shall see in Section
\ref{sec:intersectingbranes} that what one obtains upon oxidation of the
``stainless'' solutions in Figure \ref{fig:nbscan} falls into the interesting
class of ``intersecting branes'' built from four basic ``elemental'' solutions of
$D=11$ supergravity.

\clearpage

\subsection{The geometry of $(D-3)$-branes}\label{ssec:dm3branes}

     The process of vertical dimensional reduction described in the
previous subsection proceeds uneventfully until one makes
the  reduction from a $(D,d=D-3)$ solution to a $(D-1,d=D-3)$ 
solution.\footnote{Solutions with worldvolume dimension two less than the
spacetime dimension will be referred to generally as $(D-3)$-branes,
irrespective of whether the spacetime dimension is $D$ or not.} In this
step, the integral (\ref{stackint}) contains an additive divergence and
needs to be renormalized. This is easily handled by putting finite limits
$\pm L$ on the integral, which becomes $\int_{-L}^L d\tilde z(r^2 +\tilde
z^2)^{-1/2}$, and then by subtracting a divergent term $2\ln L$ before
taking the limit $L\rightarrow\infty$. Then the integral gives the expected
$\ln\hat r$ harmonic function appropriate to two transverse dimensions.

     Before proceeding any further with vertical dimensional reduction, let us
consider some of the specific properties of $(D-3)$-branes that make the next
vertical step down problematic. Firstly, the asymptotic metric of a
$(D-3)$-brane is not a globally flat space, but only a locally flat space.
This distinction means that there is in general a deficit solid angle at
transverse infinity, which is related to the total mass density of
the $(D-3)$-brane.\cite{d-3br} This means that any attempt to stack up
$(D-3)$-branes within a standard supergravity theory will soon consume the
entire solid angle at transverse infinity, thus destroying the asymptotic
spacetime in the construction.

     In order to understand the global structure of the $(D-3)$-branes in
some more detail, consider the supersymmetric string in $D=4$
dimensions.\cite{dghrr} In $D=4$, one may dualize the 2-form $A_{\mu\nu}$
field to a pseudoscalar, or axion, field $\chi$, so such strings are also
solutions to dilaton-axion gravity. The $p$-brane ansatz gives a spacetime of
the form
${\cal M}^4={\cal M}^2\times\Sigma^2$, where ${\cal M}^2$ is $D=2$
Minkowski space. Supporting this string solution, one has the 2-form gauge
field $A_{\mu\nu}$ and the dilaton $\phi$. These fields give rise to a
field stress tensor of the form
\bea
T_{\mu\nu}(A,\phi) &=&
-\ft1{16}(a^2+4)\partial_mK\partial_mK\eta_{\mu\nu}\nonumber\\
T_{mn}(A,\phi) &=& \ft18(a^2-4)(\partial_mK\partial_nK-\ft12
\delta_{mn}(\partial_pK\partial_pK))\ ,\label{stresstens}
\eea
where $a$ is as usual the dilaton coupling parameter and $e^{-K} = H =
1-8GT\ln(r)$, with $r=\sqrt{y^my^m}$, $m=2,\ 3$. If one now puts in an
elementary string source action, with the string aligned along the
$\mu,\nu=0,1$ subspace, so that $T_{mn\,({\rm source})}=0$, then one has the
source stress tensor
\be
T_{\mu\nu\,({\rm source})} = {-T\over\sqrt{-g}}\int
d^2\xi\sqrt{-\gamma}\gamma^{ij}
\partial_iX_\mu\partial_jX_\nu e^{-\fft12a\phi}\delta(x-X)\
.\label{stringsource}
\ee
By inspection of the field solution, one has $T_{mn}(A,\phi)=0$, while the
contributions to $T_{\mu\nu}$ from the $A_{\mu\nu}$ and $\phi$ fields and
also from the source (\ref{stringsource}) are both of the form ${\rm
diag}(\rho,-\rho)$. Thus, the overall stress tensor is of the form $T_{MN}={\rm
diag}(\rho,-\rho,0,0)$.

     Consequently, the Einstein equation in the transverse $m$, $n$ indices
becomes $R_{mn}-\ft12g_{mn}R=0$, since the transverse stress tensor components
vanish. This equation is naturally satisfied for a metric satisfying
the $p$-brane ansatz, because, as one can see from (\ref{riccicomps}) with
$\tilde d=A'=0$, this causes the transverse components of the Ricci tensor to
be equal to the Ricci tensor of a $D=2$ spacetime, for which
$R_{mn}-\ft12g_{mn}R\equiv 0$ is an {\it identity,} corresponding to the fact
that the usual Einstein action, $\sqrt{-g}R$, is a topological invariant in $D=2$.
Accordingly, in the transverse directions, the equations are satisfied simply by
by $0=0$.

     In the world-sheet directions, the equations become
\be
-\ft12R\eta_{\mu\nu}=-8\pi G\rho\eta_{\mu\nu}\ ,\nonumber
\ee
or just
\be
R=16\pi G\rho\ ,\label{stringeq}
\ee
and as we have already noted, $R=R_{mm}$. Owing to the fact that the $D=2$
Weyl tensor vanishes, the transverse space $\Sigma^2$ is conformally flat;
Eq.\ (\ref{stringeq}) gives its conformal factor. Thus, although there is
no sensible Einstein action in the transverse $D=2$, space, a usual form of the
Einstein equation nonetheless applies to that space as a result of the
symmetries of the $p$-brane ansatz.

     The above supersymmetric string solution may be compared to the cosmic
strings arising in gauge theories with spontaneous symmetry breaking.
There, the Higgs fields contributing to the energy density of the string
are displaced from their usual vacuum values to unbroken-symmetry
configurations at a stationary point of the Higgs potential, within a very
small transverse-space region that may be considered to be the string ``core.''
Approximating this by a delta function in the transverse space, the Ricci
tensor and hence the full curvature {\it vanish} outside the string core,
so that one obtains a {\it conical} spacetime, which is flat except at the
location of the string core. The total energy is given by the {\it
deficit angle} $8\pi GT$ of the conical spacetime. In contrast, the
supersymmetric string has a field stress tensor $T_{\mu\nu}(A,\phi)$ which is
not just concentrated at the string core but instead is smeared out over
spacetime. The difference arises from the absence of a potential for the fields
$A_{\mu\nu}$, $\phi$ supporting the solution in the supersymmetric case.
Nonetheless, as one can see from the behavior of the stress tensor $T_{mn}$ in
Eq.\ (\ref{stresstens}), the transverse space $\Sigma^2$ is asymptotically
locally flat (ALF), with a total energy density given by the overall deficit
angle measured at infinity. For multiple-centered string solutions, one has
\be
H=1-\sum_i8GT_i\ln|\vec y-\vec y_i|\ .\label{multistring}
\ee
Consequently, when considered within the original supergravity theory, the
indefinite stacking of supersymmetric strings leads to a destruction of
the transverse asymptotic space.

     A second problem with any attempt to produce $(D-2)$-branes in ordinary
supergravity theories is simply stated: starting from the $p$-brane
ansatz (\ref{ansatz},\,\ref{magans}) and searching for $(D-2)$ branes in
ordinary massless supergravity theories, one simply doesn't find any such
solutions.

\subsection{Beyond the $(D-3)$-brane barrier: Scherk-Schwarz reduction and
domain walls}\label{ssec:domainwalls}

     Faced with the above puzzles about what sort of $(D-2)$-brane could result
by vertical reduction from a $(D-3)$-brane, one can simply decide to be brave,
and to just proceed anyway with the established mathematical procedure of
vertical dimensional reduction and see what one gets. In the next step of
vertical dimensional reduction, one again encounters an additive
divergence: the integral $\int_{-L}^L dz \ln(y^2 + z^2)$ needs to be
renormalized by subtracting a divergent term $4L(\ln L -1)$. Upon subsequently
performing the integral, the harmonic function $H(y)$ becomes linear in the one
remaining transverse coordinate.

     While the mathematical procedure of vertical dimensional
reduction so as to produce some sort of $(D-2)$-brane proceeds apparently
without serious complication, an analysis of the physics of the situation needs
some care.\cite{domain} Consider the reduction from a $(D,d=D-2)$ solution (a
$p=(D-3)$ brane) to a $(D-1,d=D-3)$ solution (a $p=(D-2)$ brane). Note that both
the $(D-3)$ brane and its descendant $(D-2)$-brane have harmonic functions
$H(y)$ that blow up at infinity. For the $(D-3)$-brane, however this is not 
in itself particularly remarkable, because, as one can see by inspection of
(\ref{pbranesol}) for this case, the metric asymptotically tends 
to a locally flat space as $r\rightarrow\infty$, and also in this limit the
antisymmetric-tensor one-form field strength
\be
F_m = -\epsilon_{mn}\partial_n H\label{oneformfs}
\ee
tends asymptotically to zero, while the dilatonic scalar $\phi$ tends to its modulus
value $\phi_\infty$ (set to zero for simplicity in (\ref{pbranesol})).  The expression
(\ref{oneformfs}) for the field strength, however, shows that the next reduction step
down to the $(D-1,d=D-2)$ solution has a significant new feature: upon stacking up
$(D-3)$ branes prior to the vertical reduction, thus producing a linear harmonic
function in the transverse coordinate $y$,
\be
H(y) = \mbox{const.} + my\ ,\label{linharmonic}
\ee
the field strength (\ref{oneformfs}) acquires a {\em constant} component along
the stacking axis $\leftrightarrow$ reduction direction $z$,
\be
F_z = -\epsilon_{zy}\partial_y H = m\ ,\label{constF}
\ee
which implies an unavoidable dependence\,\footnote{Note that this
vertical reduction from a $(D-3)$-brane to a $(D-2)$-brane is the first
case in which one is {\em forced} to accept a dependence on the reduction
coordinate $z$; in all higher-dimensional vertical reductions, such $z$
dependence can be removed by a gauge transformation. The zero-form gauge
potential in (\ref{zerofmred}) does not have the needed gauge symmetry,
however.} of the corresponding zero-form gauge potential on the reduction
coordinate:
\be
A_{[0]}(x,y,z) = mz + \chi(x,y)\ .\label{zerofmred}
\ee

     From a Kaluza-Klein point of view, the unavoidable linear
dependence of a gauge potential on the reduction coordinate given in
(\ref{zerofmred}) appears to be problematic. Throughout this review,
we have dealt only with {\em consistent} Kaluza-Klein reductions, for
which solutions of the reduced theory are also solutions of the
unreduced theory. Generally, retaining any dependence on a reduction
coordinate will lead to an {\em inconsistent truncation} of the
theory: attempting to impose a $z$ dependence of the form given in
(\ref{zerofmred}) prior to varying the Lagrangian will give a result
different from that obtained by imposing this dependence in the field
equations after variation.

     The resolution of this difficulty is that in performing a Kaluza-Klein reduction
with an ansatz like (\ref{zerofmred}), one ends up outside the standard set of massless
supergravity theories. In order to understand this, let us again focus on the problem
of consistency of the Kaluza-Klein reduction. As we have seen, consistency of any
restriction means that the restriction may either be imposed on the field variables in
the original action prior to variation so as to derive the equations of motion, or
instead may be imposed on the field variables in the equations of motion after
variation, with an equal effect. In this case, solutions obeying the restriction will
also be solutions of the general unrestricted equations of motion.

     The most usual guarantee of consistency in Kaluza-Klein dimensional
reduction is obtained by restricting the field variables to carry zero
charge with respect to some conserved current, {\it e.g.}\ momentum in the
reduction dimension. But this is not the only way in which consistency may
be achieved. In the present case, retaining a linear dependence on the
reduction coordinate as in (\ref{zerofmred}) would clearly produce
an inconsistent truncation if the reduction coordinate were to
appear explicitly in any of the field equations. But this does not imply
that a truncation is necessarily inconsistent just because a gauge potential
contains a term linear in the reduction coordinate. Inconsistency of a
Kaluza-Klein truncation occurs when the original, unrestricted, field
equations imply a condition that is inconsistent with the reduction
ansatz. If a particular gauge potential appears in the action
{\em only through its derivative,} \ie through its field strength,
then a consistent truncation may be achieved provided that the
restriction on the gauge potential implies that the field strength is
independent of the reduction coordinate. A zero-form gauge potential on
which such a reduction may be carried out, occurring in the action only
through its derivative, will be referred to as an {\em axion.}

     Requiring axionic field strengths to be independent of the reduction
coordinate amounts to extending the Kaluza-Klein reduction framework so as to
allow for {\em linear} dependence of an axionic zero-form potential on the
reduction coordinate, precisely of the form occurring in (\ref{zerofmred}).
So, provided $A_{[0]}$ is an axion, the reduction (\ref{zerofmred}) turns
out to be consistent after all. This extension of the Kaluza-Klein ansatz
is in fact an instance of {\em Scherk-Schwarz} reduction.\cite{ss,bdgpk}
The basic idea of Scherk-Schwarz reduction is to use an Abelian rigid symmetry of a
system of equations in order to generalize the reduction ansatz by allowing a linear
dependence on the reduction coordinate in the parameter of this Abelian symmetry.
Consistency is guaranteed by cancellations orchestrated by the
Abelian symmetry in field-equation terms where the parameter does not
get differentiated. When it does get differentiated, it contributes
only a term that is itself independent of the reduction coordinate.
In the present case, the Abelian symmetry guaranteeing consistency of
(\ref{zerofmred}) is a simple shift symmetry
$A_{[0]}\rightarrow A_{[0]}+\mbox{const.}$ 

     Unlike the original implementation of the Scherk-Schwarz reduction
idea,\cite{ss} which used an Abelian $U(1)$ phase symmetry acting on
spinors, the Abelian shift symmetry used here commutes with supersymmetry,
and hence the reduction does not spontaneously break supersymmetry.
Instead, gauge symmetries for some of the antisymmetric tensors will be
broken, with a corresponding appearance of mass terms. As with all
examples of vertical dimensional reduction, the $\Delta$ value
corresponding to a given field strength is also preserved. Thus,
$p$-brane solutions related by vertical dimensional reduction, even in the
enlarged Scherk-Schwarz sense, preserve the same amount of unbroken
supersymmetry and have the same value of $\Delta$.    

     It may be necessary to make several redefinitions and integrations by
parts in order to reveal the axionic property of a given zero-form, and
thus to prepare the theory for a reduction like (\ref{zerofmred}).
This is most easily explained by an example, so let us consider the first
possible Scherk-Schwarz reduction\,\footnote{A higher-dimensional
Scherk-Schwarz reduction is possible~\cite{bdgpk} starting from type IIB
supergravity in $D=10$, using the axion appearing in the
\ffrac{$\rm{SL}(2,\R)$}{${\rm SO}(2)$} scalar sector of that theory.} in
the sequence of theories descending from (\ref{D11act}), starting in
$D=9$ where the first axion field appears.\cite{domain} The Lagrangian for
massless $D=9$ maximal supergravity is obtained by specializing the general
dimensionally-reduced action (\ref{I_D}) given in
Section \ref{sec:pbraneans} to this case:
\cramp\bea
{\cal L}_9 &=& \sqrt{-g}\Big[R -\ft12(\partial\phi_1)^2 -\ft12
(\partial\phi_2)^2 -
\ft12 e^{-\ft32\phi_1+{\sqrt7\over2}\phi_2} (\partial\chi)^2 -\ft1{48}
e^{\vec a\cdot \vec\phi} (F_{[4]})^2\nonumber\\
&&- \ft12 e^{\vec a_1\cdot \vec\phi} (F_{[3]}^{(1)})^2 -\ft12
e^{\vec a_2\cdot \vec\phi} (F_{[3]}^{(2)})^2 
-\ft14 e^{\vec a_{12}\cdot \vec\phi} (F_{[2]}^{(12)})^2 -\ft14
e^{\vec b_1\cdot\vec\phi} ({\cal F}_{[2]}^{(1)})^2\nonumber\\
&&- \ft14 e^{\vec b_2\cdot \vec\phi} ({\cal F}_{[2]}^{(2)})^2\Big]
-\ft12 \tilde F_{[4]}\wedge \tilde F_{[4]} \wedge A_{[1]}^{(12)} -
\tilde F_{[3]}^{(1)} \wedge \tilde F_{[3]}^{(2)} \wedge A_{[3]}\
,\label{d9lag}
\eea\uncramp
where $\chi={\cal A}_{[0]}^{(12)}$ and $\vec\phi=(\phi_1,\phi_2)$.

     Within the scalar sector $(\vec\phi,\chi)$ of (\ref{d9lag}), the
dilaton coupling has been made explicit; in the rest of the Lagrangian,
the dilaton vectors have the general structure given in
(\ref{dilvecrels},\,\ref{dilvecprods}). The scalar sector of (\ref{d9lag})
forms a nonlinear $\sigma$-model for the manifold \ffrac{${\rm
GL}(2,\R)$}{${\rm SO}(2)$}. This already makes it appear that one may
identify $\chi$ as an axion available for Scherk-Schwarz reduction.
However, account must still be taken of the Chern-Simons structure lurking
inside the field strengths in (\ref{Fred},\,\ref{d9lag}). In detail, the field
strengths are given by
\begin{subeqnarray} 
F_{[4]}&=&\tilde F_{[4]} - \tilde F_{[3]}^{(1)}\wedge {\cal A}_{[1]}^{(1)}
- \tilde F_{[3]}^{(2)}\wedge {\cal A}_{[1]}^{(2)}\nonumber\\
&&\quad\quad +\chi \tilde F_{[3]}^{(1)} \wedge {\cal A}_{[1]}^{(2)} -
\tilde F_{[2]}^{(12)}
\wedge {\cal A}_{[1]}^{(1)}
\wedge {\cal A}_{[1]}^{(2)}\\ 
F^{(1)}_{[3]} &=& \tilde F^{(1)}_{[3]} - \tilde F_{[2]}^{(12)} 
\wedge {\cal A}_{[1]}^{(2)}\\ 
F_{[3]}^{(2)} &=& \tilde F_{[3]}^{(2)} + \tilde F_{[2]}^{(12)}\wedge
{\cal A}_{[1]}^{(1)} -
\chi \tilde F^{(1)}_{[3]}\\ 
F_{[2]}^{(12)} &=& \tilde F^{(12)}_{[2]}\quad\quad
{\cal F}_{[2]}^{(1)} = \tilde{\cal F}_{[2]}^{(1)} -d\chi\wedge
{\cal A}_{[1]}^{(2)}\\
{\cal F}_{[2]}^{(2)} &=& \tilde {\cal F}_{[2]}^{(2)}
\quad\quad {\cal F}_{[1]}^{(12)} =  d\chi\ ,\label{csterms}
\end{subeqnarray}
where the field strengths carrying tildes are the na\"{\i}ve
expressions without Chern-Simons corrections, {\it i.e.}\ 
$\tilde F_{n]}=dA_{[n-1]}$. Now the appearance of undifferentiated $\chi$
factors in (\ref{csterms}a,c) makes it appear that a Scherk-Schwarz
reduction would be inconsistent. However, one may eliminate these
undifferentiated factors by making the field redefinition
\be
A_{[2]}^{(2)}\longrightarrow A_{[2]}^{(2)} +
\chi A_{[2]}^{(1)}\ ,\label{Aredef}
\ee
after which the field strengths (\ref{csterms}a,c) become
\begin{subeqnarray}
F_{[4]}&=&\tilde F_{[4]} - \tilde F_{[3]}^{(1)}\wedge {\cal
A}_{[1]}^{(1)} - \tilde F_{[3]}^{(2)}\wedge {\cal A}_{[1]}^{(2)}\nonumber\\
&&\quad\quad -d\chi \wedge A_{[2]}^{(1)}\wedge {\cal A}_{[1]}^{(2)} -
\tilde F_{[2]}^{(12)}\wedge {\cal A}_{[1]}^{(1)}\wedge {\cal
A}_{[1]}^{(2)}\\
\stepcounter{subequation}
F_{[3]}^{(2)} &=& \tilde F_{[3]}^{(2)} + F_{[2]}^{(12)}\wedge {\cal
A}_{[1]}^{(1)} +
d\chi \wedge A_{[2]}^{(1)}\label{newcsterms}\ ,
\end{subeqnarray}
the rest of (\ref{csterms}) remaining unchanged.

     After the field redefinitions (\ref{Aredef}), the axion field
$\chi={\cal A}_{[0]}^{(12)}$ is now ready for application of the Scherk-Schwarz
reduction ansatz (\ref{zerofmred}). The coefficient of the term linear in
the reduction coordinate $z$ has been denoted $m$ because it carries the
dimensions of mass, and correspondingly its effect on the reduced action
is to cause the appearance of mass terms. Applying (\ref{zerofmred}) to
the $D=9$ Lagrangian, one obtains a $D=8$ reduced
Lagrangian\,\footnote{I am grateful to Marcus Bremer for help in correcting 
some errors in the original expression of Eq.\ (\ref{massived8}) given 
in Ref.\,\cite{domain}}
\bea
\rlap{${\cal L}_{8\,{\rm ss}} ~~=~~$}&&\nonumber\\
&&\sqrt{-g}\Big[R -\ft12 
(\partial\phi_1)^2 -\ft12 (\partial\phi_2)^2  -\ft12 
(\partial\phi_3)^2 \nonumber\\
&&- \ft12 e^{\vec b_{12}\cdot \vec \phi}(\partial \chi - m 
{\cal A}_{[1]}^{(3)})^2 -\ft12 e^{\vec b_{13}\cdot \vec \phi}
(\partial {\cal A}_{[0]}^{(13)} - \partial\chi{\cal A}_{[0]}^{(23)} + m 
{\cal A}_{[1]}^{(2)})^2 \nonumber\\
&&-\ft12 e^{\vec b_{23}\cdot \vec \phi}(\partial {\cal A}_{[0]}^{(23)})^2 
-\ft12 e^{\vec a_{123}\cdot \vec \phi}(\partial A_{[0]}^{(123)})^2\nonumber\\
&&-\ft1{48} e^{\vec a\cdot\vec \phi} (F_{[4]}- m A_{[2]}^{(1)} \wedge
{\cal A}_{[1]}^{(2)}\wedge {\cal A}_{[1]}^{(3)})^2 
-\ft1{12} e^{\vec a_1\cdot
\vec\phi}(F_{[3]}^{(1)})^2 \nonumber\\
&&-\ft1{12} e^{\vec a_2\cdot
\vec\phi}(F_{[3]}^{(2)}- m A_{[2]}^{(1)}\wedge {\cal A}_{[1]}^{(3)})^2 
-\ft1{12} e^{\vec a_3\cdot
\vec\phi}(F_{[3]}^{(3)} + m A_{[2]}^{(1)} \wedge {\cal A}_{[1]}^{(2)} 
)^2\nonumber\\
&&-\ft14 e^{\vec a_{12}\cdot \vec \phi}(F_{[2]}^{(12)})^2 
-\ft14 e^{\vec a_{13}\cdot \vec \phi}(F_{[2]}^{(13)})^2
-\ft14 e^{\vec a_{23}\cdot \vec \phi}(F_{[2]}^{(23)}+m A_{[2]}^{(1)})^2
\nonumber\\
&&-\ft14 e^{\vec b_{1}\cdot \vec \phi}({\cal F}_{[2]}^{(1)}
- m {\cal A}_{[1]}^{(2)} \wedge {\cal A}_{[1]}^{(3)})^2
-\ft14 e^{\vec b_{2}\cdot \vec \phi}({\cal F}_{[2]}^{(2)})^2
-\ft14 e^{\vec b_{3}\cdot \vec \phi}({\cal F}_{[2]}^{(3)})^2 \nonumber\\
&& - \ft12 m^2 e^{\vec b_{123}\cdot\vec\phi}\Big] + mF_{[3]}^{(1)}\wedge 
A_{[2]}^{(1)}\wedge A_{[3]} + {\cal L}_{FFA}\ ,
\label{massived8}
\eea
where the dilaton vectors are now those appropriate for $D=8$; the term ${\cal
L}_{FFA}$ contains only $m$-independent terms. 

     It is apparent from
(\ref{massived8}) that the fields ${\cal A}_{[1]}^{(3)}$, ${\cal A}_{[1]}^{(2)}$
and $A_{[2]}^{(1)}$ have become massive. Moreover, there are field redefinitions
under which the fields $\chi$, ${\cal A}_{[0]}^{(13)}$ and $A_{[1]}^{(23)}$ may be
absorbed. One way to see how this absorption happens is to notice that
the action obtained from (\ref{massived8}) has a set of three
Stueckelberg-type gauge transformations under which 
${\cal A}_{[1]}^{(3)}$, ${\cal A}_{[1]}^{(2)}$ and $A_{[2]}^{(1)}$ transform
according to their standard gauge transformation laws. These three
transformations are accompanied, however, by various
compensating transformations necessitated by the Chern-Simons
corrections present in (\ref{massived8}) as well as by $m$-dependent
shift transformations of $\chi$, ${\cal A}_{[0]}^{(13)}$ and
$A_{[1]}^{(23)}$, respectively.  Owing to the presence of these local
shift terms in the three Stueckelberg symmetries, the fields $\chi$, ${\cal
A}_{[0]}^{(13)}$ and $A_{[1]}^{(23)}$ may be gauged to zero. After gauging these
three fields to zero, one has a clean set of mass terms in (\ref{massived8}) for
the fields ${\cal A}_{[1]}^{(3)}$, ${\cal A}_{[1]}^{(2)}$ and $A_{[2]}^{(1)}$.

     As one descends through the available spacetime dimensions for
supergravity theories, the number of axionic scalars available for a
Scherk-Schwarz reduction step increases. The numbers of axions are given
in the following Table:

\vspace{-.4cm}
\begin{table}[ht]
\centering
\caption{Supergravity axions versus spacetime dimension.\label{tab:axions}}
\vspace{0.4cm}
\begin{tabular}{|l|c|c|c|c|c|c|}
\hline
$D$&9&8&7&6&5&4\\
\hline
$N_{\rm axions}$&1&4&10&20&36&63\\
\hline
\end{tabular}
\end{table}

     Each of these axions gives rise to a {\em distinct} massive
supergravity theory upon Scherk-Schwarz reduction,\cite{domain} and each
of these reduced theories has its own pattern of mass generation. In
addition, once a Scherk-Schwarz reduction step has been performed, the
resulting theory can be further reduced using ordinary Kaluza-Klein
reduction. Moreover, the Scherk-Schwarz and ordinary Kaluza-Klein
processes do not commute, so the number of theories obtained by
the various combinations of Scherk-Schwarz and ordinary dimensional
reduction is cumulative. In addition, there are numerous possibilities of
performing Scherk-Schwarz reduction simultaneously on a number of axions. This can
be done either by arranging to cover a number of axions simultaneously with
derivatives, or by further Scherk-Schwarz generalisations of the Kaluza-Klein
reduction process.\cite{kkm} For further details on the panoply of Scherk-Schwarz
reduction possibilities, we refer the reader to Refs.\,\cite{domain,kkm}

     The single-step procedure of Scherk-Schwarz dimensional reduction described
above may be generalised to a procedure exploiting the various cohomology classes
of a multi-dimensional compactification manifold.\cite{domtop} The key to this
link between the Scherk-Schwarz generalised dimensional reduction and the topology
of the internal Kaluza-Klein manifold $\cal K$ is to recognise that the
single-step reduction ansatz (\ref{zerofmred}) may be generalised to
\be
A_{[n-1]}(x,y,z) = \omega_{[n-1]} + A_{[n-1]}(x,y)\ ,\label{n-1fmred}
\ee
where $\omega_{[n-1]}$ is an $(n-1)$ form defined locally on $\cal K$, whose
exterior derivative $\Omega_{[n]}=d\omega_{[n-1]}$ is an element of the cohomology
class
$H^n({\cal K},\R)$. For example, in the case of a single-step generalised
reduction on a circle $S^1$, one has $\Omega_{[1]}=mdz\in H^1(S^1,\R)$,
reproducing our earlier single-step reduction (\ref{zerofmred}). 

     As another example, consider a generalised reduction on a 4-torus $T^4$
starting in $D=11$, setting $A_{[3]}(x,y,z)\makebox[0pt]{\phantom{\Huge|}} =
\omega_{[3]} + A_{[3]}(x,y)$ with $\Omega_{[4]}=d\omega_{[3]}=mdz_1\wedge
dz_2\wedge dz_3\wedge dz_4\in H^4(T^4,\R)\makebox[0pt]{\phantom{\Big|}}$. In this
example, one may choose to write $\omega_{[3]}$ locally as
$\omega_{[3]}=mz_1dz_2\wedge dz_3\wedge dz_4$. All of the other fields are reduced
using the standard Kaluza-Klein ansatz, with no dependence on any of the $z_i$
coordinates. The theory resulting from this
$T^4$ reduction is a $D=7$ massive supergravity with a cosmological potential,
analogous to the $D=8$ theory (\ref{massived8}). The same theory (up to field
redefinitions) can also be obtained~\cite{domain} by first making an ordinary
Kaluza-Klein reduction from
$D=11$ down to $D=8$ on a 3-torus $T^3$, then making an $S^1$ single-step
generalised Scherk-Schwarz reduction (\ref{zerofmred}) from
$D=8$ to $D=7$. Although the $T^4$ reduction example simply reproduces a massive
$D=7$ theory that can also be obtained {\it via} the single-step ansatz
(\ref{zerofmred}), the recognition that one can use any of the $H^n({\cal K},\R)$
cohomology classes of the compactification manifold $\cal K$ significantly
extends the scope of the generalised reduction procedure. For example, it allows
one to make generalised reductions on manifolds such as K3 or on Calabi-Yau
manifolds.\cite{domtop}

     For our present purposes, the important feature of theories obtained
by Scherk-Schwarz reduction is the appearance of cosmological potential
terms such as the penultimate term in Eq.\ (\ref{massived8}). Such terms
may be considered within the context of our simplified action
(\ref{igen}) by letting the rank $n$ of the field strength take the value
zero. Accordingly, by consistent truncation of (\ref{massived8}) or of one
of the many theories obtained by Scherk-Schwarz reduction in lower
dimensions, one may arrive at the simple Lagrangian
\be
{\cal L}= \sqrt{-g}\Big[R -\ft12\nabla_{\sst M}\phi\nabla^{\sst M}\phi
-\ft12 m^2 e^{a \phi}\Big]\ .\label{coslag}
\ee
Since the rank of the form here is $n=0$, the elementary/electric type
of solution would have worldvolume dimension $d=-1$, which is not
very sensible, but the solitonic/magnetic solution has $\tilde
d=D-1$, corresponding to a $p=D-2$ brane, or {\em domain
wall,} as expected. Relating the parameter $a$ in (\ref{coslag}) to the
reduction-invariant parameter $\Delta$ by the standard formula
(\ref{Delta}) gives $\Delta=a^2-2(D-1)/(D-2)$; taking the corresponding
$p=D-2$ brane solution from (\ref{pbranesol}), one finds
\bsea
&ds^2 = H^{\ft4{\Delta(D-2)}} \, \eta_{\mu\nu}\, dx^\mu dx^\nu 
+ H^{\ft{4(D-1)}{\Delta(D-2)}} \, dy^2&\\
&e^\phi = H^{-2a/\Delta} \ ,&\label{dwsol}
\esea
where the harmonic function $H(y)$ is now a linear function of the single
transverse coordinate, in accordance with (\ref{linharmonic}).\footnote{Domain
walls solutions such as (\ref{dwsol}) in supergravity theories were found
for the $D=4$ case in Ref.\,\cite{c} and a review of them has
been given in Ref.\,\cite{cs}} The
curvature of the metric (\ref{dwsol}a) tends to zero at large values of
$|y|$, but it diverges if $H$ tends to zero. This latter singularity can be
avoided by taking $H$  to be
\be
H=\mbox{const.} + M|y|\label{dwH}
\ee
where $M=\ft12 m\sqrt\Delta$. With the choice (\ref{dwH}), there is just a
delta-function singularity at the location of the domain wall at $y=0$,
corresponding to the discontinuity in the gradient of $H$.

     The domain-wall solution (\ref{dwsol},\,\ref{dwH}) has the peculiarity
of tending asymptotically to flat space as $|y|\rightarrow\infty$,
within a theory that does not naturally admit flat space as a solution (by
``naturally,'' we are excluding the case $a\phi\rightarrow-\infty$). Moreover,
the theory (\ref{coslag}) does not even admit a non-flat maximally-symmetric
solution, owing to the complication of the cosmological potential. The domain-wall
solution (\ref{dwsol},\,\ref{dwH}), however, manages to ``cancel'' this potential
at transverse infinity, allowing at least asymptotic flatness for this solution.

     This brings us back to the other facets of the consistency problem for
vertical dimensional reduction down to $(D-2)$-branes as discussed in subsection
\ref{ssec:dm3branes}. There is no inconsistency between the existence of
domain-wall solutions like (\ref{dwsol},\,\ref{dwH}) and the inability to find
such solutions in standard supergravity theories, or with the conical-spacetime
character of $(D-3)$-branes, because these domain walls exist only in {\em massive}
supergravity theories like (\ref{massived8}), with a vacuum structure different
from that of standard massless supergravities. Because the Scherk-Schwarz
generalised dimensional reduction used to obtain them was a consistent truncation,
such domain walls can be oxidised back to solutions of higher-dimensional massless
supergravities, but in that case, they have the form of stacked solutions prepared
for vertical reduction, with non-zero field strengths in the reduction directions,
as in our example (\ref{constF}).

\section{Intersecting branes, scattering
branes}\label{sec:intersectingbranes}
\subsection{Multiple component solutions}\label{ssec:multicomps}

     Given the existence of solutions (\ref{Fprop}) with several active field
strengths $F^{\alpha}_{[n]}$, but with coincident charge centers, it is
natural to try to find solutions where the charge centers for the
different $F^{\alpha}_{[n]}$ are separated.\cite{boundstates} This will lead us
to a better understanding of the $\Delta\ne4$ solutions shown in Figure
\ref{fig:nbscan}. Consider a number of field strengths that individually
have $\Delta=4$ couplings, but now look for a solution where $\ell$ of
these field strengths are active, with centers $\vec y_\alpha$,
$\alpha=1,\ldots,\ell$. Let the charge parameter for $F^{\alpha}$ be
$\lambda^\alpha$. Thus, for example, in the magnetic case, one sets
\be
F^{\alpha}_{m_1,\ldots,m_n} =
\lambda^\alpha\epsilon_{m_1,\ldots,m_np}{y^p\over|\vec y-\vec
y_\alpha|^{n+1}}\ .\label{Falpha}
\ee
In both the electric and the magnetic cases, the $\lambda^\alpha$ are
related to the integration constants $k^\alpha$ appearing in the metric by
$k^\alpha=\lambda^\alpha/\tilde d$. Letting $\varsigma=\pm 1$ in the
electric/magnetic cases as before, the solution for the metric and the active
dilatonic combinations $e^{\varsigma\vec a_\alpha\cdot\vec\phi}$ is given by
\bea
ds^2 &=& \prod_{\alpha=1}^\ell H_\alpha^{-\tilde d\over D-2}dx^\mu dx_\mu +
\prod_{\alpha_1}^\ell H_\alpha^{d\over D-2}dy^mdy^m\nonumber\\
e^{\varsigma\vec a_\alpha\cdot\vec\phi} &=& H_\alpha^2\sum_{\beta_1}^\ell
H_\beta^{-d\tilde d\over D-2} \label{multichargemultifs}\\
H_\alpha &=& 1+{k^\alpha\over|\vec y-\vec y_\alpha|^{\tilde d}}\ .\nonumber
\eea
The non-trivial step in verifying the validity of this solution is the
check that the non-linear terms still cancel in the Einstein equations,
even with the multiple centers.\cite{boundstates}

     Now consider a solution with two field strengths ($F^1_{[n]}$,
$F^2_{[n]}$) in which the two charge parameters are taken to be the same,
$\lambda_\alpha=\lambda$, while the charge centers are allowed to coalesce.
When the charge centers have coalesced, the resulting solution may be
viewed as a single-field-strength solution for a field strength rotated by
$\pi/4$ in the space of field strengths ($F^1_{[n]}$, $F^2_{[n]}$). Since
the charges add {\it vectorially,} the net charge parameter in this case
will be $\lambda=\sqrt2\lambda$, and the net charge density will be
$U=\sqrt2\lambda\Omega_{D-d-1}/4$. On the other hand, the total mass density
will add as a {\it scalar} quantity, so ${\cal E}={\cal E}_1+{\cal
E}_2=2\lambda\Omega_{D-d-1}/4=\sqrt2U$. Thus, the coalesced solution satisfies
${\cal E}={2U/\sqrt\Delta}$ with $\Delta=2$. Direct comparison with our general
$p$-brane solution (\ref{pbranesol}) shows that the coalesced solution agrees
precisely with the single-field-strength $\Delta=2$ solution. Generalizing this
construction to a case with $N$ separate $\Delta=4$ components, one finds in the
coincident limit a $\Delta=4/N$ supersymmetric solution from the
single-field-strength analysis. In the next subsection, we shall see that as one
adds new components, each one separately charged with respect to a different
$\Delta=4$ field strength, one progressively breaks more and more supersymmetry.
For example, the above solution (\ref{multichargemultifs}) leaves unbroken 1/4 of
the original supersymmetry. Since the $\Delta=4/N$ solutions may in this way be
separated into $\Delta=4$ components while still preserving some degree of unbroken
supersymmetry, and without producing any relative forces to disturb their
equilibrium, they may be considered to be ``bound states at
threshold.''\,\cite{boundstates} We shall shortly see that the zero-force
property of such multiple-component solutions is related to their managing
still to preserve unbroken a certain portion of rigid supersymmetry, even
though this portion is reduced with respect to the half-preservation
characterising single-component
$\Delta=4$ solutions.

\subsection{Intersecting branes and the four elements in
$D=11$}\label{ssec:fourelements}

     The multiple-charge-center solutions (\ref{multichargemultifs}) to the
dimensionally reduced theory (\ref{I_D}) may automatically be interpreted as
solutions of any one of the higher-dimensional theories descending from the
$D=11$ theory (\ref{D11act}). This automatic ``oxidation'' is possible because
we have insisted throughout on considering only consistent truncations.
Although all lower-dimensional solutions may automatically be oxidised in
this way into solutions of higher-dimensional supergravity theories, it is
not guaranteed that these oxidised branes always fall into the class of
isotropic $p$-brane solutions that we have mainly been discussing. For example,
in $D=9$, one has a two-black-hole solution of the form
(\ref{multichargemultifs}), supported by a 1-form gauge potential
$A^{12}_{[1]}$ descending from the $D=11$ gauge potential $A_{[3]}$ and also by
another 1-form gauge potential, \eg ${\cal A}^2_{[1]}$, emerging from the
metric as a Kaluza-Klein vector field. Upon oxidising the two-black-hole
solution back to $D=11$, one finds the solution\crampest
\bea
&\makebox[0pt]{$ds_{11}^2$}&=
H_1^{1\over3}(y)\left[H_1^{-1}(y)\{-dt^2+
d\rho^2+d\sigma^2+(H_2(y)-1)(dt+d\rho)^2\} + dy^mdy^m\right]\nn\\
&\makebox[0pt]{$A_{[3]}$}&= H_1^{-1}(y)dt\wedge d\rho\wedge
d\sigma\ ,\qquad
m=3,\ldots,10,\qquad\mbox{\underline{wave$\Vert$2-brane}}\label{wave2brane}
\eea\uncramp
which depends on two independent harmonic functions $H_1(y)$ and $H_2(y)$,
where the $y^m$ are an 8-dimensional set of ``overall transverse'' coordinates.

     Although the solution (\ref{wave2brane}) clearly falls outside the class of
$p$-brane or multiple $p$-brane solutions that we have considered so far, it
nonetheless has two clearly recognisable elements, associated to the two
harmonic functions $H_1(y)$ and $H_2(y)$. In order to identify these two
elements, we may use the freedom to trivialise one or the other of these
harmonic functions by setting it equal to unity. Thus, setting $H_2=1$, one
recovers
\bea
ds_{11}^2 &=&
H^{1\over3}(y)\left[H^{-1}(y)\{-dt^2+
d\rho^2+d\sigma^2 + dy^mdy^m\right]\nn\\
A_{[3]} &=& H^{-1}(y)dt\wedge d\rho\wedge d\sigma,\qquad
m=3,\ldots,10,\hspace{1cm}\underline{\hbox{2-brane}}\label{2brane}
\eea
which one may recognised as simply a certain style of organising the
harmonic-function factors in the $D=11$ membrane solution~\cite{ds}
(\ref{isoel2br}), generalised to an arbitrary harmonic function
$H(y)\leftrightarrow H_1(y)$ in the membrane's transverse space.

     Setting $H_1=1$ in (\ref{wave2brane}), on the other hand, produces a
solution of $D=11$ supergravity that is not a $p$-brane (\ie it is not a
Poincar\'e-invariant hyperplane solution). What one finds for $H_1=1$ is a
classic solution of General Relativity found originally in 1923 by
Brinkmann,\cite{brinkmann} the {\em pp wave}:
\bea
ds_{11}^2 &=&
\{-dt^2+d\rho^2+(H(y)-1)(dt+d\rho)^2\} + dy^mdy^m\nn\\
A_{[3]} &=& 0\ ,\qquad
m=2,\ldots,10,\hspace{3cm}\underline{\hbox{pp wave}}\label{wave}
\eea
where for a general wave solution, $H(y)$ could be harmonic in the 9
dimensions $y^m$ transverse to the two lightplane dimensions $\{t,\rho\}$ in
which the wave propagates; for the specific case obtained by setting $H_1=1$ in
(\ref{wave2brane}), $H(y)\leftrightarrow H_2(y)$ is constant in one of these 9
directions, corresponding to the coordinate $\sigma$ in (\ref{wave2brane}).

     The solution (\ref{wave2brane}) thus may be viewed as a $D=11$ pp wave
superposed on a membrane. Owing to the fact that the harmonic function $H_2(y)$
depends only on the overall transverse coordinates $y^m$, $m=3,\ldots,10$, the
wave is actually ``delocalised'' in the third membrane worldvolume direction,
\ie the solution (\ref{wave2brane}) is independent of $\sigma$ as well as of
its own lightplane coordinates. Of course, this delocalisation of the wave
in the $\sigma$ direction is just what makes it possible to perform a dimensional
reduction of (\ref{wave2brane}) on the $\{\rho,\sigma\}$ coordinates down to a
$D=9$ configuration of two particles of the sort considered in
(\ref{multichargemultifs}), \ie the wave in (\ref{wave2brane}) has already been
stacked up in the $\sigma$ direction as is necessary in preparation for a
vertical dimensional reduction. Another point to note about (\ref{wave2brane})
is that the charge centers of the two harmonic functions $H_1$ and $H_2$ may be
chosen completely independently in the overall transverse space. Thus, although
this is an example of an ``intersecting'' brane configuration, it should be
understood that the two components of (\ref{wave2brane}) need not actually
overlap on any specific subspace of spacetime. The term ``intersecting'' is
generally taken to mean that there are shared worldvolume coordinates, in this
case the $\{t,\rho\}$ overlap between the membrane worldvolume and the
lightplane coordinates.\cite{intersecting}

     A very striking feature of the family of multiple-component $p$-brane
solutions is that their oxidations up to $D=11$ involve combinations of only 4
basic ``elemental'' $D=11$ solutions. Two of these we have just met in the
oxidised solution (\ref{wave2brane}): the membrane and the pp wave. The two
others are the ``duals'' of these: the 5-brane~\cite{guv} and a solution
describing the oxidation to $D=11$ of the ``Kaluza-Klein
monopole.''\,\cite{kkmonopole} The 5-brane may be written in a style similar to
that of the membrane (\ref{2brane}):
\bea
ds_{11}^2 &=&
H^{2\over3}(y)\left[H^{-1}(y)\{-dt^2+dx_1^2+\ldots+dx_5^2\} +
dy^mdy^m\right]\nn\\
F_{[4]} &=& {}^\ast\!dH(y),\qquad
m=6,\ldots,10,\hspace{2cm}\underline{\hbox{5-brane}}\label{5brane}
\eea
where the $H(y)$ is a general harmonic function in the 5-dimensional transverse
space.

     The Kaluza-Klein monopole oxidised up to $D=11$ is the solution
\bsea
ds_{11}^2 &=&
-dt^2+dx_1^2+\ldots+dx_6^2+ds^2_{\rm TN}(y)\nn\\
A_{[3]} &=& 0\\
ds^2_{\rm TN} &=& H(y)dy^idy^i+H^{-1}(y)(d\psi+V_i(y)dy^i)^2
\ ,\qquad i=1,2,3,\nn\\
\vec\nabla\times\vec V &=& \vec\nabla H\
,\hspace{5cm}\underline{\hbox{Taub-NUT}}\label{nut}
\esea
where $ds^2_{\rm TN}$ is the Taub-NUT metric, a familiar four-dimensional
Euclidean gravitational instanton. The harmonic function $H$ in (\ref{nut}) is
a function only of the 3 coordinates $y^i$, and not of the coordinate $\psi$,
which plays a special r\^ole. Generally, the solution (\ref{nut}) has a conical
singularity on the hyperplane $y^i=0$, but this becomes a mere coordinate
singularity, similar to that for flat space in polar coordinates, providing the
coordinate $\psi$ is periodically identified. For a single-center harmonic
function $H(y)=1+k/(|y|)$, the appropriate identification period for $\psi$ is
$4\pi k$. 

     Thus, the Taub-NUT solution naturally invites interpretation as a
compactified solution in one less dimension, after reduction on $\psi$. In the
case of the original Kaluza-Klein monopole,\cite{kkmonopole} the starting solution
had 4+1 dimensions, giving rise after compactification to a magnetically-charged
particle in $D=4$ dimensions. The solution (\ref{nut}) has an additional 6
spacelike worldvolume dimensions $x_1,\ldots,x_6$, so after reduction on the
$\psi$ coordinate one has a magnetically-charged 6-brane solution in $D=10$.

     The relation $\Delta\psi=4\pi k$ between the compactification period of
$\psi$ and the charge-determining integration constant $k$ in the harmonic
function $H$ of the solution (\ref{nut}) gives rise to a quantisation condition at
the quantum level involving the magnetic charge of the dimensionally-reduced $D=10$
6-brane descending from (\ref{nut}) and the electric charge of the
extreme black hole particle obtained by reducing the pp wave (\ref{wave}). This
quantisation condition is nothing other than an ordinary quantisation of momentum
for Fourier wave components on a compact space, in this case the compact $\psi$
direction. In terms of the electric and magnetic charges
$U$ and $V$ of the dimensionally reduced particle and 6-brane, one finds
$UV=2\pi\kappa_{10}^2n$, with $n\in\Z$ (where
$\kappa_{10}^2$ occurs because the charges $U$ and $V$ as defined in
(\ref{electric},\,\ref{magnetic}) are not dimensionless). This is precisely of
the form expected for a Dirac charge quantisation condition. In Section
\ref{sec:dualities} we shall return to the subject of charge quantisation
conditions more generally for the charges carried by $p$-branes.

     Let us now return to the question of supersymmetry preservation and
enquire whether intersecting branes like (\ref{wave2brane}) can also preserve
some portion of unbroken rigid supersymmetry. All four of the elemental $D=11$
solutions (\ref{2brane} -- \ref{nut}) preserve half the
$D=11$ rigid supersymmetry. We have already seen this for the membrane
solution in subsection \ref{ssect:supersymmetry}. As another example, one may
consider the supersymmetry preservation conditions for the pp wave solution
(\ref{wave}). We shall skip over points 1) and 2) of the discussion analogous to
that of subsection \ref{ssect:supersymmetry} and shall instead concentrate just on
the projection conditions that must be satisfied by the surviving rigid
supersymmetry parameter $\epsilon_\infty$. Analogously to our earlier abbreviated
discussion using just the supersymmetry algebra, consider this algebra in the
background of a pp wave solution (\ref{wave}) propagating in the $\{01\}$
directions of spacetime, with normalisation to unit length along the wave's
propagation direction:
\be
{1\over\hbox{length}}\{Q_\alpha,Q_\beta\} = 2{\cal E}P_{01}\qquad\qquad 
P_{01} = \ft12(\oneone + \Gamma^{01})\ ,\label{waveproj}
\ee
where $P_{01}$ is again a projection operator with half of its eigenvalues zero,
half unity. Consequently, the pp wave solution (\ref{wave}) preserves half of
the $D=11$ rigid supersymmetry.

     Now let us apply the projection-operator analysis to the
wave$\Vert$2-brane solution (\ref{wave2brane}). Supersymmetry preservation in a
membrane background oriented parallel to the $\{012\}$ hyperplane requires the
projection condition $P_{012}\epsilon_\infty=0$ (\ref{projepsilon}), while
supersymmetry preservation in a pp wave background with a $\{01\}$
lightplane requires $P_{01}\epsilon_\infty=0$. Imposing these two conditions
simultaneously is consistent because these projectors commute,
\be
[P_{012},P_{01}]=0\ .\label{projconsistency}
\ee
Since ${\rm tr}(P_{012}P_{01})=\ft14\cdot32$, the imposition of both projection
conditions on $\epsilon_\infty$ cuts the preserved portion of rigid
$D=11$ supersymmetry down to $\frac14$.

     Now, let's consider another example of an intersecting-brane solution,
containing as elements a $D=11$ membrane, 5-brane pair. The
solution is\goodbreak
\bsea
\hspace{.5cm}ds^2 &=&
H_1^{\fft13}(y)H_2^{\fft23}(y)\big[H_1^{-1}(y)H_2^{-1}(y)(-dt^2 +
dx_1^2)\hspace{2cm}\mbox{\underline{$2\perp5(1)$}}\nn\\ &&\hspace{1cm} +
H_1^{-1}(y)(dx_2^2)+H_2^{-1}(y)(dx_3^2+\ldots+dx_6^2)\nn\\
&&\hspace{2cm}+dy^mdy^m\big]\qquad\qquad m=7,\ldots,10\\
&&F_{m012}=\partial_m(H_1^{-1})\qquad\qquad
F_{2mnp}=-\epsilon_{mnpq}\partial_qH_2\ ,
\label{2perp5}
\esea
where as in the wave-on-a-membrane solution (\ref{wave2brane}), the harmonic
functions $H_1(y)$ and $H_2(y)$ depend only on the overall transverse
coordinates. By considering special cases where $H_2=1$ or $H_1=1$, one
identifies the membrane and 5-brane elements of the solution (\ref{2perp5});
as before, these elements are delocalised in ({\it i.e.,} independent of) the
``relative transverse'' directions, by which one means the directions
transverse to one element's worldvolume but belonging to the worldvolume of
the other element, \ie the directions $\{2;3,\ldots,6\}$ for the
solution (\ref{2perp5}). Note that both the membrane and 5-brane elements share
the worldvolume directions $\{01\}$; these are accordingly called ``overall
worldvolume'' directions. Considering this ``intersection'' to be a string (but
recall, however, that the overall-transverse charge centers of $H_1$ and
$H_2$ need not coincide, so there is not necessarily a true string overlap), the
solution (\ref{2perp5}) is denoted $2\perp5(1)$.

     The forms of the wave$\Vert$2-brane solution (\ref{wave2brane})
and the $2\perp5(1)$ solution (\ref{2perp5}) illustrate the general structure
of intersecting-brane solutions. For a two-element solution, there are four
sectors among the coordinates: overall worldvolume, two relative transverse
sectors and the overall transverse sector. One may make a sketch of these
relations for the $2\perp5(1)$ solution (\ref{2perp5}):

\smallskip
\begin{center}
\begin{tabular}{|cc|c|cccc|cccc|}
\multicolumn{1}{c}{$0$}&\multicolumn{1}{c}{$1$}&\multicolumn{1}{c}{$2$}&{$3$}&
{$4$}&{$5$}&\multicolumn{1}{c}{$6$}&{$7$}&{$8$}&{$9$}&\multicolumn{1}{c}{$10$}\\
\hline
x&x&x&&&&&&&&\\
\hline
x&x&&x&x&x&x&&&&\\
\hline
\multicolumn{2}{c}{\makebox[0pt]{\phantom{\Huge|}W25\phantom{\Huge|}}\ \ }&
\multicolumn{1}{c}{\makebox[0pt]{W2T5}}&
\multicolumn{4}{c}{W5T2}&
\multicolumn{4}{c}{T25}
\end{tabular}
\end{center}

\smallskip\noindent The character of each coordinate is indicated in this
sketch: W2 and W5 indicating worldvolume coordinates with respect to each of
the two elements and T2 and T5 indicating transverse coordinates with
respect to each of the two elements. Thus, the overall worldvolume coordinates
are the W25 coordinates and the overall transverse coordinates are the T25
coordinates. Having established this coordinate classification, the general
structure of the intersecting brane metric is as follows. For each element, one
puts an overall conformal factor $H_i^{d/(D-2)}(y)$ for the whole metric, and then
in addition one puts a factor $H_i^{-1}(y)$ in front of each $dx^2$ term belonging
to the worldvolume of the $i^{\rm th}$ element. One may verify this pattern in the
structure of (\ref{2perp5}). This pattern has been termed the {\em harmonic
function rule.}\cite{intersecting}

     This summary of the structure of intersecting brane solutions
does not replace a full check that the supergravity equations of motion are
solved, and in addition one needs to establish which combinations of the $D=11$
elements may be present in a given solution. For a fuller review on this subject,
we refer the reader to Ref.\,\cite{gauntlett} For now, let us just check point 3)
in the supersymmetry-preservation analysis for the $2\perp5(1)$ solution
(\ref{2perp5}). For each of the two elements, one has a projection condition on
the surviving rigid supersymmetry parameter
$\epsilon_\infty$:
$P_{012}\epsilon_\infty=0$ for the membrane and $P_{013456}\epsilon_\infty=0$
for the 5-brane. These may be consistently imposed at the same time, because
$[P_{012},P_{013456}]=0$, similarly to our discussion of the wave$\Vert$2-brane
solution. The amount of surviving supersymmetry in the $2\perp5(1)$ solution is
$\frac14$, because ${\rm tr}(P_{012}P_{013456})=\ft14\cdot32$.
     
\subsection{Brane probes, scattering branes and
modulus $\sigma$-model geometry}\label{ssec:scatterbranes}

     The existence of static configurations such as the wave$\Vert$2-brane
solution (\ref{wave2brane}) or the $2\perp5(1)$ intersecting-brane solution
(\ref{2perp5}) derives from the properties of the transverse-space Laplace
equation (\ref{laplace}) arising in the process of solving the
supergravity equations subject to the $p$-brane ans\"atze
(\ref{ansatz}--\ref{magans}). The Laplace equation has the well-known property
of admitting multi-center solutions, which we have already encountered in Eq.\
(\ref{multiharmonic}). Physically, the existence of such multi-center
solutions corresponds either to a cancellation of attractive gravitational and
dilatonic forces against repulsive antisymmetric-tensor forces, or to the fact
that one brane couples to the background supergravity fields with a conformal
factor that wipes out the effects of the other brane. In order to see such
cancellations more explicitly, one may use a source coupling analogous to the
$D=11$ bosonic supermembrane action (\ref{smemb}) in order to treat the
limiting  problem of a light brane probe moving in the background of a heavy
brane.\cite{ds,tseytlinprobe} In this limit, one may ignore the deformation of the
heavy-brane background caused by the light brane. The use of the brane-probe
coupling is a simple way to approximately treat time-dependent brane
configurations. For a $p$-brane probe of this sort coupled to a $D$-dimensional
supergravity background, the probe action is
\bsea
I_{\rm probe} &=& -T_\alpha\int d^{p+1}\xi\left(-\det(\partial_\mu
x^m\partial_\nu x^n g_{mn}(x)\right)^\fft12e^{\fft12\varsigma^{\rm pr}\vec
a_\alpha\cdot\vec\phi} + Q_\alpha\int \tilde A^\alpha_{[p+1]}\nn\\
\\
\nn\\
\tilde A^\alpha_{[p+1]} &=& [(p+1)!]^{-1}\partial_{\mu_1}
x^{m_1}\cdots\partial_{\mu_{p+1}}x^{m_{p+1}} A^\alpha_{m_1\cdots m_{p+1}}
d\xi^{\mu_1}\wedge\cdots\wedge d\xi^{\mu_{p+1}}\ .\nn\\
\label{probeact}
\esea
The dilaton coupling in (\ref{probeact}a) occurs because one needs to have the
correct source for the $D$-dimensional Einstein frame, \ie the conformal frame
in which the $D$-dimensional Einstein-Hilbert action is free from dilatonic
scalar factors. Requiring that the source match correctly to a $p$-brane
(probe) solution demands the presence of the dilaton coupling
$e^{\fft12\varsigma^{\rm pr}\vec a_\alpha\cdot\vec\phi}$, where
$\varsigma^{\rm pr}=\pm1$ according to whether the $p$-brane probe is of
electric or magnetic type and where $\vec a_\alpha$ is the dilaton vector
appearing in the kinetic-term dilaton coupling in (\ref{I_D}) for the gauge
potential $A^\alpha_{[p+1]}$, to which the $p$ brane probe couples.

     As a simple initial example of such a brane probe, one may take a light $D=11$
membrane probe in the background of a parallel and similarly-oriented heavy
membrane.\cite{ds} In this case, the brane-probe action (\ref{probeact})
becomes just the $D=11$ supermembrane action (\ref{smemb}). If one takes the
form of the heavy-membrane background from the electric ansatz
(\ref{ansatz},\,\ref{elans}), and if one chooses the ``static'' worldvolume
gauge $\xi^\mu=x^\mu$, $\mu=0,1,2,$ then the bosonic probe action becomes
\be
I_{\rm probe} = -T\int d^3\xi\left(\sqrt{-\det(e^{2A(y)}\eta_{\mu\nu} +
e^{2B(y)}\partial_\mu y^m\partial_\nu y^m)} - e^{C(y)}\right )\
.\label{d11probe}
\ee
Expanding the square root in (\ref{d11probe}), one finds at order $(\partial
y)^0$ the effective potential
\be
V_{\rm probe} = T(e^{3A(y)}-e^C)\ .\label{membprobepot}
\ee
Recalling the condition (\ref{crel}), which becomes just $e^{3A(y)}=e^{C(y)}$
for the membrane background, one has directly $V_{\rm probe}=0$, confirming the
absence of static forces between the two membrane components.

     Continuing on in the expansion of (\ref{d11probe}) to order $(\partial
y)^2$ in the probe velocity, one has the effective probe $\sigma$-model 
\be
I^{(2)}_{\rm probe}=-{T\over2}\int d^3\xi e^{3A(y)}e^{2(B(y)-A(y)}\partial_\mu
y^m\partial_\nu y^m\eta_{\mu\nu}\ ,\label{probesigma}
\ee
but, recalling the supersymmetry-preservation condition (\ref{phirel})
characterising the heavy-membrane background, the probe
$\sigma$-model metric in (\ref{probesigma}) reduces simply to 
\be
\gamma^{mn} = e^{A(y) + 2B(y)}\delta^{mn}=\delta^{mn}\ ,\label{flatsigma}
\ee {\it i.e.,} the
membrane-probe $\sigma$-model metric is {\em flat}.

     The flatness of the membrane-probe $\sigma$-model metric (\ref{flatsigma})
accords precisely with the degree of rigid supersymmetry that survives in the
underlying supergravity solution with two parallel, similarly oriented $D=11$
membranes, which we found in subsection \ref{ssect:supersymmetry} to have
$\ft12\cdot32$ components, \ie $d=2+1$, $N=8$ probe-worldvolume supersymmetry.
This high degree of surviving supersymmetry is too restrictive in its
constraints on the form of the $\sigma$-model to allow for anything other than
a flat metric, precisely as one finds in (\ref{flatsigma}). Continuing on with
the expansion of (\ref{d11probe}), one first finds a nontrivial interaction
between the probe and the heavy membrane background at order
$(\partial y)^4$ (odd powers being ruled out by time-reversal invariance of
the $D=11$ supergravity equations).

     Now consider a brane-probe configuration with less surviving
supersymmetry, and with correspondingly weaker constraints on the probe
worldvolume $\sigma$-model. Corresponding to the wave$\Vert$2-brane solution
(\ref{wave2brane}), one has, after dimensional reduction down to $D=9$
dimensions, a system of two black holes supported by {\em different}
$\Delta=4$, $D=9$ vector fields: one descending from the $D=11$ 3-form gauge
potential and one descending from the metric.

     Now repeat the brane-probe analysis for the two-black-hole configuration,
again choosing a static gauge on the probe worldvolume, which in the present
case just becomes $\xi^0=t$. Again expand the determinant of the induced
metric in (\ref{probeact}). At order $(\partial y)^0$, this now gives $V_{\rm
probe}=e^Ae^{-{3\over2\sqrt7}}\phi$, but this potential turns out to be just a
constant because the heavy-brane background satisfies $A={3\over2\sqrt7}\phi$.
Thus, we confirm the expected static zero-force condition for the
$\ft14$ supersymmetric two-black-hole configuration descending from the
wave$\Vert$2-brane solution (\ref{wave2brane}). This zero-force condition
arises not so much as a result of a cancellation between different forces but
as a result of the probe's coupling to the background  with a dilatonic factor in
(\ref{probeact}) that wipes out the conformal factor occurring in the heavy
brane background metric.

     Proceeding on to $({\rm velocity})^2$ order, one now obtains a
non-trivial probe $\sigma$-model, with metric
\be
\gamma^{mn} = H_{\rm back}(y)\delta^{mn}\ ,\label{modulusmetric}
\ee
where $H_{\rm back}$ is the harmonic function controlling the heavy brane's
background fields; for the case of two black holes in $D=9$, the harmonic
function $H_{\rm back}$ has the structure $(1+k/r^6)$. 

     The above test-brane analysis for two $D=9$ black holes
is confirmed by a more detailed study of the low-velocity scattering of
supersymmetric black holes performed by Shiraishi.\cite{shiraishi} The
procedure is a standard one in soliton physics: one promotes the moduli of
a static solution to time-dependent functions and then substitutes the
resulting generalized field configuration back into the original field
equations. This leads to a set of differential equations on the modulus
variables which may be viewed as effective equations for the moduli. In the
general case of multiple black hole scattering, the resulting system of
differential equations may be quite complicated. The system of equations,
however, simplifies dramatically in cases corresponding to the scattering of
supersymmetric black holes, \eg the above pair of $D=9$ black holes,
where the result turns out to involve only 2-body forces. These two-body forces
may be derived from an effective action involving the position vectors of the
two black holes. Separating the center-of-mass motion from the relative motion,
one obtains the same modulus metric (\ref{modulusmetric}) as that found in the
brane-probe analysis above, except for a rescaling which replaces the
brane-probe mass by the reduced mass of the two-black-hole system.

     Now we should resolve a puzzle of how this non-trivial $d=1$
scattering modulus $\sigma$-model turns out to be consistent with the
surviving supersymme\-try.\cite{gps} The modulus variables of the two-black-hole
system are fields in one dimension, \ie time. The $N$-extended supersymmetry
algebra in
$d=1$ is
\be
\{Q^I,Q^J\}=2\delta^{IJ}\hat H\hskip1cm I=1,\ldots,N\ ,\label{1dimsusy}
\ee
where $\hat H$ is the Hamiltonian. A $d=1$, $N=1$ $\sigma$-model is
specified by a triple $({\cal M},\gamma,A_{[3]})$, where $\cal M$ is the
Riemannian $\sigma$-model manifold, $\gamma$ is the metric on $\cal M$ and
$A_{[3]}$ is a 3-form on $\cal M$ which plays the r\^ole of torsion in the
derivative operator acting on fermions, $\nabla_t^{(+)} =
\partial_tx^i\nabla^{(+)}_i$, where $\nabla^{(+)}_i\lambda^j =
\nabla_i\lambda^j+\ft12A^j{}_{ik}\lambda^k$. The $\sigma$-model
action may be written using $N=1$ superfields $x^i(t,\theta)$
(where $x^i(t)=x^i\for_{\theta=0}$, $\lambda^i(t)=Dx^i\for_{\theta=0}$) as
\be
I = - \ft12\int dtd\theta(\im\gamma_{ij}Dx^i{d\over dt}x^j + {1\over
3!}A_{ijk}Dx^iDx^jDx^k)\ .\label{ssaction}
\ee
One may additionally~\cite{cp} have a set of spinorial $N=1$ superfields
$\psi^a$, with Lagrangian $-\ft12 h_{ab}\psi^a\nabla_t\psi^b$, where $h_{ab}$
is a fibre metric and $\nabla_t$ is constructed using an appropriate connection
for the fibre corresponding to the $\psi^a$. However, in the present
case we shall not include this extra superfield. In order to have extended
supersymmetry in (\ref{ssaction}), one starts by positing a second set of
supersymmetry transformations of the form $\delta x^i = \eta I^i{}_jDx^j$,
and then requires these transformations to close to form the
$N=2$ algebra (\ref{1dimsusy}); then one also requires that the
action (\ref{ssaction}) be invariant. In this way, one obtains the equations
\bsea
I^2 &=& -\oneone\hspace{2cm}\\
N^i_{jk}\equiv I^i_{[j,k]} &=& 0\\
\gamma_{kl}I^k{}_iI^l{}_j &=& \gamma_{ij}\\
\nabla^{(+)}_{(i}I^k{}_{j)} &=& 0\\
\partial_{[i}(I^m{}_jA_{\vert m\vert
kl]})-2I^m{}_{[i}\partial_{[m}A_{jkl]]}) &=& 0\ ,\label{susyconds}
\esea
where (\ref{susyconds}a,b) follow from requiring the closure of the algebra
(\ref{1dimsusy}) and (\ref{susyconds}c-e) follow from requiring invariance of
the action (\ref{ssaction}). Conditions (\ref{susyconds}a,b) imply that $\cal
M$ is a complex manifold, with $I^i{}_j$ as its complex structure.

     The structure of the conditions (\ref{susyconds}) is more complicated than
might have been expected. Experience with $d=1+1$ extended supersymmetry~\cite{cp}
might have lead one to expect, by simple dimensional reduction, just the condition
$\nabla^{(+)}_iI^j{}_k=0$. Certainly, solutions of this condition also satisfy
(\ref{susyconds}c--e), but the converse is not true, \ie the $d=1$ extended
supersymmetry conditions are ``weaker'' than those obtained by dimensional
reduction from $d=1+1$, even though the $d=1+1$ minimal spinors are, as in $d=1$,
just real single-component objects. Conversely, the $d=1+1$ theory implies a
``stronger'' condition; the difference is explained by $d=1+1$ Lorentz
invariance: not all $d=1$ theories can be ``oxidized'' up to Lorentz-invariant
$d=1+1$ theories. In the present case with two $D=9$ black holes, this is
reflected in the circumstance that after even one dimensional oxidation
from $D=9$ up to $D=10$, the solution already contains a pp wave element (so
that we have a $D=10$ ``wave-on-a-string'' solution), with a lightplane metric
that is not Poincar\'e invariant.

     Note also that the $d=1$ ``torsion'' $A_{[3]}$ is not required to be
closed in (\ref{susyconds}). $d=1$ supersymmetric theories satisfying
(\ref{susyconds}) are {\em analogous} to (2,0) chiral supersymmetric
theories in $d=1+1$, but the weaker conditions (\ref{susyconds}) warrant a
different notation for this wider class of models; one may call them 2b
supersymmetric $\sigma$-models.\cite{gps} Such models are characterized by a
K\"ahler geometry with torsion.

     Continuing on to $N=8$, $d=1$ supersymmetry, one finds an 8b
generalization~\cite{gps} of the conditions (\ref{susyconds}), with 7 independent
complex structures built using the octonionic structure constants\,\footnote{We
let the conventional octonionic ``$0$'' index be replaced by ``$8$'' here in
order to avoid confusion with a timelike index; the $\E^8$ transverse space is of
Euclidean signature.} $\varphi_{ab}{}^c$: $\delta x^i=\eta^aI_a{}^i{}_jDx^j$,
$a=1,\ldots7$, with
$(I_a)^8{}_b=\delta_{ab}$, $(I_a)^b{}_8=-\delta^b{}_a$,
$(I_a)^b{}_c=\varphi_a{}^b{}_c$, where the octonion multiplication rule is
$e_ae_b=-\delta_{ab}+\varphi_{ab}{}^ce_c$. Models satisfying such conditions have
an ``octonionic K\"ahler geometry with torsion,'' and are called OKT
models.\cite{gps} 

     Now, are there any non-trivial solutions to these conditions? Evidently,
from the brane-probe and Shiraishi analyses, there must be. For our two $D=9$
black holes with a $D=8$ transverse space, one may start from the ansatz
$ds^2=H(y)ds^2(\E^8)$, $A_{ijk}=\Omega_{ijk}{}^\ell\partial_\ell H$, where
$\Omega$ is a 4-form on $\E^8$. Then, from the 8b generalization of condition
(\ref{susyconds}d) one learns $\Omega_{8abc}=\varphi_{abc}$ and
$\Omega_{abcd}=-{}^\ast \varphi_{abcd}$; from the 8b generalization of condition
(\ref{susyconds}e) one learns $\delta^{ij}\partial_i\partial_j H=0$. Thus we
recover the familiar dependence of p-brane solutions on transverse-space
harmonic functions, and we reobtain the brane-probe or Shiraishi
structure of the black-hole modulus scattering metric with
\be
H_{\rm relative}=1+{k_{\rm red}\over|y_1-y_2|^6}\ ,
\ee
where $k_{\rm red}$ determines the reduced mass of the two black holes.

\section{Duality symmetries and charge quantisation}\label{sec:dualities}

     As one can see from our discussion of Kaluza-Klein dimensional
reduction in Section \ref{sec:kkred}, progression down to lower dimensions
$D$ causes the number of dilatonic scalars $\vec\phi$ and also the number
of zero-form potentials of 1-form field strengths to proliferate. When one
reaches $D=4$, for example, a total of 70 such spin-zero fields has
accumulated. In $D=4$, the maximal $(N=8)$ supergravity equations of
motion have a linearly-realized $H={\rm SU}(8)$ symmetry; this is also the
automorphism symmetry of the $D=4$, $N=8$ supersymmetry algebra relevant
to the (self-conjugate) supergravity multiplet. In formulating this
symmetry, it is necessary to consider {\em complex} self-dual
and anti-self-dual combinations of the 2-form field strengths, which are
the highest-rank field strengths occurring in $D=4$, higher ranks having been
eliminated in the reduction or by dualization. Using two-component
notation for the $D=4$ spinors, these combinations transform as
$F_{\alpha\beta}^{[ij]}$ and $\bar F_{\dot\alpha\dot\beta\,[ij]}$,
$i,j=1,\ldots,8$, {\it i.e.}\ as a complex 28-dimensional dimensional
representation of ${\rm SU}(8)$. Since this complex representation can be
carried only by the complex field-strength combinations and not by the
1-form gauge potentials, it cannot be locally formulated at the level of
the gauge potentials or of the action, where only an ${\rm SO}(8)$
symmetry is apparent.

     Taking all the spin-zero fields together, one finds that they form a
rather impressive nonlinear $\sigma$-model with a 70-dimensional manifold.
Anticipating that this manifold must be a coset space with $H={\rm SU}(8)$
as the linearly-realized denominator group, Cremmer and Julia~\cite{cj}
deduced that it had to be the manifold \ffrac{${\rm E}_{7(+7)}$}{${\rm
SU}(8)$}; since the dimension of ${\rm E}_7$ is 133 and that of ${\rm
SU}(8)$ is 63, this gives a 70-dimensional manifold. Correspondingly, a
nonlinearly-realized ${\rm E}_{7(+7)}$ symmetry also appears as an
invariance of the $D=4$, $N=8$ maximal supergravity equations of motion.
Such nonlinearly-realized symmetries of supergravity theories have always
had a somewhat mysterious character. They arise in part out of general
covariance in the higher dimensions, from which supergravities arise by
dimensional reduction, but this is not enough: such symmetries act
transitively on the $\sigma$-model manifolds, mixing both fields arising from
the metric and also from the reduction of the $D=11$ 3-form potential
$A_{[3]}$ in (\ref{D11act}).

     In dimensions $4\le D\le9$, maximal supergravity has the
sets of $\sigma$-model nonlinear $G$ and linear $H$ symmetries shown in
Table \ref{tab:sugrasyms}. In all cases, the spin-zero fields take their
values in ``target'' manifolds \ffrac{$G$}{$H$}. Just as the asymptotic
value at infinity of the metric defines the reference, or ``vacuum''
spacetime with respect to which integrated charges and energy/momentum are
defined, so do the asymptotic values of the spin-zero fields define the
``scalar vacuum.'' These asymptotic values are referred to as the {\em
moduli} of the solution. In string theory, these moduli acquire
interpretations as the {\em coupling constants} and vacuum {\em
$\theta$-angles} of the theory. Once these are determined for a given
``vacuum,'' the {\em classification symmetry} that organizes the distinct
solutions of the theory into multiplets with the same energy must be a
subgroup of the {\em little group}, or {\em isotropy group,} of the
vacuum. In ordinary General Relativity with asymptotically flat
spacetimes, the analogous group is the spacetime Poincar\'e group times
the appropriate ``internal'' classifying symmetry, \eg the group of rigid (\ie
constant-parameter) Yang-Mills gauge transformations.

\vspace{-.4cm}
\begin{table}[ht]
\centering
\caption{Supergravity $\sigma$-model symmetries.\label{tab:sugrasyms}}
\vspace{0.4cm}
\begin{tabular}{|c|c|c|}
\hline
$D$&$G$&$H$\\
\hline\hline
9&${\rm GL}(2,\R)$&${\rm SO}(2)$\\
\hline
8&${\rm SL}(3,\R)\times{\rm SL}(2,\R)$&${\rm SO}(3)\times{\rm
SO}(2)$\\
\hline
7&${\rm SL}(5,\R)$&${\rm SO}(5)$\\
\hline
6&${\rm SO}(5,5)$&${\rm SO}(5)\times{\rm SO}(5)$\\
\hline
5&${\rm E}_{6(+6)}$&${\rm USP}(8)$\\
\hline
4&${\rm E}_{7(+7)}$&${\rm SU}(8)$\\
\hline
3&${\rm E}_{8(+8)}$&${\rm SO}(16)$\\
\hline
\end{tabular}
\end{table}

     The isotropy group of any point on a coset manifold \ffrac{$G$}{$H$}
is just $H$, so this is the classical ``internal'' classifying symmetry for
multiplets of supergravity solutions.

\subsection{An example of duality symmetry: $D=8$ supergravity}
\label{ssec:d8sugra}

     In maximal $D=8$ supergravity, one sees from Table
\ref{tab:sugrasyms} that $G={\rm SL}(3,\R)\times{\rm SL}(2,\R)$ and the
isotropy group is $H={\rm SO}(3)\times{\rm SO}(2)$. We have an $(11-3=8)$
vector of dilatonic scalars as well as a singlet $F_{[1]}^{ijk}$ and a
triplet ${\cal F}_{[1]}^{ij}$ $(i,j,k = 1,2,3)$ of 1-form field strengths
for zero-form potentials. Taken all together, we have a manifold of
dimension 7, which fits in precisely with the dimension of the
\ffrac{$({\rm SL}(3,\R)\times{\rm SL}(2,\R))$}{$({\rm SO}(3)\times{\rm
SO}(2))$} coset-space manifold: $8+3-(3+1)=7$.

     Owing to the direct-product structure, we may for the time being
drop the 5-dimensional \ffrac{${\rm SL}(3,\R)$}{${\rm SO}(3)$}
sector and consider for simplicity just the 2-dimensional \ffrac{${\rm
SL}(2,\R)$}{${\rm SO}(2)$} sector. Here is the relevant part of the
action:\,\cite{ilpt}
\bea
I_8^{{\rm SL}(2)} &=& \int d^8x\sqrt{-g}\Big[R-\ft12\nabla_{\sst
M}\sigma\nabla^{\sst M}\sigma - \ft12e^{-2\sigma}\nabla_{\sst
M}\chi\nabla^{\sst M}\chi\nonumber\\
&&\hspace{3cm} -
{1\over2\cdot4!}e^\sigma(F_{[4]})^2 - {1\over2\cdot4!}\chi
F_{[4]}{}^\ast\!F_{[4]}\Big]\label{d8sl2}
\eea
where ${}^\ast\!F^{\sst MNPQ}=1/(4!\sqrt{-g})\epsilon^{{\sst
MNPQ}x_1x_2x_3x_4}F_{x_1x_2x_3x_4}$ (the $\epsilon^{[8]}$ is a density, hence
purely numerical).

     On the scalar fields $(\sigma,\chi)$, the ${\rm SL}(2,\R)$ symmetry
acts as follows: let $\lambda=\chi+\im e^\sigma$; then 
\be
\Lambda=\pmatrix{a&b\cr c&d}
\ee
with $ab-cd=1$ is an
element of ${\rm SL}(2,\R)$ and acts on $\lambda$ by the
fractional-linear transformation
\be
\lambda\longrightarrow {a\lambda + b\over c\lambda + d}\ .\label{fractlin}
\ee

     The action of the ${\rm SL}(2,\R)$ symmetry on the 4-form field
strength gives us an example of a symmetry of the equations of motion that
is not a symmetry of the action. The field strength $F_{[4]}$ forms an
${\rm SL}(2,\R)$ {\em doublet} together with
\be
G_{[4]}=e^\sigma{}^\ast\!F_{[4]} - \chi F_{[4]}\ ,
\ee
{\it i.e.,}
\be
\pmatrix{F_{[4]}\cr G_{[4]}} \longrightarrow (\Lambda^{\rm
T})^{-1}\pmatrix{F_{[4]}\cr G_{[4]}}\ .\label{F4transf}
\ee
One may check that these transform the $F_{[4]}$ field equation
\be
\nabla_{\sst M}(e^\sigma F^{\sst MNPQ} + \chi {}^\ast\!F^{\sst MNPQ}) = 0
\label{F4eq}
\ee
into the corresponding Bianchi identity,
\be
\nabla_{\sst M}{}^\ast\!F^{\sst MNPQ} = 0\ .\label{F4bianchi}
\ee
Since the field equations may be expressed purely in terms of $F_{[4]}$,
we have a genuine symmetry of the field equations in the transformation
(\ref{F4transf}), but since this transformation cannot be expressed
locally in terms of the gauge potential $A_{[3]}$, this is not a
local symmetry of the action. The transformation
(\ref{fractlin},\,\ref{F4transf}) is a $D=8$ analogue of ordinary Maxwell duality
transformation in the presence of scalar fields. Accordingly, we shall refer
generally to the supergravity $\sigma$-model symmetries as duality symmetries.

     The $F_{[4]}$ field strength of the $D=8$ theory supports
elementary/electric $p$-brane solutions with $p=4-2=2$, \ie
membranes, which have a $d=3$ dimensional worldvolume. The corresponding
solitonic/magnetic solutions in $D=8$ have worldvolume dimension $\tilde
d=8-3-2=3$ also. So in this case, $F_{[4]}$ supports {\em both} electric
and magnetic membranes. It is also possible in this case to have solutions
generalizing the purely electric or magnetic solutions considered so far to
solutions that carry both types of charge, {\it i.e.}\  {\em dyons}.\cite{ilpt}
This possibility is also reflected in the combined Bogomol'ny bound\,\footnote{In
comparing (\ref{d8bogbound}) to the single-charge bounds (\ref{bogbounds}), one
should take note that for
$F_{[4]}$ in (\ref{d8sl2}) we have $\Delta=4$, so $2/\sqrt\Delta = 1$.} for
this situation, which generalizes the single-charge bounds (\ref{bogbounds}):
\be
{\cal E}^2 \ge e^{-\sigma_\infty}(U + \chi_\infty V)^2 +
e^{\sigma_\infty} V^2\ ,\label{d8bogbound}
\ee
where $U$ and $V$ are the electric and magnetic charges and
$\sigma_\infty$ and $\chi_\infty$ are the moduli, {\it
i.e.}\ the constant asymptotic values of the scalar fields $\sigma(x)$ and
$\chi(x)$. The bound (\ref{d8bogbound}) is itself ${\rm SL}(2,\R)$ invariant,
provided that one transforms both the moduli $(\sigma_\infty,\chi_\infty)$
(according to (\ref{fractlin})) and also the charges $(U,V)$. For the simple
case with $\sigma_\infty=\chi_\infty=0$ that we have mainly chosen in order to
simplify the writing of explicit solutions, the bound (\ref{d8bogbound}) reduces to
${\cal E}^2\ge U^2 + V^2$, which is invariant under an obvious isotropy group
$H={\rm SO}(2)$.

\subsection{$p$-form charge quantisation
conditions}\label{ssec:chargequantisation}

     So far, we have discussed the structure of $p$-brane solutions at a
purely classical level. At the classical level, a given supergravity theory can
have a continuous spectrum of electrically and magnetically charged
solutions with respect to any one of the $n$-form field strengths that can
support the solution. At the quantum level, however, an important
restriction on this spectrum of solutions enters into force: the
Dirac-Schwinger-Zwanziger (DSZ) quantisation conditions for particles with
electric or magnetic or dyonic charges.\cite{nep,teit} As we have seen,
however, the electric and magnetic charges carried by branes and appearing in
the supersymmetry algebra (\ref{susyalg}) are {\em forms,} and the
study of their charge-quantisation properties involves some special features
not seen in the $D=4$ Maxwell case.\cite{blps}

     We shall first review a Wu-Yang  style of argument,\cite{nep} (for
a Dirac-string argument, see Ref.\,\cite{teit,dght}) considering a closed
sequence $\cal W$ of deformations of one $p$-brane, say an electric one, in
the background fields set up by a dual, magnetic, $\hat p=D-p-4$ brane. After
such a sequence of deformations, one sees from the supermembrane action
(\ref{smemb}) that the electric
$p$-brane wavefunction picks up a phase factor
\be
\exp\left({\im Q_{\rm e}\over(p+1)!}\oint_{\cal W}A_{M_1\ldots
M_{p+1}}dx^M_1\wedge\ldots\wedge dx^{M_{p+1}}\right)\
,\label{phasefactor}
\ee
where $A_{[p+1]}$ is the gauge potential set up (locally) by the
magnetic $\hat p$-brane background.

     A number of differences arise in this problem with respect to the
ordinary Dirac quantisation condition for $D=4$ particles. One of these is
that, as we have seen in subsection \ref{ssec:charges}, objects carrying
$p$-form charges appearing in the supersymmetry algebra (\ref{susyalg}) are
necessarily either infinite or are wrapped around compact spacetime
dimensions. For infinite $p$-branes, some deformation sequences $\cal W$ will
lead to a divergent integral in the exponent in (\ref{phasefactor}); such
deformations would also require an infinite amount of energy, and so
should be excluded from consideration. In particular, this excludes
deformations that involve rigid rotations of an entire infinite brane. Thus,
at least the {\em asymptotic} orientation of the electric brane must be
preserved throughout the sequence of deformations. Another way of viewing this
restriction on the deformations is to note that the asymptotic orientation
of a brane is encoded into the electric $p$-form charge, and so one should
not consider changing this $p$-form in the course of the deformation any more
than one should consider changing the magnitude of the electric charge in the
ordinary $D=4$ Maxwell case.

     We shall see shortly that another difference with respect to the
ordinary $D=4$ Dirac quantisation of particles in Maxwell theory
will be the existence of ``Dirac-insensitive'' configurations, for which the
phase in (\ref{phasefactor}) vanishes.

     Restricting attention to deformations that give non-divergent phases,
one may use Stoke's theorem to rewrite the integral in (\ref{phasefactor}):
\bea
&&{Q_{\rm e}\over(p+1)!}\oint_{\cal W}A_{M_1\ldots
M_{p+1}}dx^M_1\wedge\ldots\wedge dx^{M_{p+1}} =\nn\\
&&\qquad
{Q_{\rm e}\over(p+2)!}\int_{{\cal M}_{\cal W}}F_{M_1\ldots
M_{p+2}}dx^{M_1}\wedge\ldots\wedge dx^{M_{p+2}} = Q_{\rm e}\Phi_{{\cal M}_{\cal
W}}\ ,\qquad\label{caphase}
\eea
where ${\cal M}_{\cal W}$ is any surface ``capping'' the closed
surface $\cal W$, \ie a surface such that $\partial{{\cal M}_{\cal
W}}={\cal W}$; $\Phi_{{\cal M}_{\cal W}}$ is then the flux through the
cap ${\cal M}_{\cal W}$. Choosing the capping surface in two different
ways, one can find a flux discrepancy $\Phi_{{\cal M}_1}-\Phi_{{\cal
M}_2}=\Phi_{{\cal M}_1\cap{\cal M}_2}=\Phi_{{\cal M}_{\rm total}}$
(taking into account the orientation sensitivity of the flux integral).
Then if ${\cal M}_{\rm total}={\cal M}_1\cap{\cal M}_2$ ``captures'' the
magnetic $\hat p$-brane, the flux $\Phi_{{\cal M}_{\rm total}}$ will
equal the magnetic charge $Q_{\rm m}$ of the $\hat p$-brane; thus the
discrepancy in the phase factor (\ref{phasefactor}) will be simply $\exp(\im Q_{\rm
e}Q_{\rm m})$. Requiring this to equal unity gives,\cite{nep} in
strict analogy to the ordinary case of electric and magnetic particles in
$D=4$, the Dirac quantisation condition
\be
Q_{\rm e}Q_{\rm m}=2\pi n\ ,\qquad n\in\Z\ .\label{simpledirac}
\ee

     The charge quantisation condition (\ref{simpledirac}) is almost, but
not quite, the full story. In deriving (\ref{simpledirac}), we have not
taken into account the $p$-form character of the charges. Taking this
into account shows that the phase in (\ref{phasefactor}) vanishes for a
measure-zero set of configurations of the electric and magnetic
branes.\cite{blps} This is easiest to explain in a simplified case
where the electric and magnetic branes are kept in static flat
configurations, with the electric $p$-brane oriented along the directions
$\{x^{M_1}\ldots x^{M_p}\}$. The phase factor
(\ref{phasefactor}) then becomes $\exp(\im Q_{\rm e}\oint_{\cal
W}A_{M_1\ldots M_pR}\ffrac{$\partial x^R$}{$\partial\sigma$})$,
where $\sigma$ is an ordering parameter for the closed sequence of
deformations $\cal W$. In making this deformation sequence, we recall
from the above discussion that one should restrict the deformations to
preserve the asymptotic orientation of the deformed $p$-brane. For
simplicity, one may simply consider moving the electric $p$-brane by parallel
transport around the magnetic $\hat p$-brane in a closed loop. The accrued
phase factor is invariant under gauge transformations of the potential
$A_{[p+1]}$. This makes it possible to simplify the discussion by making use
of a specially chosen gauge. Note that magnetic $\hat p$-branes have purely
transverse field strengths like (\ref{fieldstrengths}b); there is accordingly
a gauge in which the gauge potential $A_{[p+1]}$ is also purely transverse,
\ie it vanishes whenever any of its indices point along a worldvolume
direction of the magnetic $\hat p$-brane. Consideration of more general
deformation sequences yields the same result.\cite{blps}

     Now one can see how the Dirac-insensitive configurations arise: the
phase in (\ref{phasefactor}) vanishes whenever there is even a partial
alignment between the electric and the magnetic branes, \ie when there
are shared worldvolume directions between the two branes. This
measure-zero set of Dirac-insensitive configurations may be simply
characterised in terms of the $p$ and $\hat p$ charges themselves by the
condition $Q^{\rm el}_{[p]}\wedge Q^{\rm mag}_{[\hat p]}=0$. For such
configurations, one obtains no Dirac quantisation condition. To summarise, one
may incorporate this orientation restriction into the Dirac quantisation condition
(\ref{simpledirac}) by writing a $(p+\hat p)$-form quantisation condition
\be
Q^{\rm el}_{[p]}\wedge Q^{\rm mag}_{[\hat p]} = 2\pi n{Q^{\rm
el}_{[p]}\wedge Q^{\rm mag}_{[\hat p]}\over|Q^{\rm el}_{[p]}|\,
|Q^{\rm mag}_{[\hat p]}|}\ ,\qquad n\in\Z\ ,\label{formdirac}
\ee
which reduces to (\ref{simpledirac}) for all except the
Dirac-insensitive set of configurations.

\subsection{Charge quantisation conditions and dimensional
reduction}\label{ssec:dirac}

     The existence of Dirac-insensitive configurations may seem to be of only
peripheral importance, given that they constitute only a measure-zero subset of
the total set of asymptotic brane configurations. However, their relevance
becomes more clear when one considers the relations existing between the $p$-form
charges under dimensional reduction. Let us recall the relations (\ref{Fred})
between the field strengths in different dimensions. Now, the electric and
magnetic charges carried by branes in $D$ dimensions take the forms
\bsea
Q_{\rm e} &=& \int\left(e^{\vec c\cdot\vec\phi}{\,}^\ast\!F + \kappa(A)\right)\\
Q_{\rm m} &=& \int\tilde F\ ,\label{redcharges}
\esea
where $\tilde F=dA$, $F=\tilde F + \hbox{(Chern-Simons modifications)}$ (\ie
modifications involving lower-order forms arising in the dimensional reduction
similar to those in the $D=9$ case (\ref{csterms})) and
$\vec c$ is the dilaton vector corresponding to
$F$ in the dimensionally-reduced action (\ref{I_D}). The term $\kappa(A)$ in
(\ref{redcharges}$a$) is the analogue of the term $\ft12 A_{[3]}\wedge
F_{[4]}$ in (\ref{electric}). From the expressions (\ref{Fred}) for the
reduced field strengths and their duals, one obtains the following
relations between the original charges in $D=11$ and those in the reduced
theory:

\vspace{-.4cm}
\begin{table}[ht]
\centering
\caption{Relations between $Q^{11}$ and $Q^{\sst D}$\label{tab:redcharges}}
\vspace{0.4cm}
\begin{tabular}{|c|c|c|c|c|}\hline
\phantom{\Big[}& $F_{[4]}$ & $F_{[3]}^i$ & $F_{[2]}^{ij}$ & $F_{[1]}^{ijk}$
\\ \hline\hline
\phantom{\Big[} Electric $Q^{11}_{\rm e}=$ & $Q^{\sst D}_{\rm e}\, V$ &
$Q^{\sst D}_{\rm e}\, 
\fft{V}{L_i}$ & $Q^{\sst D}_{\rm e}\, \fft{V}{L_i\, L_j}$ & 
$Q^{\sst D}_{\rm e}\, \fft{V}{L_i\, L_j\, L_k}$\\ \hline
\phantom{\Big[} Magnetic $Q^{11}_{\rm m}=$ & $Q^{\sst D}_{\rm m}$ & 
$Q^{\sst D}_{\rm m}\,  {L_i}$
& $Q^{\sst D}_{\rm m}\, {L_i\, L_j}$ & $Q^{\sst D}_{\rm m}\, {L_i\,
L_j\, L_k}$\\ \hline
\end{tabular}
\end{table}
\noindent where $L_i=\int dz^i$ is the compactification period of the
reduction coordinate $z^i$ and $V=\int d^{11-D}z=\prod^{11-D}_{i=1}L_i$ is
the total compactification volume. Note that the factors of $L_i$ cancel out
in the various products of electric and magnetic charges only for
charges belonging to the same field strength in the reduced dimension $D$.

     Now consider the quantisation conditions obtained between the
various dimensionally reduced charges shown in Table \ref{tab:redcharges}. We
need to consider the various schemes possible for dimensional reduction of
dual pairs of ($p$,$\hat p$) branes. We have seen that for single-element
brane solutions, there are two basic schemes, as explained in Section
\ref{sec:kkred}: diagonal, which involves reduction on a worldvolume
coordinate, and vertical, which involves reduction on a transverse coordinate
after preparation by ``stacking up'' single-center solutions so as to generate a
transverse-space translation invariance needed for the
dimensional reduction.

     For the dimensional reduction of a solution containing two elements, there
are then four possible schemes, depending on whether the reduction coordinate
$z$ belongs to the worldvolume or to the transverse space of each brane. For an
electric/magnetic pair, we have the following four reduction possibilities:
diagonal/diagonal, diagonal/vertical, vertical/diagonal and vertical/vertical.
Only the mixed cases will turn out to preserve Dirac sensitivity in the lower
dimension after reduction. 

     This is most easily illustrated by considering the diagonal/diagonal
case, for which $z$ belongs to the worldvolumes of both branes. With such a
shared worldvolume direction, one has clearly fallen into the measure-zero
set of Dirac-insensitive configurations with $Q^{\rm el}_{[p]}\wedge Q^{\rm
mag}_{[\hat p]}=0$ in the higher dimension $D$. Correspondingly, in $(D-1)$
dimensions one finds that the diagonally reduced electric $(p-1)$ brane
is supported by an $n=p+1$ form field strength, but the diagonally
reduced magnetic $(\hat p-1)$ brane is supported by an $n=p+2$ form;
since only branes supported by the same field strength can have a Dirac
quantisation condition, this diagonal/diagonal reduction properly
corresponds to a Dirac-insensitive configuration.

     Now consider the mixed reductions, \eg diagonal/vertical. In performing
a vertical reduction of a magnetic $\hat p$-brane by stacking up an
infinite deck of single-center branes in order to create the $\R$
translational invariance necessary for the reduction, the total magnetic
charge will clearly diverge. Thus, in a vertical reduction it is
necessary to reinterpret the magnetic charge $Q_{\rm m}$ as a charge
density per unit $z$ compactification length. Before obtaining the Dirac
quantisation condition in the lower dimension, it is necessary to
restore a gravitational-constant factor of $\kappa^2$ that should properly have
appeared in the quantisation conditions (\ref{simpledirac},\,\ref{formdirac}). As
one may verify, the electric and magnetic charges as defined in
(\ref{electric},\,\ref{magnetic}) are not dimensionless. Thus,
(\ref{simpledirac}) in $D=11$ should properly have been written $Q_{\rm
e}Q_{\rm m}=2\pi\kappa_{11}^2 n$. If one lets the compactification length be
denoted by $L$ in the $D$-dimensional theory prior to dimensional reduction, 
then one obtains a Dirac phase $\exp(\im\kappa_{D-1}^{-2}Q_{\rm e}Q_{\rm m}L)$.
This fits precisely, however, with another aspect of dimensional reduction: the
gravitational constants in dimensions $D$ and $D-1$ are related by
$\kappa^2_D=L\kappa^2_{D-1}$. Thus, in dimension $D-1$ one obtains the
expected quantisation condition $Q_{\rm e}Q_{\rm m}=2\pi\kappa_{D-1}^2
n$. Note, correspondingly, that upon making a mixed diagonal/vertical
reduction, the electric and magnetic branes remain dual to each other in the
lower dimension, supported by the same $n=p-1+2=p+1$ form field strength. The
opposite mixed vertical/diagonal reduction case goes similarly, except that 
the dual branes are then supported by the same $n=p+2$ form field
strength.

     In the final case of vertical/vertical reduction, Dirac sensitivity is
lost in the reduction, not owing to the orientation of the branes, but
because in this case both the electric and the magnetic charges need to be
interpreted as densities per unit compactification length, and so one obtains
a phase $\exp(\im\kappa_{D}^{-2}Q_{\rm e}Q_{\rm m}L^2)$. Only one factor of $L$
is absorbed into $\kappa^2_{D-1}$, and one has ${\rm lim}_{L\rightarrow
0}L^2/\kappa^2_D=0$. Correspondingly, the two dimensionally reduced branes in
the lower dimension are supported by different field strengths: an $n=p+2$ form
for the electric brane and an $n=p+1$ form for the magnetic brane.

     Thus, there is a perfect accord between the structure of the Dirac
quantisation conditions for $p$-form charges in the various
supergravity theories related by dimensional reduction. The existence of
Dirac-insensitive configurations plays a central r\^ole in establishing this
accord, even though they represent only a subset of measure zero from the point
of view of the higher-dimensional theory.

     Another indication of the relevance of the Dirac-insensitive
configurations is the observation~\cite{blps} that all the 
intersecting-brane solutions with some degree of preserved supersymmetry, as
considered in Section \ref{sec:intersectingbranes},  
correspond to Dirac-insensitive configurations. This may immediately be seen
in such solutions as the $2\perp5(1)$ solution (\ref{2perp5}), but it is also
true for solutions involving pp wave and Taub-NUT elements.

\subsection{Counting $p$-branes}\label{ssec:weyl}

     As we have seen at the classical level, the classifying symmetry for
solutions in a given scalar vacuum, specified by the values of the scalar
moduli, is the linearly-realized isotropy symmetry $H$ given in Table
\ref{tab:sugrasyms}. When one takes into account the Dirac quantisation
condition, this classifying symmetry becomes restricted to a discrete
group, which clearly must be a subgroup of the corresponding $G(\Z)$ duality group,
so in general one seeks to identify the group $G(\Z)\cap H$. The value of
this intersection is modulus-dependent, showing that the homogeneity of
the \ffrac{$G$}{$H$} coset space is broken at the quantum level by the
quantisation condition. Classically, of course, the particular point on
the vacuum manifold \ffrac{$G$}{$H$} corresponding to the scalar moduli can
be changed by application of a transitively-acting $G$ transformation, for
example with a group element $g$. Correspondingly, the isotropy subgroup $H$
moves by conjugation with $g$,
\be
H\longrightarrow gHg^{-1}\ .\label{Hconj}
\ee

     The discretized duality group $G(\Z)$, on the other hand, does not depend upon
the moduli. This is because the modulus dependence cancels out in the ``canonical''
charges that we have defined in Eq.\ (\ref{redcharges}). One way to see this is to
use the relations between charges in different dimensions given in Table
\ref{tab:redcharges}, noting that there are no scalar moduli in $D=11$, so the
modulus-independent relations of Table \ref{tab:redcharges} imply that the
lower-dimensional charges (\ref{redcharges}) do not depend on the
moduli.\footnote{Note that the compactification periods $L_i$ appearing in Table
\ref{tab:redcharges} have values that may be adjusted by convention. These should
not be thought of as determining the geometry of the compactifying internal
manifold, which is determined instead by the scalar moduli. Thus, the relations of
Table \ref{tab:redcharges} imply the independence of the canonically-defined
charges from the physically relevant moduli.} 

     Another way to understand this is by comparison with ordinary Maxwell
electrodynamics, where an analogous charge would be that derived from the action
$I_{\rm Max}=-\ffrac1{$\st(4e^2)$}\int F_{\mu\nu}^{\rm can}F^{{\rm can}\,\mu\nu}$,
corresponding to a covariant derivative $D_\mu=\partial_\mu+\im A_\mu^{\rm can}$.
This is analogous to our dimensionally reduced action (\ref{I_D}) from which the
charges (\ref{redcharges}) are derived, because the modulus factors $e^{\vec
c\cdot\vec\phi_\infty}$ appearing in (\ref{I_D}) (together with the rest of the
$\vec\phi$ dilatonic scalar dependence) play the r\^oles of coupling constant
factors like $e^{-2}$. If one wants to compare this to the ``conventional''
charges defined with respect to a conventional gauge potential $A_\mu^{\rm
conv}=e^{-1}A_\mu^{\rm can}$, for which the action is $-\ffrac14\int
F_{\mu\nu}^{\rm conv}F^{{\rm conv}\,\mu\nu}$, then the canonical and conventional
charges obtained {\it via} Gauss's law surface integrals are related by
\be
Q_{\rm can}={1\over2e^2}\int d^2\Sigma^{ij}\epsilon_{ijk}F^{{\rm can}\,0k} =
{1\over2e}\int d^2\Sigma^{ij}\epsilon_{ijk}F^{{\rm conv}\,0k}={1\over e}Q_{\rm
conv}\ .\label{canconvrel}
\ee
Thus, in the Maxwell electrodynamics case, the dependence on the electric
charge unit $e$ drops out in $Q_{\rm can}$, although the conventional charge
$Q_{\rm conv}$ scales proportionally to $e$. The modulus independence of the
charges (\ref{redcharges}) works in a similar fashion. Then, given that the
discretised quantum duality group $G(\Z)$ is {\em defined} by the requirement that
it map the set of Dirac-allowed charges onto itself, it is evident that the group
$G(\Z)$, referred to the canonical charges (\ref{redcharges}), does not depend on
the moduli.

     As a consequence of the different modulus dependences of $H$ and of $G(\Z)$,
it follows that the size of the intersection group $G(\Z)\cap H$ is dependent on
the moduli. The analogous feature in ordinary Maxwell theory is that a true duality
symmetry of the theory only arises when the electric charge takes the value
$e=1$ (in appropriate units), since the duality transformation maps $e\rightarrow
e^{-1}$. Thus, the value $e=1$ is a distinguished value.

     The distinguished point on the scalar vacuum manifold for general
supergravity theories is the one where all the scalar moduli vanish. This is
the point where $G(\Z)\cap H$ is maximal. Let us return to our $D=8$ example to
help identify what this group is. In that case, for the scalars $(\sigma,\chi)$,
we may write out the transformation in detail using (\ref{fractlin}):
\bea
e^{-\sigma}&\longrightarrow&(d+c\chi)^2e^{-\sigma} + c^2e^\sigma\nonumber\\
\chi e^{-\sigma}&\longrightarrow& (d + c\chi)(b + a\chi)e^{-\sigma} +
ace^\sigma\ .\label{sigmachitransfs}
\eea
Requiring $a,b,c,d \in \Z$ and also that the modulus point
$\sigma_\infty=\chi_\infty=0$ be left invariant, we find only two
transformations: the identity and the transformation $a=d=0, b=-1, c=1$, which
maps $\sigma$ and $\chi$ according to
\bea
e^{-\sigma}&\longrightarrow&e^\sigma+\chi^2e^{-\sigma}\nonumber\\
\chi e^{-\sigma}&\longrightarrow&-\chi e^{-\sigma}\
.\label{discretesigmachitransfs}
\eea
Thus, for our truncated system, we find just an $S_2$ discrete symmetry as
the quantum isotropy subgroup of ${\rm SL}(2,\Z)$ at the distinguished
point on the scalar vacuum manifold. This $S_2$ is the natural analogue of the
$S_2$ symmetry that appears in Maxwell theory when $e=1$.

     In order to aid in identifying the pattern behind this $D=8$
example, suppose that the zero-form gauge potential $\chi$ is small, and
consider the $S_2$ transformation to lowest order in $\chi$. To this
order, the transformation just flips the signs of $\sigma$ and $\chi$.
Acting on the field strengths $(F_{[4]},G_{[4]})$, one finds
\be
(F_{[4]},G_{[4]})\longrightarrow (-G_{[4]},F_{[4]})\ .\label{infchitransf}
\ee
One may again check (in fact to all orders, not just to lowest order in
$\chi$) that (\ref{infchitransf}) maps the field equation for $F_{[4]}$
into the corresponding Bianchi identity:
\be
\nabla_{\sst M}(e^\sigma F^{\sst MNPQ} + \chi{}^\ast\!F^{\sst
MNPQ})\longrightarrow - \nabla_{\sst M}{}^\ast\!F^{\sst MNPQ}\ .
\ee
Considering this $S_2$ transformation to lowest order in the zero-form
$\chi$ has the advantage that the sign-flip of $\phi$ may be ``impressed''
upon the $\vec a$ dilaton vector for $F_{[4]}$: $\vec a\rightarrow -\vec
a$. The general structure of such $G(\Z)\cap H$ transformations will be
found by considering the impressed action of this group on the dilaton
vectors.

     Now consider the \ffrac{${\rm SL}(3,\R)$}{${\rm SO}(3)$} sector of the
$D=8$ scalar manifold, again with the moduli set to the distinguished point
on the scalar manifold. To lowest order in zero-form gauge
potentials, the action of ${\rm SL}(3,\Z)\cap H$ may similarly by impressed
upon the 3-form dilaton vectors, causing in this case a {\em permutation}
of the $\vec a_i$, generating for the $D=8$ case overall the discrete group
$S_3\times S_2$. Now that we have a bit more structure to contemplate, we
can notice that the $G(\Z)\cap H$ transformations {\em leave the $(\vec
a,\vec a_i)$ dot products invariant.}\cite{weyl}

     The invariance of the dilaton vectors' dot products prompts one to
return to the algebra (\ref{dilvecprods}) of these dot products and see
what else we may recognize in it. Noting that the duality groups given in
Table \ref{tab:sugrasyms} for the higher dimensions $D$ involve
${\rm SL}(N,\R)$ groups, we recall that the {\em weight vectors} $\vec
h_i$ of the fundamental representation of ${\rm SL}(N,\R)$ satisfy
\be
\vec h_i\cdot\vec h_j = \delta_{ij}-{1\over N}\ ,\qquad\qquad \sum_{i=1}^N
\vec h_i = 0\ .\label{weightprods}
\ee
These relations are precisely those satisfied by ${\pm 1\over\sqrt2}\vec
a$ and ${1\over\sqrt2}\vec a_i$, corresponding to the cases $N=2$ and
$N=3$. This suggests that the action of the maximal $G(\Z)\cap H$
group (\ie for scalar moduli set to the distinguished point on
the scalar manifold) may be identified in general with the symmetry
group of the set of fundamental weights for the corresponding supergravity
duality group $G$ as given in Table \ref{tab:sugrasyms}. The symmetry
group of the fundamental weights is the {\em Weyl group}~\cite{weyl} of
$G$, so the action of the maximal $G(\Z)\cap H$ $p$-brane classifying
symmetry is identified with that of the Weyl group of $G$.

     As one proceeds down through the lower-dimensional cases, where the
supergravity symmetry groups shown in Table \ref{tab:sugrasyms} grow in
complexity, the above pattern persists:\,\cite{weyl} in all cases, the
action of the maximal classifying symmetry $G(\Z)\cap H$ may be identified
with the Weyl group of $G$. This is then the group that {\em counts} the
distinct p-brane solutions\,\footnote{Of course, these solutions must also
fall into supermultiplets with respect to the unbroken supersymmetry; the
corresponding supermultiplet structures have been discussed in
Ref.\,\cite{dr}} of a given type (\ref{admenerg}), subject to the
Dirac quantisation condition and referred to the distinguished point on the
scalar modulus manifold. For example, in $D=7$, where from Table
\ref{tab:sugrasyms} one sees that $G={\rm SL}(5,\R)$ and $H={\rm SO}(5)$,
one finds that the action of $G(\Z)\cap H$ is equivalent to that of the
discrete group $S_5$, which is the Weyl group of ${\rm SL}(5,\R)$. In the
lower-dimensional cases shown in Table \ref{tab:sugrasyms}, the  discrete
group $G(\Z)\cap H$ becomes less familiar, and is most simply described
as the Weyl group of $G$.

     From the analysis of the Weyl-group duality multiplets, one may
tabulate~\cite{weyl} the multiplicities of $p$-branes residing at each
point of the plot given in Figure \ref{fig:nbscan}. For supersymmetric
$p$-branes arising from a set of $N$ participating field strengths $F_{[n]}$,
corresponding to $\Delta=4/N$ for the dilatonic scalar coupling, one finds the
multiplicities given in Table \ref{tab:weylmults}. By combining these
duality multiplets together with the diagonal and vertical dimensional
reduction families discussed in Sections \ref{sec:kkred} and
\ref{ssec:vertical}, the full set of $p\le(D-3)$ branes shown in
Figure \ref{fig:nbscan} becomes ``welded'' together into one overall
symmetrical structure.

\begin{table}[ht]
\centering
\caption{Examples of $p$-brane Weyl-group
multiplicities\label{tab:weylmults}}
\vspace{0.4cm}
\begin{tabular}{|c|c||c|c|c|c|c|c|c|}
\cline{3-9}
\multicolumn{2}{c|}{}&\multicolumn{7}{|c|}{$D$}\\
\cline{3-9}\hline
$F_{[n]}$&$\Delta$&10&9&8&7&6&5&4\\
\hline\hline
$F_{[4]}$&4&1&1&2&\multicolumn{4}{c|}{}\\
\hline
$F_{[3]}$&4&1&2&3&5&10&\multicolumn{2}{c|}{}\\
\hline
&4&1&1+2&6&10&16&27&56\\
\cline{2-9}
$F_{[2]}$&2&&2&6&15&40&135&756\\
\cline{2-9}
&\ffrac43&\multicolumn{5}{c|}{}&45&2520\\
\hline
&4&&2&8&20&40&72&126\\
\cline{2-9}
$F_{[1]}$&2&\multicolumn{2}{c|}{}&12&60&280&1080&3780\\
\cline{2-9}
&\ffrac43&\multicolumn{4}{c|}{}&480&4320&30240+2520\\
\hline
\end{tabular}
\end{table}

\subsection{The charge lattice}\label{ssec:chargelattice}

     For the electric and magnetic BPS brane solutions supported by a given field
strength, we have seen above that the Dirac charge quantisation condition
(\ref{formdirac}) implies that, given a certain minimum ``electric'' charge 
(\ref{redcharges}a), the allowed set of magnetic charges is determined. Then,
taking the minimum magnetic charge from this set, the argument may be turned
around to show that the set of allowed electric charges is given by integer
multiples of the minimum electric charge. This argument does not directly
establish, however, what the minimum electric charge is, \ie the value of the
charge unit. This cannot be established by use of the Dirac quantisation condition
alone. 

     There are other tools, however, that one can use to fix the charge
lattice completely. To do so, we shall need to exploit the existence of certain
special ``unit-setting'' brane types, and also to exploit fully the consequences
of the assumption that the $G(\Z)$ duality symmetry remains exactly valid at the
quantum level. We have already encountered one example of a ``unit-setting'' brane
in subsection \ref{ssec:fourelements}, where we encountered the pp wave/Taub-NUT
pair of $D=11$ solutions. We saw there that the Taub-NUT solution (\ref{nut})
is nonsingular provided that the coordinate $\psi$ is periodically identified with
period $L=4\pi k$, where $k$ is the charge-determining parameter in the
3-dimensional harmonic function $H(y)=1+k/(|y|)$. Upon dimensional reduction down
to $D=10$, one obtains a magnetic 6-brane solution, with a charge {\em
classically} discretised to take a value in the set
\be
Q_{\rm m}=rL\ ,\qquad r\in\Z\ .\label{6brmaglat}
\ee
Given these values for the magnetic charge, the $D=10$ Dirac quantisation
condition 
\be
Q_{\rm e}Q_{\rm m}=2\pi\kappa_{10}^2n\ ,\qquad n\in\Z\ ,\label{d10dirac}
\ee
or, equivalently, as we saw in subsection \ref{ssec:fourelements}, the quantisation
of $D=11$ pp wave momentum in the compact $\psi$ direction, gives an allowed
set of electric charges
\be
Q_{\rm e}={2\pi\kappa_{10}^2\over L}n\ ,\qquad n\in\Z\ .\label{6brelat}
\ee
Thus, the requirement that magnetic $D=10$ 6-branes oxidise up to non-singular
Taub-NUT solutions in $D=11$ fully determines the 6-brane electric and magnetic
charge units and not just the product of them which occurs in the Dirac
quantisation condition 

     If one assumes that the $G(\Z)$ duality symmetries remain strictly unbroken at
the quantum level, then one may relate the 6-brane charge units to those of other
BPS brane types.\footnote{For details of the duality relations between charge
units for different $p$-branes, see Ref.\,\cite{blps}} In doing so, one must
exploit the fact that brane solutions with Poincar\'e worldvolume symmetries may
be dimensionally reduced down to lower dimensions, where the duality groups shown
in Table \ref{tab:sugrasyms} grow larger. In a given dimension $D$, the $G(\Z)$
duality symmetries only rotate between $p$-branes of the same worldvolume
dimension, supported by the same kind of field strength, as we have seen from our
discussion of the Weyl-group action on $p$-branes given in subsection
\ref{ssec:weyl}. Upon reduction down to dimensions $D_{\rm red}<D$, however, the
solutions descending from an original $p$-brane in $D$ dimensions are subject to a
larger $G(\Z)$ duality symmetry, and this can be used to rotate a descendant brane
into descendants of $p'$-branes for various values of $p'$. Dimensional oxidation
back up to $D$ dimensions then completes the link, establishing relations {\it
via} the duality symmetries between various BPS brane types which can be supported
by different field strengths, including field strengths of different
rank.\cite{mpower} This link may be used to establish relations between the charge
units for the various $p$-form charges of differing rank, even though the
corresponding solutions are Dirac-insensitive to each each other.

     Another charge-unit-setting BPS brane species occurs in the $D=10$ type IIB
theory. This theory has a well-known difficulty with the formulation of a
satisfactory action, although its field equations are perfectly well-defined. The
difficulty in formulating an action arise from the presence of a self-dual 5-form
field strength, $H_{[5]} = {}^\ast\!H_{[5]}$. The corresponding electrically and
magnetically charged BPS solutions are 3-branes, and, owing to the self-duality
condition, these solutions are actually {\em dyons}, with a charge vector at $45^\circ$
to the electric axis. We shall consider the type IIB theory in some
more detail in Section \ref{sec:activelocal}; for now, it will be sufficient for us
to note that the dyonic 3-branes of $D=10$ type IIB theory are also a unit-setting
brane species.\cite{blps} The unit-setting property arises because of a
characteristic property of the Dirac-Schwinger-Zwanziger quantisation condition for
dyons in dimensions $D=4r+2$: for dyons $(Q_{\rm e}^{(1)},Q_{\rm m}^{(1)})$,
$(Q_{\rm e}^{(2)},Q_{\rm m}^{(2)})$, this condition is {\em
symmetric:}\,\cite{blps}
\be
Q_{\rm e}^{(1)}Q_{\rm m}^{(2)} + Q_{\rm e}^{(2)}Q_{\rm
m}^{(1)}=2\pi\kappa^2_{4r+2}n\ ,\qquad n\in\Z\ ,\label{d10diracdyon}
\ee
unlike the more familiar antisymmetric DSZ condition that is obtained in
dimensions $D=4r$. The symmetric nature of (\ref{d10diracdyon}) means that
dyons may be Dirac-sensitive to others of their own type,\footnote{They will be
Dirac-sensitive provided one of them is slightly rotated so as to avoid having any
common worldvolume directions with the other, in order to avoid having a
Dirac-insensitive configuration as discussed in subsection \ref{ssec:dirac}.}
quite differently from the antisymmetric cases in $D=4r$ dimensions. For the
$45^\circ$ dyonic 3-branes, one thus obtains the quantisation condition
\be
|Q_{[3]}|=n\sqrt\pi\kappa_{\rm\sst II B}\ ,\qquad n\in\Z\label{sd3brunit}
\ee
where $\kappa_{\rm\sst IIB}$ is the gravitational constant for the type IIB theory.
Then, using duality symmetries, one may relate the $\sqrt\pi\kappa_{\rm\sst II B}$
charge unit to those of other supergravity R--R charges.

     Thus, using duality symmetries together with the pp wave/Taub-NUT and
self-dual 3-brane charge scales, one may determine the charge-lattice units for
all BPS brane types.\cite{mpower,blps}. It is easiest to express the units of
the resulting overall charge lattice by making a specific choice for the
compactification periods. If one lets all the compactification periods $L_i$ be
equal,
\be
L_i=L_{\rm IIB}=L=(2\pi\kappa_{11}^2)^{\fft19}\ ,\label{periodrels}
\ee
then the electric and magnetic charge-lattice units for rank-$n$
field strengths in dimension $D$ are determined to be~\cite{blps}
\be
\Delta Q_{\rm e}=L^{D-n-1}\ ,\qquad \Delta Q_{\rm m}=L^{n-1}\ .\label{chargeunits}
\ee

\section{Local versus active dualities}\label{sec:activelocal}

     The proper interpretation of the discretised Cremmer-Julia
$G(\Z)$ duality symmetry at the level of supergravity theory is subject to a
certain amount of debate, but at the level of string theory the situation becomes
more clear. In any dimension $D$, there is a subgroup of $G(\Z)$ that corresponds
to {\em T duality,} which is a perturbative symmetry holding order-by-order in the
string loop expansion. T duality~\cite{tduality} consists of transformations
that invert the radii of a toroidal compactification, under which
quantised string oscillator modes and string winding modes become interchanged.
Aside from such a relabeling, however, the overall string spectrum remains
unchanged. Hence, T duality needs to be viewed as a {\em local} symmetry in string
theory, \ie string configurations on compact manifolds related by T duality are
{\em identified.} Depending on whether one considers $(D-3)$ branes to be an
unavoidable component of the spectrum, the same has also been argued to be the
case at the level of the supergravity effective field theory.\cite{sen} The
well-founded basis, in string theory at least, for a local interpretation of the T
duality subgroup of $G(\Z)$ has led subsequently to the
hypothesis~\cite{ht,wittvarious} that the full duality group $G(\Z)$ should be
given a local interpretation: sets of string solutions and moduli related by
$G(\Z)$ transformations are to be treated as equivalent descriptions of a single
state. This local interpretation of the $G(\Z)$ duality transformations is similar
to that adopted for general coordinate transformations viewed {\em passively,}
according to which, {\it e.g.,} flat space in Cartesian or in Rindler coordinates
is viewed as one and the same solution.

     As with general coordinate transformations, however, duality symmetries may
occur in several different guises that are not always clearly distinguished. As
one can see from the charge lattice discussed in subsection
\ref{ssec:chargelattice}, there is also a $G(\Z)$ covariance of the set of charge
vectors for {\em physically inequivalent} BPS brane solutions. In the
discussion of subsection \ref{ssec:chargelattice}, we did not consider in
detail the action of $G(\Z)$ on the moduli, because, as we saw in subsection
\ref{ssec:weyl}, the canonically-defined charges (\ref{redcharges}) are in fact
modulus-independent. 

     Since the dilatonic and axionic scalar moduli determine the
coupling constants and vacuum $\theta$-angles of the theory, these quantities
should be fixed when quantising about a given vacuum state of the theory. This is
similar to the treatment of asymptotically flat spacetime in gravity, where the
choice of a particular asymptotic geometry is necessary in order to establish the
``vacuum'' with respect to which quantised fluctuations can be considered. 

     Thus, in considering physically-inequivalent solutions, one should compare
solutions with the same asymptotic values of the scalar fields. When this is done,
one finds that solutions carrying charges (\ref{redcharges}) related by $G(\Z)$
transformations generally have differing mass densities. Since the standard
Cremmer-Julia duality transformations, such as those of our $D=8$ example in
subsection \ref{ssec:d8sugra}, commute with $P^0$ time translations and so
necessarily preserve mass densities, it is clear that the BPS spectrum at fixed
scalar moduli cannot form a multiplet under the standard Cremmer-Julia $G(\Z)$ 
duality symmetry. This conclusion is in any case unavoidable, given
the local interpretation adopted for the standard duality transformations as
discussed above: once one has identified solution/modulus sets under
the standard $G(\Z)$ duality transformations, one cannot then turn
around and use the same $G(\Z)$ transformations to generate inequivalent solutions.

     Thus, the question arises: is there any spectrum-generating symmetry lying
behind the apparently $G(\Z)$ invariant charge lattices of inequivalent solutions
that we saw in subsection \ref{ssec:chargelattice}? At least at the
classical level, and for single-charge (\ie $\Delta=4$) solutions, the
answer~\cite{clps} turns out to be `yes.' We shall illustrate the point using
type IIB supergravity as an example.\footnote{For a detailed discussion of ${\rm
SL}(2,\R)$ duality in type IIB supergravity, see Ref.\,\cite{schwarz}}

\subsection{The symmetries of type IIB supergravity}\label{ssec:IIB}

     Aside from the difficulties arising from the self-duality condition for the
5-form field strength $H_{[5]}$, the equations of motion of the bosonic fields of
the IIB theory may be derived from the action
\crampest
\bea
I^{{\rm IIB}}_{10} &=& \int d^{10}x[eR +\ft14 e\, {\rm
tr}(\nabla_\mu{\cal M}^{-1}\,\nabla^\mu{\cal M}) -\ft1{12}e\, H_{[3]}^T\,
{\cal M}\, H_{[3]} -\ft1{240} e\, H_{[5]}^2
\nonumber\\ &&\phantom{\int d^{10}x[eR +\ft14}- \ft1{2\sqrt2}
\epsilon_{ij}{\,}^\ast (B_{[4]}\wedge  dA_{[2]}^{(i)}\wedge dA_{[2]}^{(j)})]\ .
\label{2blag}
\eea\uncramp
The 5-form self-duality condition
\be
H_{[5]} = {}^\ast\!H_{[5]}\label{H5sd}
\ee 
may be handled in the fashion of Ref.\,\cite{bbo}, being imposed by hand as an
extra constraint on the field equations obtained by varying (\ref{2blag}). This
somewhat hybrid procedure will be sufficient for our present purposes.

     The matrix $\cal M$ in (\ref{2blag}) contains two scalar fields: a dilatonic
scalar $\phi$ which occurs nonlinearly through its exponential, and an axionic
scalar $\chi$, which may also be considered to be a zero-form gauge potential;
explicitly, one has
\be
{\cal M} = \pmatrix{e^{-\phi} + \chi^2\, e^\phi & \chi\, e^{\phi} \cr
                    \chi\, e^\phi & e^\phi}\ . \label{scalmat}
\ee
The doublet $H_{[3]}$ contains the field strengths of the 2-form gauge
potentials $A_{[2]}$:
\be
H_{[3]} = \pmatrix{dA_{[2]}^{(1)}\cr dA_{[2]}^{(2)} }\ .\label{h3}
\ee

     The action (\ref{2blag}) is invariant under the ${\rm SL}(2,\R)$
transformations
\be
H_{[3]}\longrightarrow (\Lambda^T)^{-1}\, H_{[3]}\ ,\qquad 
{\cal M}\longrightarrow \Lambda\, {\cal M}\, \Lambda^T\ ,\label{tran}
\ee
where the ${\rm SL}(2,\R)$ parameter matrix is
\be
\Lambda = \pmatrix{a & b \cr c & d}\ ,\label{lmat}
\ee
and the ${\rm SL}(2,\R)$ constraint is $ad-bc =1$. If one defines the complex
scalar field $\tau=\chi + i\,  e^{-\phi}$, then the transformation on ${\cal
M}$ can be rewritten as the fractional linear transformation
\be
\tau \longrightarrow \fft{a \tau + b}{c\tau + d}\ .
\ee
Note that since $H_{[5]}$ is a singlet under ${\rm SL}(2,\R)$, the
self-duality  constraint (\ref{H5sd}), which is imposed by hand, also preserves 
the ${\rm SL}(2,\R)$ symmetry. Since this ${\rm SL}(2,\R)$ transformation rotates
the doublet $A_{2]}$ of electric 2-form potentials amongst themselves, this is
an ``electric-electric'' duality, as opposed to the ``electric-magnetic''
duality discussed in the $D=8$ example of subsection \ref{ssec:d8sugra}.
Nonetheless, similar issues concerning duality multiplets for a fixed scalar
vacuum arise in both cases.

     There is one more symmetry of the equations of motion following
from the action (\ref{2blag}). This is a rather humble symmetry that is not often
remarked upon, but which will play an important role in constructing active
${\rm SL}(2,\R)$ duality transformations for the physically distinct BPS string and
5-brane multiplets of the theory. As for pure source-free Einstein theory, the
action (\ref{2blag}) transforms homogeneously as $\lambda^3$ under the following
scaling transformations:
\be
g_{\mu\nu}\longrightarrow \lambda^2\, g_{\mu\nu}\ ,\qquad
A_{[2]}^{(i)}\longrightarrow \lambda^2\, A_{[2]}^{(i)}\ ,\qquad
H_{[5]}\longrightarrow \lambda^4\, H_{[5]}\ ;\label{trombone}
\ee
note that the power of $\lambda$ in each field's transformation is equal to
the number of indices it carries, and, accordingly, the scalars $\phi$ and
$\chi$ are not transformed. Although the transformation (\ref{trombone}) does not
leave the action (\ref{2blag}) invariant, the $\lambda^3$ homogeneity of this
scaling for all terms in the action is sufficient to produce a
symmetry of the IIB equations of motion. It should be noted that the ${\rm
SL}(2,\R)$ electric-magnetic duality of the $D=8$ example given in subsection
\ref{ssec:d8sugra} shares with the transformation (\ref{trombone}) the feature of
being a symmetry only of the equations of motion, and not of the action.

     The $SL(2,\R)$ transformations map solutions of (\ref{2blag}) into other
solutions. We shall need to consider in particular the action of these
transformations on the  charges carried by solutions. From the equations of motion
of the 3-form field strength $H_{[3]}$ in (\ref{2blag}),
\be
d{\,}^\ast (MH_{[3]}) = -{1\over\sqrt2}H_{[5]}\wedge\Omega H_{[3]}\ ,\label{h3eq}
\ee
where $\Omega$ is the $SL(2,\R)$-invariant tensor
\be
\Omega = \pmatrix{0&1\cr -1&0}\ ,\label{omega}
\ee
one finds that the following two-component quantity is conserved:
\be
Q_{\rm e} = \int\left({}^\ast(MH_{[3]}) + {1\over3\sqrt2}\Omega(2B_{[4]}\wedge
H_{[3]} - H_{[5]}\wedge A_{[2]})\right)\ .\label{qel}
\ee
Under an $SL(2,\R)$ transformation, $Q_{\rm e}$ transforms covariantly as a
doublet:
$Q_{\rm e} \rightarrow \Lambda Q_{\rm e}$.

     By virtue of the Bianchi identities for the 3-form field strength, one
has in addition a topologically-conserved magnetic charge doublet,
\be
Q_{\rm m} = \int H_{[3]}\ ,\label{qmag}
\ee
which transforms under $SL(2,\R)$ as $Q_{\rm m}\rightarrow (\Lambda^{\rm
T})^{-1}Q_{\rm m}$, \ie contragrediently to $Q_{\rm e}$. The transformation
properties of the electric and magnetic charge doublets are just such as to ensure
that the Dirac quantization condition $Q_{\rm m}^{\rm T}Q_{\rm e}\in
2\pi\kappa_{\rm\sst II B}^2\Z$ is $SL(2,\R)$ invariant.

     The overall effect of this standard ${\rm SL}(2,\R)$ symmetry
on type IIB supergravity solutions may be expressed in terms of its action on the
solutions' charges and on the scalar moduli. This group action may be viewed as an
automorphism of a vector bundle, with the scalar fields' \ffrac{${\rm
SL}(2,\R)$}{${\rm SO}(2)$} target manifold as the base space, and the charge
vector space as the fiber.

     We have seen in our general discussion of charge lattices in subsection
\ref{ssec:chargelattice} that the continuous classical Cremmer-Julia symmetries $G$
break down to discretised $G(\Z)$ symmetries that map between states on the
quantum charge lattice. In the present type IIB case, the classical ${\rm
SL}(2,\R)$ symmetry breaks down to ${\rm SL}(2,\Z)$ at the quantum level. Taking
the basis states of the IIB charge lattice to be
\be 
\bmath{e}_1=\pmatrix{1\cr0}\qquad\qquad\bmath{e}_2=\pmatrix{0\cr1}\
,\label{basistates}
\ee
the surviving ${\rm SL}(2,\Z)$ group will be represented by ${\rm SL}(2,\R)$
matrices with integral entries.

     As we have discussed above, the discretised duality symmetries $G(\Z)$ are
given a local interpretation in string theory. In the case of the type IIB theory,
this is a hypothesis rather than a demonstrated result, because the ${\rm
SL}(2,\Z)$ transformations map between NS--NS and R--R states, and this is a
distinctly non-perturbative transformation. Adopting this hypothesis nonetheless,
an orbit of the standard ${\rm SL}(2,\Z)$ transformation reduces to a single point;
after making the corresponding identifications, the scalar modulus space becomes
the double coset space
$\leavevmode\lower.25ex\hbox{\the\scriptfont0
${\rm SL}(2,\Z)$}\backslash\kern.05em\raise.5ex\hbox{\the\scriptfont0
${\rm SL}(2,\R)$}/\kern.05em
\lower.25ex\hbox{\the\scriptfont0 ${\rm SO}(2)$}$.

\subsection{Active duality symmetries}\label{ssec:activedualities}

     Now let us see how duality multiplets of the physically inequivalent BPS
states can occur,  even though they will contain states with different mass
densities. This latter fact alone tells us that we must include some transformation
that acts on the metric. We shall continue with our exploration of the continuous
classical ${\rm SL}(2,\R)$ symmetry of the type IIB theory. Finding the
surviving quantum-level ${\rm SL}(2,\Z)$ later on will be a straightforward matter
of restricting the transformations to a subgroup. The procedure starts with a
standard ${\rm SL}(2,\R)$ transformation, which transforms the doublet charges
(\ref{qel}) in a straightforwardly linear fashion, but which also transforms in an
unwanted way the scalar moduli. Subsequent compensating transformations will then
have the task of eliminating the unwanted transformation of the scalar moduli, but
without changing the ``already final'' values of the charges. Let us suppose that
this initial transformation, with parameter $\Lambda$, maps the charge vector and
complex scalar modulus $(Q,\tau_\infty)$ to new values $(Q',\tau'_\infty)$.

     After this initial $\Lambda$ transformation, one wishes to return the
complex scalar modulus $\tau'_\infty$ to its original value $\tau_\infty$, in order
to obtain an overall transformation that does not in the end disturb the complex
modulus. To do this, notice that within ${\rm SL}(2,\R)$ there is a
subgroup that leaves a doublet charge vector $Q'$ invariant up to an overall
rescaling. This projective stability group of $Q'$ is isomorphic to the {\bf
Borel} subgroup of ${\rm SL}(2,\R)$:
\be
{\bf Borel} = \left\{\pmatrix{a&b\cr0&a^{-1}}\ \hbox{$\Big|$}\ a,b
\in\R\right\}\ .\label{borel}
\ee
This standard representation of the ${\rm SL}(2,\R)$ Borel subgroup clearly
leaves the basis charge vector $\bmath{e}_1$ of Eq.\ (\ref{basistates}) invariant
up to scaling by $a$. For a general charge vector $Q'$, there will exist a
corresponding projective stability subgroup which is isomorphic to (\ref{borel}),
but obtained by conjugation of (\ref{borel}) with an element of $H\isomorphic {\rm
SO}(2)$. The importance of the Borel subgroup for our present purposes is that it
acts {\em transitively} on the
$\ffrac{$G$}{$H$} = \ffrac{${\rm SL}(2,\R)$}{${\rm SO}(2)$}$ coset space in which
the scalar fields take their values, so this transformation may be used to return
the scalar moduli to the original values they had before the $\Lambda$
transformation.

     The next step in the construction is to correct for the unwanted scaling
$Q'\rightarrow aQ'$ which occurs as a result of the Borel compensating
transformation, by use of a further compensating scaling of the form
(\ref{trombone}), $aQ'\rightarrow\lambda^2aQ'$, in which one picks the rigid
parameter $\lambda$ such that $\lambda^2a=1$. This almost completes the
construction of the active ${\rm SL}(2,\R)$. For the final step, note that the
transformation (\ref{trombone}) also scales the metric,
$g_{\mu\nu}\rightarrow\lambda^2g_{\mu\nu}=a^{-1}g_{\mu\nu}$. Since one does
not want to alter the asymptotic metric at infinity, one needs to compensate for
this scaling by a final general coordinate transformation,
$x^\mu\rightarrow x'^\mu=a^{-1/2}x^\mu$.

     The overall active ${\rm SL}(2,\R)$ duality package constructed in this way
transforms the charges in a linear fashion, $Q\rightarrow\lambda Q'$, in exactly
the same way as the standard supergravity Cremmer-Julia ${\rm SL}(2,\R)$ duality,
but now leaving the complex scalar modulus $\tau_\infty$ unchanged. This is
achieved by a net construction that acts upon the field variables of the theory in
a quite nonlinear fashion. This net transformation may be explicitly written by
noting that for ${\rm SL}(2,\R)$ there is an Iwasawa decomposition
\be
\Lambda = \tilde b h\ ,\label{iwasawasl2}
\ee  
where $\tilde b\in {\bf Borel}_{Q'}$ is an element of the projective
stability group of the final charge vector $Q'$ and where $h\in{\rm
H}_{\tau_\infty}$ is an element of the stability subgroup of $\tau_\infty$.
Clearly, the Borel transformation that is needed in this construction is just
$b=(\tilde b)^{-1}$, leaving thus a transformation $h\in{\rm H}_{\tau_\infty}$
which does not change the complex modulus $\tau_\infty$. The compensating scaling
transformation $t$ of the form (\ref{trombone}) and the associated general
coordinate transformation also leave the scalar moduli unchanged. The net active
${\rm SL}(2,\R)$ transformation thus is just $bt\Lambda=th$. Specifically, for
$\tau_\infty=\chi_\infty+\im e^{-\phi\infty}$ and a transformation $\Lambda$
mapping $Q_{\rm i}=\pmatrix{p_{\rm i}\cr q_{\rm i}}$ to
$Q_{\rm f}=\pmatrix{p_{\rm f}\cr q_{\rm f}}=\Lambda Q_{\rm i}$, the
$h\in H_{\tau_\infty}$ group element is
\bea h_{\rm f{}i} = V_\infty\pmatrix{\cos\tilde\theta_{\rm
f{}i}&\sin\tilde\theta_{\rm f{}i}\cr -\sin\tilde\theta_{\rm
f{}i}&\cos\tilde\theta_{\rm f{}i}}V_\infty^{-1}\ ,\phantom{\matrix{I\cr I\cr
I}}\hspace{.3cm}&&
\tilde\theta_{\rm f{}i}=\tilde\theta_{\rm f}-\tilde\theta_{\rm i}\nonumber\\
\tan\tilde\theta_{\rm i}=e^{\phi_\infty}(\tan\theta_{\rm i}-\chi_\infty)\
,\hspace{1.3cm}&&\tan\theta_{\rm i}=p_{\rm i}/q_{\rm i}\ ,\label{hfi}
\eea 
where the matrix $V_\infty$ is an element of {\bf Borel} that has the effect
of moving the scalar modulus from the point $\tau=\im$ to the point $\tau_\infty$:
\be 
V_\infty=e^{-\phi_\infty/2}\pmatrix{1&e^{\phi_\infty}\chi_\infty\cr
0&e^{\phi_\infty}}\ .
\ee 
The matrix $V_\infty$ appearing here is also the asymptotic limit of a matrix
$V(\phi,\chi)$ that serves to factorize the matrix $\cal M$ given in
(\ref{scalmat}), $M=VV^{\rm T}$. This factorization makes plain the transitive
action of the Borel subgroup on the ${\rm SL}(2,\R)/{\rm SO}(2)$ coset space
in which the scalar fields take their values. Note that the matrix $\cal M$
determines both the scalar kinetic terms and also their interactions with the
various antisymmetric-tensor gauge fields appearing in the action (\ref{2blag}).

     The scaling-transformation part of the net active ${\rm SL}(2,\R)$
construction is simply expressed as a ratio of mass densities, 
\be
t_{\rm f{}i}={m_{\rm f}\over m_{\rm i}}\ , \hspace{1cm} m_{\rm i}^2=Q_{\rm i}^{\rm
T}{\cal M}_\infty^{-1}Q_{\rm i}\ .\label{trombfi}
\ee
This expression reflects the fact that the scaling symmetry (\ref{trombone}) acts
on the metric and thus enables the active ${\rm SL}(2,\R)$ transformation to relate
solutions at different mass-density levels $m_{\rm i,f}$. Since, by contrast, the
mass-density levels are invariant under the action of the standard ${\rm
SL}(2,\R)$, it is clear that the two realizations of this group are distinctly
different. Mapping between different mass levels, referred to a given scalar vacuum
determined by the complex modulus $\tau_\infty$, can only be achieved by
including the scaling transformation (\ref{trombfi}).

     The group composition property of the active ${\rm SL}(2,\R)$ symmetry needs
to be checked in the same fashion as for nonlinear realizations generally, \ie
one needs to check that a group operation ${\cal O}(\Lambda,Q)=th$ acting on an
initial state characterized by a charge doublet $Q$ combines with a second group
operation according to the rule
\be
{\cal O}(\Lambda_2,\Lambda_1Q){\cal O}(\Lambda_1,Q)={\cal
O}(\Lambda_2\Lambda_1,Q)\ .
\ee
One may verify directly that the nonlinear realization given by
(\ref{hfi},\,\ref{trombfi}) does in fact satisfy this composition law, when acting
on any of the fields of the type IIB theory.

     At the quantum level, the Dirac quantization
condition restricts the allowed states of the theory to a discrete
charge lattice, as we have seen. The standard ${\rm SL}(2,\R)$
symmetry thus becomes restricted to a discrete ${\rm SL}(2,\Z)$ subgroup
in order to respect this charge lattice, and the active ${\rm SL}(2,\R)$
constructed above likewise becomes restricted to an ${\rm SL}(2,\Z)$ subgroup. This
quantum-level discretised group of active transformations is obtained simply by
restricting the matrix parameters $\Lambda$ for a classical active ${\rm
SL}(2,\R)$ transformation so as to lie in ${\rm SL}(2,\Z)$.

     In lower-dimensional spacetime, the supergravity duality
groups $G$ shown in Table \ref{tab:sugrasyms} grow in rank and the structure of the
charge orbits becomes progressively more and more complicated, but the above story
is basically repeated for an important class of $p$-brane solutions. This is the
class of single-charge solutions, for which the charges $Q$ fall into {\em
highest-weight} representations of $G$. The duality groups shown in Table
\ref{tab:sugrasyms} are all maximally noncompact, and possess an Iwasawa
decomposition generalizing the
${\rm SL}(2,\R)$ case (\ref{iwasawasl2}):
\be
\Lambda=\tilde bh\;\hspace{1cm}\tilde b\in{\bf Borel}_Q\ ,\ h\in H_{\rm moduli}\ ,
\ee
where ${\bf Borel}_Q$ is isomorphic to the Borel subgroup of $G$. Once again,
this subgroup acts transitively on the coset space \ffrac{$G$}{$H$} in which the
scalar fields take their values, so this is the correct subgroup to use for a
compensating transformation to restore the moduli to their original values
in a given scalar vacuum. As in the ${\rm SL}(2,\R)$ example of the type IIB
theory, one may see that this group action is transitive by noting that the matrix
$\cal M$ (\ref{scalmat}) which governs the scalar kinetic terms and
interactions can be parameterized in the form ${\cal M}=VV^\#$, where $V$ is an
element of the Borel subgroup. The operation $\#$ here depends on the groups $G$
and $H$ in question; in spacetime dimensions $D\ge4$ we have
\be
V^\# =\cases{V^{\rm T},&for $H$ orthogonal\cr
V^{\dag},&for $H$ unitary\cr
\Omega V^{\dag},&for $H$ a $\rm USp$ group.}
\ee
(The $D=3$ case in which $G=E_{8(+8)}$ and $H={\rm SO}(16)$ needs to be treated as
a special case.\cite{cjlp})

     Given the above group-theoretical structure, the construction of active $G$
symmetry transformations that preserve the scalar moduli proceeds in strict analogy
with the type IIB ${\rm SL}(2,\R)$ example that we have presented. This
construction depends upon the existence of a projective stability
group~\cite{clps,cjlp} of the charge $Q$ that is isomorphic to the Borel subgroup
of $G$. This is the case whenever $Q$ transforms according to a highest-weight
representation of $G$. The BPS brane solutions with this property are the
single-charge solutions with
$\Delta=4$. As we have seen in Section \ref{sec:intersectingbranes}, BPS brane
solutions with $\Delta=4/N$ can be interpreted as coincident-charge-center cases of
intersecting-brane solutions with $N$ elements, each of which would separately be
a $\Delta=4$ solution on its own. The construction of active duality
symmetries for such multiple-charge solutions remains an open problem, for they 
have a larger class of integration constants, representing relative positions and
phases of the charge components. Only the asymptotic scalar moduli can be moved
transitively by the Borel subgroup of $G$ and, correspondingly, the 
representations carried by the charges in such multi-charge cases are not of
highest-weight type.

     The active $G(\Z)$ duality constructions work straightforwardly enough at the
classical level, but their dependence on symmetries of field equations that are
not symmetries of the corresponding actions gives a reason for caution about their
quantum durability. This may be a subject where string theory needs to
intervene with its famed ``miracles.'' Some of these miracles can be seen in
supergravity-level analyses of the persistence of BPS solutions with
arbitrary mass scales, despite the presence of apparently threatening quantum
corrections,\cite{clps} but a systematic way to understand the remarkable
identities making this possible is not known. Thus, there still remain some areas
where string theory appears to be more clever than supergravity.

\section{Non-compact $\sigma$-models, null geodesics, and harmonic
maps}\label{sec:harmonicmaps}

     A complementary approach~\cite{neukr,bgm,gal} to the analysis of brane
solutions in terms of the four $D=11$ elemental solutions presented in Section
\ref{sec:intersectingbranes} is to make a dimensional reduction until only
overall-transverse dimensions remain, and then to consider the resulting nonlinear
$\sigma$-model supporting the solution. In such a reduction, all of the worldvolume
and relative-transverse coordinates are eliminated, {\em including the time
coordinate,} which is possible because the BPS solutions are all time independent.
The two complementary approaches to the analysis of BPS brane solutions may thus be
characterised as oxidation up to the top of Figure \ref{fig:nbscan}, or reduction
down to the left edge Figure \ref{fig:nbscan}, \ie reduction down to BPS
``instantons,'' or $p=-1$ branes, with worldvolume dimension $d=0$.

     The $d=0$ instanton solutions are supported by 1-form field strengths, \ie the
derivatives of axionic scalars, $F_{[1]}=d\chi$. Taken together with the dilatonic
scalars accumulated in the process of dimensional reduction, these form a {\em
noncompact nonlinear $\sigma$-model} with a target manifold \ffrac{$G$}{$H'$},
where $G$ is the usual supergravity symmetry group shown in Table
\ref{tab:sugrasyms} for the corresponding (reduced) dimension $D$ but
$H'$ is a noncompact form of the modulus little group $H$ shown in Table
\ref{tab:sugrasyms}. The difference between the groups $H'$ and $H$ arises because
dimensional reduction on the time coordinate introduces extra minus signs, with
respect to the usual spatial-coordinate Kaluza-Klein reduction, in ``kinetic''
terms for scalars descending from vector fields in the $(D+1)$ dimensional theory
including the time dimension. Scalars descending from scalars or from the
metric in $(D+1)$ dimensions do not acquire extra minus signs. The change to the
little group $H'$ is also needed for the transformation of field strengths of
higher rank, but these need not be considered for our discussion of the BPS
instantons. The relevant groups\,\footnote{The author would like to thank Chris
Pope and Hong L\"u for pointing out an error in an earlier version of this
table.} for the noncompact $\sigma$-models in dimensions
$9\ge D\ge3$ are given in Table \ref{tab:ncompactsyms}. These should be compared to
the standard Cremmer-Julia groups given in Table \ref{tab:sugrasyms}. 

\begin{table}[ht]
\centering
\caption{Symmetries for BPS instanton $\sigma$-models.\label{tab:ncompactsyms}}
\vspace{0.4cm}
\begin{tabular}{|c|c|c|}
\hline
$D$&$G$&$H'$\\
\hline\hline
9&${\rm GL}(2,\R)$&${\rm SO}(1,1)$\\
\hline
8&${\rm SL}(3,\R)\times{\rm SL}(2,\R)$&${\rm SO}(2,1)\times{\rm
SO}(1,1)$\\
\hline
7&${\rm SL}(5,\R)$&${\rm SO}(3,2)$\\
\hline
6&${\rm SO}(5,5)$&${\rm SO}(5,\C)$\\
\hline
5&${\rm E}_{6(+6)}$&${\rm USP}(4,4)$\\
\hline
4&${\rm E}_{7(+7)}$&${\rm SU}^\ast(8)$\\
\hline
3&${\rm E}_{8(+8)}$&${\rm SO}^\ast(16)$\\
\hline
\end{tabular}
\end{table}

     The sector of dimensionally-reduced supergravity that is relevant for the
instanton solutions consists just of the transverse-space Euclidean-signature
metric and the \ffrac{$G$}{$H'$} $\sigma$-model, with an action
\be
I_\sigma=\int d^Dy\sqrt g\left(R-\ft12G_{\sst AB}(\phi)\partial_i\phi^{\sst
A}\partial_j\phi^{\sst B}g^{ij}\right)\ ,\label{ncsigma}
\ee
where the $\phi^{\sst A}$ are $\sigma$-model fields taking values in the
\ffrac{$G$}{$H'$} target space, $G_{\sst AB}$ is the target-space metric and
$g^{ij}(y)$ is the Euclidean-signature metric for the $\sigma$-model domain space.
The equations of motion following from (\ref{ncsigma}) are
\bsea
&{1\over\sqrt g}\nabla_i(\sqrt gg^{ij}G_{\sst AB}(\phi)\partial_j\phi^{\sst B}) =
0&\\ 
&R_{ij} = \ft12G_{\sst AB}(\phi)\partial_i\phi^{\sst A}\partial_j\phi^{\sst
B}\ ,&\label{sigmaeqns}
\esea
where $\nabla_i$ is a covariant derivative; when acting on a target-space vector
$V_{\sst A}$, it is given by
\be
\nabla_i V_{\sst A} = \partial_iV_{\sst A}-\Gamma^{\sst D}_{\sst
AE}(G)\partial_i\phi^{\sst E}V_{\sst D}\ ,\label{nabla}
\ee
in which $\Gamma^{\sst A}_{\sst BC}(G)$ is the Christoffel connection for the
target-space metric $G_{\sst AB}$. The action (\ref{ncsigma}) and the field equations
(\ref{sigmaeqns}) are covariant with respect to general-coordinate transformations on
the $\sigma$-model target manifold
\ffrac{$G$}{$H'$}. The action (\ref{ncsigma}) and the field equations (\ref{sigmaeqns})
are also covariant with respect to general $y^i\rightarrow y'^i$ coordinate
transformations of the domain space. These two types of general coordinate
transformations are quite different, however, in that the domain-space
transformations constitute a true gauge symmetry of the dynamical system
(\ref{ncsigma}), while the $\sigma$-model target-space transformations generally
change the metric $G_{\sst AB}(\phi^{\sst A})$ and so correspond to an actual
symmetry of (\ref{ncsigma}) only for the finite-parameter group
$G$ of target-space {\em isometries}.

     As in our original search for $p$-brane solutions given in Section
\ref{sec:pbraneans}, it is appropriate to adopt an {\em ansatz} in order to focus
the search for solutions. In the search for instanton solutions, the metric
ansatz can take a particularly simple form:
\be
g^{ij}=\delta^{ij}\ ,\label{sigmetans}
\ee
in which the domain-space metric is assumed to be {\em flat.} The $\sigma$-model
equations and domain-space gravity equations for the flat metric (\ref{sigmetans})
then become
\bsea
&\nabla^i(G_{\sst AB}(\phi)\partial_i\phi^{\sst B}) = 0&\\
&R_{ij}=\ft12G_{\sst AB}(\phi)\partial_i\phi^{\sst A}\partial_j\phi^{\sst B} = 0&\
.\label{flatsigeqns}
\esea

     Now comes the key step~\cite{neukr} in finding instanton solutions to the
specialised equations (\ref{flatsigeqns}): for single-charge solutions, one supposes
that the $\sigma$-model fields $\phi^{\sst A}$ depend on the domain-space
coordinates $y^i$ only through some intermediate scalar functions $\sigma(y)$, \ie
\be
\phi^{\sst A}(y) = \phi^{\sst A}(\sigma(y))\ .\label{harmap}
\ee
After making this assumption, the $\sigma$-model $\phi^{\sst A}$ equations
(\ref{flatsigeqns}a) become
\be
\nabla^2\sigma{d\phi^{\sst A}\over d\sigma} +
(\partial_i\sigma)(\partial_i\sigma)\left[{d^2\phi^{\sst A}\over d\sigma^2} +
\Gamma^{\sst A}_{\sst BC}(G){d\phi^{\sst B}\over d\sigma}{d\phi^{\sst C}\over d\sigma}
\right] = 0\ ,\label{intsigeqn}
\ee
while the gravitational equation (\ref{flatsigeqns}b) becomes the {\em constraint}
\be
G_{\sst AB}(\phi){d\phi^{\sst A}\over d\sigma}{d\phi^{\sst B}\over d\sigma}=0\
.\label{redconstr}
\ee
An important class of solutions to (\ref{intsigeqn}) is obtained by taking
\bsea
&\displaystyle{\nabla^2\sigma = 0}&\\
&\displaystyle{{d^2\phi^{\sst A}\over d\sigma^2} + \Gamma^{\sst A}_{\sst
BC}(G){d\phi^{\sst B}\over d\sigma}{d\phi^{\sst C}\over d\sigma} = 0}\
.&\label{redsigeqn}
\esea

\begin{figure}[ht]
\leavevmode\centering
\epsfbox{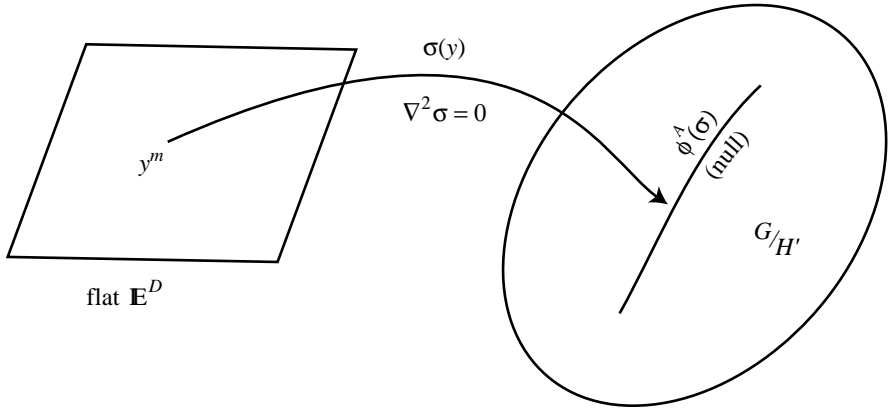}
\caption{Harmonic map from $\E^D$ to a null geodesic in
\ffrac{$G$}{$H'$}.\label{fig:harmapgeom}}
\end{figure}

     At this point, one can give a picture of the $\sigma$-model maps involved in
the system of equations (\ref{redconstr},\,\ref{redsigeqn}), noting that
(\ref{redsigeqn}a) is just Laplace's equation and that (\ref{redsigeqn}b) is the
geodesic equation on \ffrac{$G$}{$H'$}, while the constraint (\ref{redconstr})
requires the tangent vector to a geodesic to be a null vector. The
intermediate function $\sigma(y)$ is required by (\ref{redsigeqn}a) to be a {\em
harmonic} function mapping from the flat (\ref{sigmetans}) Euclidean domain space
onto a {\em null geodesic} on the target space \ffrac{$G$}{$H'$}. Clearly, the
harmonic map $\sigma(y)$ should be identified with the harmonic function $H(y)$
that controls the single-charge brane solutions (\ref{pbranesol}). On the geodesic
in \ffrac{$G$}{$H'$}, on the other hand, $\sigma$ plays the role of an affine
parameter. The importance of the noncompact structure of the target space manifold
\ffrac{$G$}{$H'$}, for the groups $G$ and $H'$ given in Table
\ref{tab:ncompactsyms}, now becomes clear: only on such a noncompact manifold does
one have nontrivial null geodesics as required by the gravitational constraint
(\ref{redconstr}). The $\sigma$-model solution (\ref{harmap}) oxidises back up to
one of the single-charge brane solutions shown in Figure
\ref{fig:nbscan}, and, conversely, any solution shown in Figure \ref{fig:nbscan}
may be reduced down to a corresponding noncompact $\sigma$-model solution of this
type. This sequence of $\sigma$-model maps is sketched in Figure
\ref{fig:harmapgeom}.

     An extension~\cite{bgm,gal} of this $\sigma$-model picture allows for
solutions involving multiple harmonic maps $\sigma_a(y)$. In that case, one deals
not with a single geodesic, but with a totally geodesic submanifold of
\ffrac{$G$}{$H'$}, and, moreover, the geodesics generated by any curve in the
intermediate $\sigma_a$ parameter space must be null. This is the $\sigma$-model
construction that generates multi-charge solutions, giving rise to
intersecting-brane solutions of the types discussed in Section
\ref{sec:intersectingbranes}. As with the intersecting-brane solutions, however,
there are important compatibility conditions that must be satisfied in order for
such multi-charge solutions to exist. We saw in subsection \ref{ssec:fourelements}
that, in order for some portion of the rigid supersymmetry to remain unbroken, the
projectors constraining the surviving supersymmetry parameter need to be
consistent. In the $\sigma$-model picture, a required condition is expressed in
terms of the velocity vectors for the null geodesics. If one adopts a matrix
representation $\bmath M$ for points in the coset manifold \ffrac{$G$}{$H'$},
the $\sigma$-model equations for the matrix fields ${\bmath M}(y^m)$ are simply
written
\be
\nabla^i({\bmath M}^{-1}\partial_i{\bmath M}) = 0\ .\label{matrixeqn}
\ee

     Points on the geodesic submanifold with affine parameters $\sigma_a$ may be
written 
\be
{\bmath M}={\bmath A}\exp(\sum_a{\bmath B}_a\sigma_a)\
,\label{multigeodesic}
\ee 
where the constant matrices ${\bmath B}_a$ give the velocities for the various
geodesics parametrised by the $\sigma_a$, while an initial point on these
geodesics is specified by the constant matrix $\bmath A$. The compatibility
condition between these velocities is given by the double-commutator
condition~\cite{gal}
\be
[[{\bmath B}_a,{\bmath B}_b],{\bmath B}_c]=0\ .\label{compcond}
\ee
This condition allows one to rewrite (\ref{multigeodesic}) as
\be
{\bmath M}={\bmath A}\exp\left(-\ft12\sum_{c>b}\sum_b[{\bmath
B}_b,{\bmath B}_c]\sigma_b\sigma_c\right)\prod_a\exp({\bmath B}_a\sigma_a)\
,\label{revgeodesic}
\ee
where the first factor commutes with the ${\bmath B}_a$ as a result of
(\ref{compcond}). The matrix current then becomes
\be
{\bmath M}^{-1}\partial_i{\bmath M} = \sum_a{\bmath
B}_a\partial_i\sigma_a - \ft12\sum_{c>b}\sum_b[{\bmath B}_b,{\bmath
B}_c](\sigma_b\partial_i\sigma_c-\sigma_c\partial_i\sigma_b)\ ,\label{matcurrent}
\ee
and this is then seen to be conserved provided the $\sigma_a$ satisfy
$\nabla^2\sigma_a(y)=0$, \ie they are harmonic maps from the Euclidean
overall-transverse space of the $y^m$ into the geodesic submanifold
(\ref{multigeodesic}). The constraint imposed by the gravitational equation is
\be
R_{ij}=\ft14\sum_{a,b}{\rm tr}({\bmath B}_a{\bmath
B}_b)\partial_i\sigma_a\partial_j\sigma_b=0\ ,\label{matgrav}
\ee
which is satisfied provided the geodesics parametrised by the $\sigma_a$
are null and orthogonal, \ie
\be
{\rm tr}({\bmath B}_a{\bmath B}_b)=0\ .\label{orthonull}
\ee

     The general set of stationary multi-charge brane solutions is thus
obtained in the $\sigma$-model construction by identifying the set of totally null,
totally geodesic submanifolds of \ffrac{$G$}{$H'$} such that the velocity vectors
satisfy the compatibility condition (\ref{compcond}).

     Aside from the elegance of the above $\sigma$-model picture of the
equations governing BPS brane solutions, these constructions make quite clear
the places where assumptions have been made that are more stringent than are
really necessary. One example of this is the assumption that the transverse-space
geometry is flat, Eq.\ (\ref{sigmetans}). This is clearly more restrictive than is
really necessary; one could just as well have a more general Ricci-flat
domain-space geometry, with a correspondingly covariantised constraint for the
null geodesics on the noncompact manifold \ffrac{$G$}{$H'$}. The use of more
general Ricci-flat transverse geometries is at the basis of ``generalised
$p$-brane'' solutions that have been considered in Refs.\,\cite{beckers,nonperpint}

\section{Concluding remarks}\label{sec:concl}

     In this review, we have discussed principally the structure of
classical $p$-brane solutions to supergravity theories. Some topics that deserve a
fuller treatment have only been touched upon here. For example, the worldvolume
symmetries of $p$-brane sources, and in particular the important subject of
$\kappa$ symmetry, which bridges the gap between the full target-space
supersymmetry of the ambient supergravity theory and the fractional supersymmetry
surviving in the BPS brane background, have only been touched upon. For a fuller
treatment, the reader is referred to Refs~\cite{2bror1,dkl}, or to the more recent
discussions of $\kappa$-symmetric actions for cases involving R--R sector
antisymmetric-tensor fields.\cite{kappa}

     Another aspect of the $p$-brane story, which we have only briefly presented
here in Section \ref{sec:intersectingbranes}, is the large family of intersecting
branes. These include~\cite{nonperpint} also intersections at angles other than
$90^\circ$, and can involve fractions of preserved supersymmetry other than
inverse powers of 2. For a fuller treatment of some of these subjects, the reader
is referred to Ref.\,\cite{intersecting,gauntlett}, and for the implications of
charge conservation in determining the allowed intersections to
Refs.~\cite{branesurgery}

     Yet another aspect of this subject that we have not dwelt upon here is the
intrinsically string-theoretic side, in which some of the BPS supergravity solutions
that we have discussed appear as Dirichlet surfaces on which open strings can end; for
this, we refer the reader to Ref.\,\cite{jp}

     Of course, the real fascination of this subject lies in its
connection to the emerging picture in string theory/quantum gravity,
and in particular to the r\^oles that BPS supergravity solutions play as states
stable against the effects of quantum corrections. In this emerging picture, the
duality symmetries that we have discussed in Section \ref{sec:dualities} play an
essential part, uniting the underlying type IIA, IIB, $E_8\times E_8$ and ${\rm
SO}(32)$ heterotic, and also the type I string theories into one overall theory,
which then also has a phase with $D=11$ supergravity as its field-theory limit. The
usefulness of classical supergravity considerations in probing the structure of
this emerging ``M theory'' is one of the major surprises of the subject.

\section*{Acknowledgments}
The author would like to acknowledge helpful conversations with Marcus
Bremer, Bernard de Wit, Mike Duff, Fran\c{c}ois Englert, Gary Gibbons, Hong L\"u,
George Papadopoulos, Chris Pope, and Paul Townsend. The author would like to thank
The Abdus Salam International Center for Theoretical Physics for the invitations to
give lectures at two successive summer schools, on which this review is based, and
also CERN, SISSA, the Ecole Normale Sup\'erieure, UCLA and the University of
Pennsylvania for hospitality at different periods during the writing. This work
was supported in part by the Commission of the European Communities under contract
ERBFMRX-CT96-0045.

\end{document}